\documentclass[a4paper,11pt]{article}
\usepackage{jheppub}
\usepackage{amsmath,amssymb,slashed}
\usepackage[utf8]{inputenc}
\usepackage{graphicx}
\usepackage{hyperref}
\usepackage{bm,bbold}
\usepackage{accents}
\usepackage{xcolor}
\def\!{\mskip-\thinmuskip}

\newcommand{\di}{{\rm d}}

\newcommand{\wT}{\widehat{T}}
\newcommand{\wj}{\widehat{j}}
\newcommand{\wQ}{\widehat{Q}}

\newcommand{\wO}{\widehat{O}}
\newcommand{\wrho}{\widehat{\rho}}

\newcommand{\tr}{{\rm tr}}  
\newcommand{\e}{{\rm e}}

\newcommand{\SP}{\mathfrak{S}}
\newcommand{\wS}{\widehat{S}}

\def\wO{\widehat O}
\def\wQ{\widehat Q}

\def\wS{\widehat S}

\def\opsi{\overline{\psi}}

\def\wPi{\widehat \Pi}

\newcommand{\Tr}{{\rm Tr}}

\newcommand{\ii}{{\rm i}}

\newcommand{\hnku}{{\rm H}^{(1)}}
\newcommand{\hnkd}{{\rm H}^{(2)}}

\def\ua{{\underline{a}}}
\def\ub{{\underline{b}}}

\def\uzero{{\underline{0}}}
\def\uone{{\underline{1}}}
\def\utwo{{\underline{2}}}
\def\uthree{{\underline{3}}}
\def\I{{\mathbb{1}}}
\def\fp{{\mathfrak{p}}}

\def\fr{{\mathfrak{r}}}

\arxivnumber{}
\title{ \boldmath Exact expectation values in a boost-invariant fluid of Dirac fermions with finite spin density}

\author[1]{Andrea Palermo}
\affiliation[1]{Institut Denis Poisson, CNRS UMR 7013, Universit\'e de Tours, Universit\'e d'Orl\'eans,\\ 
Parc de Grandmont, Tours, 37200, France}
\emailAdd{andrea.palermo@univ-tours.fr}

\author[2]{Daniele Roselli}
\affiliation[2]{Universit\`a degli studi di Firenze and INFN Sezione di Firenze,\\
Via G. Sansone 1, I-50019 Sesto Fiorentino (Florence), Italy}
\emailAdd{daniele.roselli@unifi.it}

\abstract{
   We study a boost-invariant, out-of-equilibrium fluid of non-interacting Dirac fermions with a finite canonical spin potential. After solving the Dirac equation in Milne coordinates, we exactly diagonalize the non-equilibrium density operator and compute the partition function and expectation values of relevant observables, including spin polarization, energy density, longitudinal and transverse pressures, spin density, and \emph{spin torque}, i.e.\ the source of spin non-conservation. We find an analytic expression for the partition function at finite spin potential, and show numerically that thermodynamic relations connecting it to thermodynamic functions hold in the system under consideration.
    We show that, in a boost-invariant system, both shear-induced polarization and the spin Hall effect are absent, and that a non-vanishing polarization can only arise from a finite spin potential in a free theory. We obtain an analytic expression for the spin polarization as a function of the spin potential in some particular cases, and otherwise compute numerically its exact expectation value at finite spin potential. Our results are discussed in the context of relativistic spin hydrodynamics and quark--gluon plasma phenomenology.
}

\begin{document}
\maketitle
\flushbottom

\section{Introduction}
The measurement of spin polarization in the Quark-Gluon Plasma (QGP) produced in heavy-ion collisions \cite{STAR:2017ckg, STAR:2018gyt} provides a unique opportunity to probe a genuinely quantum property of matter, such as spin, within a fluid-dynamical regime, see e.g. \cite{Becattini:2024uha,Niida:2024ntm} for recent reviews. For systems in thermal equilibrium, polarization is sourced by angular velocity and acceleration, which are described by the (constant) thermal vorticity tensor $\varpi_{\mu\nu}=(1/2)(\partial_\nu\beta_\mu-\partial_\mu\beta_\nu)$ \cite{Becattini:2013fla},
where $\beta_\mu=u_\mu/T$ is the four-temperature vector, with $u_\mu$ the fluid four-velocity, and $T$ the local temperature. This framework was initially very successful in describing the dependence of spin polarization on the collision energy $\sqrt{s_{NN}}$ \cite{Karpenko:2016jyx, Li:2017slc, Xie:2017upb}.
However, thermal vorticity alone cannot account for the observed momentum dependence of polarization, leading to the so-called \emph{polarization sign puzzle} \cite{Becattini:2020ngo, Xia:2018tes, Florkowski:2019voj, Wu:2019eyi, Becattini:2019ntv}. 

This puzzle has stimulated extensive theoretical work. Among the proposed solutions, the idea that the spin tensor should be included as an independent dynamical quantity in hydrodynamics has gained more and more attention \cite{Becattini:2018duy, Hattori:2019lfp, Bhadury:2020puc, Shi:2020htn, Florkowski:2017ruc,Florkowski:2018fap,Florkowski:2019qdp, Gallegos:2021bzp, Fukushima:2020ucl, Hongo:2021, Peng:2021ago, Chiarini:2024cuv}.  This idea has evolved into the framework of \emph{spin hydrodynamics}, where spin degrees of freedom are described independently of the fluid temperature and velocity (see also \cite{Huang:2024ffg} for a detailed review of the subject). This is possible thanks to an additional antisymmetric tensor,  referred to in the literature as the \emph{spin potential} $\SP_{\mu\nu}$, which is introduced as a new Lagrange multiplier associated with the spin tensor. At leading order, the spin potential evolves independently of other thermodynamic fields and eventually relaxes to the thermal vorticity. In fact, the difference $\SP_{\mu\nu}-\varpi_{\mu\nu}$ can be non-zero only out of equilibrium \cite{BecattiniBOOK}.
Although possible resolutions of the polarization sign puzzle have emerged from non-equilibrium quantum effects, such as the shear-induced polarization and the spin Hall effect \cite{Becattini:2021suc, Liu:2021uhn, Fu:2022myl}, the development of spin hydrodynamics has been remarkable, and interest in the subject continues to grow.

Nevertheless, several conceptual and practical challenges remain. Constitutive relations for expectation values are still largely unknown and are often postulated based on power-counting arguments and educated ans\"atze. The introduction of a spin potential raises questions about possible modifications of standard thermodynamic relations, which are currently under active investigation \cite{Becattini:2025oyi, Florkowski:2024bfw, Drogosz:2024gzv,Singha:2025bda,Braguta:2025ddq,Ambrus:2025dca,Florkowski:2026ofs,Armas:2026bmw}. Finally, the framework depends on the choice of energy-momentum and spin tensors, the so-called \emph{pseudogauge ambiguity}, which can significantly affect out-of-equilibrium expectation values \cite{Buzzegoli:2021wlg, Li:2020eon, Drogosz:2024rbd}.

Since a non-vanishing and independent spin potential necessarily implies a departure from equilibrium, the study of relativistic fluids with finite spin potential is highly non-trivial. In this work, we consider a quantum relativistic fluid of non-interacting Dirac fermions in the presence of a finite spin potential, within a longitudinally boost-invariant setup.
Longitudinal boost invariance has often been employed as an approximate symmetry of high-energy heavy-ion collisions since the seminal work of Bjorken \cite{Bjorken:1982qr}, and it provides an effective description of the mid-rapidity region in central ultra-relativistic collisions. Spin hydrodynamics in boost-invariant settings has been explored in several works, often as a simplified framework to test theoretical ideas \cite{Florkowski:2019qdp,Singh:2020rht,Wang:2021ngp,Biswas:2022bht,Drogosz:2024lkx,Singh:2026wvf,Singh:2026ytd}.  In this work, by imposing longitudinal boost invariance also on the density operator describing non-interacting Dirac fermions with a finite spin potential, we are able to compute exactly all relevant expectation values, including the components of the energy-momentum tensor, the spin tensor, and the spin polarization.

Our results provide, to the best of our knowledge, the first exact expression for the spin polarization at local thermodynamic equilibrium in the presence of a finite spin potential. This result offers a natural benchmark for comparison with existing formulas derived within linear response theory, and allows one to explore regimes of large spin potential. 
Furthermore, our approach enables the exact computation of local-equilibrium expectation values of both the stress-energy tensor and the spin tensor, which completely solves the thermodynamics of the system. 

The paper is organized as follows. In Section~\ref{sec: zubarev} we describe the statistical operator describing a local-equilibrium state for a boost-invariant system and introduce the physical observables that will be studied, namely the components of the conserved densities and the spin polarization vector. In Section~\ref{Sec: Dirac Milne} we solve the Dirac equation in Milne coordinates, demonstrate its equivalence to the standard formulation in Cartesian coordinates, and express thermal expectation values in terms of creation and annihilation operators of Milne modes. This is done for the stress-energy and spin tensors at local equilibrium, as well as for the spin density matrix and the spin polarization vector. 
In Sections~\ref{sec: bel} and \ref{sec: can}, we evaluate these quantities explicitly in the Belinfante and canonical pseudogauges, obtaining exact formulas for the local-equilibrium expectation values of the conserved densities as well as for the spin polarization.  We analyse our results numerically in section~\ref{Sec: Analysis}, before winding up and discussing implications of our work in Section~\ref{sec: Conclusions}.


\subsection*{Notation}
In this work we use natural units $\hbar=c=k_{\rm B}=1$. Operators on a Hilbert space are denoted with a \emph{widehat}, e.g. $\wO$, while vectors with normal unit with an \emph{hat}, e.g $\hat{n}_\mu$. The only exception is the Dirac field operator, which is denoted by $\psi$. 

Greek indices denote curvilinear, non-Cartesian, components and have values $\mu,\nu,\lambda,\dots=0,1,2,3$, while Latin indices denote only the spatial part $i,j,k\dots=1,2,3$.  
Underlined Latin indices refer to space-time Cartesian coordinates $\ua,\ub,\underline{c},\dots=\uzero,\uone,\utwo,\uthree$. The same applies for the $\gamma$ matrices: $\gamma^\ua$ refers to Cartesian and $\gamma^\mu$ to curvilinear coordinates and they are related through the vierbein $e^\mu_\ua\gamma^\ua=\gamma^\mu$. The fifth $\gamma^5$ is independent from the coordinates and thus is indicated without underline.  

For the spin connection and spin covariant derivative we adopt the +1 convention in \cite{Collas:2018jfx}. The covariant derivative for a spinor field $\psi$ is indicated with $D_\mu\psi=\partial_\mu\psi+\Gamma_\mu\psi$, with $\Gamma_\mu=(1/4)\omega^{\ \ua\ub}_\mu\gamma_\ua\gamma_\ub$ and $\omega$ is the \emph{spin-connection one form}. The usual covariant derivative of a four-vector instead is $\nabla_\mu V^\nu=\partial_\mu V^\nu-\Gamma^{\nu}_{\mu\lambda}V^\lambda$ where $\Gamma^{\nu}_{\mu\lambda}$ are the \emph{Christoffel symbols} and must not be confused with the spin connection $\Gamma_\mu$.   

With $g_{\mu\nu}$ we indicate the metric in curvilinear coordinates while with $\eta_{\ua\ub}$ the Minkowksi metric in Cartesian coordinates.
We use the mostly minus metric signature, such that in Cartesian coordinates $\eta^{\ua\ub}=\rm{diag}(1,-1,-1,-1)$. With $g$ we denote the determinant of the metric. 
The Levi-Civita pseudo-tensor is defined $\epsilon^{0123}=\frac{1}{\sqrt{-g}}\epsilon^{\uzero\uone\utwo\uthree}=+1$. 

The symbol $\Tr$ will denote the trace over the quantum states whereas the symbol $\tr$ will indicate the trace over finite spaces, for instance spin or spinor indices.

\section{Statistical operator of a longitudinally boost-invariant fluid}\label{sec: zubarev}

A quantum-relativistic system which reaches local thermodynamic equilibrium on a space-like hypersurface $\Sigma(\tau)$ is known to be described by the so-called \emph{local equilibrium operator} \cite{Becattini:2019dxo}. For a system with a finite spin density and a global conserved charge, the local equilibrium density operator is constructed employing Von Neumann procedure of maximization of entropy $S=-\Tr\left[\wrho_{\rm LE}\ln\wrho_{\rm LE}\right]$, taking into account the constraints:
\begin{equation}\label{eq: local state}
    \begin{split}
        \hat{n}_\mu\Tr\left[\wrho_{\rm LE}\wT^{\mu\nu}(x)\right]&\equiv\hat{n}_\mu\langle\wT^{\mu\nu}(x)\rangle_{\rm LE}=\hat{n}_\mu T^{\mu\nu}(x)\;,\\
        \hat{n}_\mu\Tr\left[\wrho_{\rm LE}\wj^{\mu}(x)\right]&\equiv\hat{n}_\mu\langle\wj^{\mu}(x)\rangle_{\rm LE}=\hat{n}_\mu j^{\mu}(x)\;,\\
        \hat{n}_\mu\Tr\left[\wrho_{\rm LE}\wS^{\mu,\lambda\nu}(x)\right]&\equiv\hat{n}_\mu\langle\wS^{\mu,\lambda\nu}(x)\rangle_{\rm LE}=\hat{n}_\mu S^{\mu,\lambda\nu}(x)\;,
    \end{split}
\end{equation}
where $\hat{n}_\mu$ is the orthonormal vector to the hypersurface $\Sigma$ identifying the observer and $T^{\mu\nu}(x)$, $j^\mu(x)$, and $S^{\mu,\lambda\nu}(x)$ are the local profile of energy-momentum, charge-current and spin densities.
Note that all the expectation values on the left hand side of the above identities must be formally renormalized, even in a free theory.

The above-described procedure yields the operator:
\begin{equation}\label{def: LTE1}
\begin{split}
    \wrho_{\rm LE}(\tau) &= \frac{1}{Z_{\rm LE}}e^{-\widehat{\Upsilon}(\tau)}\;,\\
    \widehat{\Upsilon}(\tau)&\equiv \int_{\Sigma(\tau)} \di\Sigma_\mu(y)\left[\wT^{\mu\nu}(y)\beta_\nu(y) -\SP_{\rho\sigma}(y)\wS^{\mu,\rho\sigma}(y)-\zeta(y)\,\wj^\mu(y)\right]\;,
\end{split}
\end{equation}
with:
\begin{equation*}
    Z_{\rm LE}=\Tr\left[\exp\left(-\widehat{\Upsilon}(\tau)\right)\right]\;,
\end{equation*}
being partition function. The thermodynamic fields $\beta$ and $\zeta$ have the familiar physical meaning of four-temperature $\beta=u/T$ and reduced chemical potential $\zeta=\mu/T$ with $T$ proper temperature and $u$ and $\mu$ four-velocity and chemical potential respectively. The additional Lagrange multiplier $\SP$ is referred to in the literature as \emph{spin potential} and it is dimensionless in natural units, and constrains the system to have a finite average spin tensor expectation value. However, for reasons that will become clear later on, we refer to $\SP$ as \emph{reduced spin potential} from now on.

The $\widehat{\Upsilon}$ operator is built in terms of the macroscopic densities which satisfy the following continuity equations:
 \begin{subequations}\label{eq: continuity}
     \begin{align}
         \nabla_\mu\wT^{\mu\nu}(x)&=0\;,\\
         \nabla_\mu\wj^\mu(x)&=0\;,\\
         \nabla_\lambda\wS^{\lambda,\mu\nu}(x)&=\wT^{\nu\mu}(x)-\wT^{\mu\nu}(x)\;,
     \end{align}
 \end{subequations}
where the usage of the covariant derivative $\nabla$ applies in the case of curvilinear coordinates.
 Both the stress-energy tensor and the four-current are conserved whereas the divergence of the spin tensor depends on the anti-symmetric part of the stress-energy tensor and is in general non vanishing.

The local equilibrium operator \eqref{def: LTE1}, despite reproducing the local thermodynamic state, does not describe the true state of the system. Indeed, eq.~\eqref{def: LTE1} explicitly depends on time $\tau$ through the hypersurface $\Sigma$, which is not the case for the true density operator in Heisenberg picture. The true non-equilibrium state for a system that is known to be at local equilibrium on some hypersurface $\Sigma(\tau_0)$ is the so-called \emph{Zubarev operator}\footnote{The difference between the true non-equilibrium operator and the local equilibrium one \eqref{def: LTE1} is that the former contains information about dissipative as well as memory effects related with the initial state \cite{Becattini:2019dxo}.}:
\begin{equation}\label{def: ZUB1}
    \wrho\equiv\wrho_{\rm LE}(\tau_0) = \frac{1}{Z_{\rm LE}(\tau_0)}e^{-\widehat{\Upsilon}(\tau_0)}\;.
\end{equation}
Being defined at some fixed time $\tau_0$, eq.~\eqref{def: ZUB1} clearly satisfies the Liouville equation. The \emph{true} non-equilibrium expectation values are thus computed with \eqref{def: ZUB1} and they constitute the r.h.s of the matching conditions \eqref{eq: local state}.

The discussion carried out so far is completely general, and without any other assumption calculations are only possible using linear response theory for small or slowly varying thermo-hydrodynamic fields ~\cite{Becattini:2019dxo, Li:2025pef}. However, additional requirements of symmetry allow for exact calculations both in flat and in curved spacetimes, see ref.~\cite{Akkelin:2018hpu, Akkelin:2020cfs,Rindori:2021quq, Becattini:2022bia, Becattini:2024vtf, Tinti:2023mtv}.
Among the possible symmetries, longitudinal\footnote{Throughout this work, the longitudinal direction will be identified with the z-axis.} boost invariance has played a prominent role as an approximate symmetry of heavy-ion collisions, in particular for very high energy central collisions. This symmetry strongly constrains the form of the non-equilibrium density operator and allows for the calculation of exact expectation values, a problem that has been extensively studied for the case of a scalar field \cite{Akkelin:2018hpu,Akkelin:2020cfs,Rindori:2021quq}. 

As it was shown in ref.~\cite{Rindori:2021quq}, imposing longitudinal boost invariance on the statistical operator forces the hypersurface, the four-temperature, and the reduced chemical potential to be invariant under such a transformation. These conditions are fulfilled if the hypersurface $\Sigma$ is defined at constant proper time $\tau=\sqrt{t^2-z^2}$, if all scalars depend on coordinates only as $T(\tau)$ and $\zeta(\tau)$, and if the four-velocity is $u=\partial_\tau$. Namely:
\begin{equation}\label{eq: boost invariant conditions}
    \Sigma:\ \tau=\text{const.}\;, \qquad \beta^\mu=\frac{u^\mu}{T(\tau)}\;, 
    \qquad u^\mu =\frac{1}{\tau}(t,0,0,z)\;, 
    \qquad\zeta=\zeta(\tau)\;.
\end{equation}
With the inclusion of the reduced spin potential $\SP$ in the density operator, we also have an additional invariance condition to be imposed:
\begin{equation*}
    \SP^{\rho\sigma}(\tau)=(\Lambda_z)^{\rho}_{\ \mu} (\Lambda_z)^{\sigma}_{\ \nu} \SP^{\mu\nu}(\tau)\;,
\end{equation*}
where $\Lambda_z$ is a longitudinal boost. In order to satisfy the above condition, the reduced spin potential can only be:
\begin{equation}\label{eq: spin potential condition}
    \SP_{\rho\sigma}=\frac{1}{2}\SP(\tau)\left(\delta^1_\rho\delta^2_\sigma-\delta^2_\rho\delta^1_\sigma\right).
\end{equation}
This is most easily seen noticing that the above reduced spin potential is proportional to the generator of rotations along the $z$-axis in the vector representation of the Lorentz group, $(\widehat{\mathrm{J}}^z)_{\mu\nu}=i\left(\delta^1_\mu\delta^2_\nu-\delta^2_\nu\delta^1_\mu\right)$, which commutes with the longitudinal boost generator $\widehat{\mathrm{K}}_z$. A term $\SP\propto \widehat{\mathrm{K}}_z$ is ruled out, since $\SP$ must be antisymmetric. 

To better deal with the longitudinal boost symmetry, it is most convenient to abandon Cartesian coordinates and adopt Milne's, which are more natural in our setting. Therefore, we perform the coordinate transformation: 
\begin{equation}\label{Milne Chart}
    \begin{alignedat}{2}
         t&= \tau\cosh\eta\;,&\qquad z&=\tau\sinh\eta\;,\\
    \tau&=\sqrt{t^2-z^2}\;,&\qquad \eta&=\frac{1}{2}\ln\frac{t+z}{t-z}\;,
\end{alignedat}
\end{equation}
where $\tau$ is the proper time and $\eta$ is the space-time rapidity. In terms of the coordinates \eqref{Milne Chart} the metric turns out to be:
\begin{equation}\label{Metrica Milne}
    \di s^2=\di\tau^2-\di {\bf x}^2_{\rm T}-\tau^2\di\eta^2\;,
\end{equation}
with ${\bf x}_{\rm T}=\left(0,x,y,0\right)$ being the transverse direction. See also appendix~\ref{app:dirac eq} for more details about Milne coordinates. The coordinates \eqref{Milne Chart} cover the future (past) light-cone of the Minkowksi space-time. From eq.~\eqref{Metrica Milne} it's immediate to see how this coordinates are not-inertial and the metric is not-static, because it explicitly depends on $\tau$. Hence is not possible to define any time-like Killing vector related to the metric \eqref{Metrica Milne}. This corresponds to the well known fact that a Bjorken flow represents an out-of equilibrium configuration. Furthermore, precisely for this reason, it will also be impossible to consider the limit of the reduced spin potential relaxing to the thermal vorticity, as this would break boost invariance: the thermal vorticity must be zero in a Bjorken flow. On the other hand, thermal shear is non vanishing.

More precisely, thermal vorticity $\varpi$ and thermal shear $\xi$ are defined respectively as:
\begin{equation}
    \varpi_{\mu\nu}
=
-\frac12
\left(
\nabla_\mu \beta_\nu
-
\nabla_\nu \beta_\mu
\right)\;,
\qquad
\xi_{\mu\nu}
=
\frac12
\left(
\nabla_\mu \beta_\nu
+
\nabla_\nu \beta_\mu
\right)\;.
\end{equation}
Using $\nabla_\mu\beta_\nu=\partial_\mu\beta_\nu-\Gamma^{\lambda}_{\mu\nu}\beta_\lambda$ with Christoffel symbols given in \eqref{eq: Christoffel} and the fact that in Milne coordinates the four-velocity is constant, $u_\mu=(1,{\bf 0})$\footnote{The assumption of boost-invariant fluid is equivalent to the request that the fluid is at rest in the Milne coordinates \cite{Bagchi:2023ysc}.}, we get that the only non-vanishing term of the gradient are $\nabla_\tau\beta_\tau$ and $\nabla_\eta\beta_\eta$ which read:
\begin{equation*}
    \nabla_\tau\beta_\tau(\tau)=\partial_\tau\beta(\tau)=-\frac{\dot{T}(\tau)}{T^2(\tau)}\;,\qquad\nabla_\eta\beta_\eta(\tau)=-\Gamma^\tau_{\eta\eta}\beta_\tau(\tau)=-\frac{\tau}{T(\tau)}\;,
\end{equation*}
where the dot denotes derivation with respect to the proper time $\tau$.
Therefore:
\begin{equation}\label{eq: varpi e xi}
    \varpi_{\mu\nu}(\tau)=0\,,\qquad\xi_{\mu\nu}(\tau)={\rm diag}\left(-\frac{\dot{T}(\tau)}{T^2(\tau)},0,0,-\frac{\tau}{T(\tau)}\right)\;.
\end{equation}

The integration measure in Milne coordinates becomes:
\begin{equation*}
    \di\Sigma_\mu=\sqrt{-g}\,\di\eta\,\di{\rm x}^2_{\rm T}\,\hat{\tau}_\mu\,=\tau\,\di\eta\,\di{\rm x}^2_{\rm T}\,\hat{\tau}_\mu\;,
\end{equation*}
with $\hat{\tau}_\mu=\delta_{\mu0}$ normal versor in the $\tau$ direction. We also note that $u_\mu=\hat{\tau}_\mu$, so the four-temperature vector is orthogonal to the hypersurface $\Sigma$.

Finally, the exponent of the density operator \eqref{def: ZUB1} can be written as:
\begin{equation*}
    \widehat{\Upsilon}(\tau)=\tau\beta(\tau)\int\di^2{\rm x}_{\rm T}\di\eta\left[\wT^{\tau\tau}-\frac{1}{2}\Omega(\tau)\left(\delta^x_\rho\delta^y_\sigma-\delta^y_\rho\delta^x_\sigma\right)\wS^{\tau,\rho\sigma}
    -\mu(\tau)\,\wj^\tau\right]\;,
\end{equation*}
and describes the most general state with boost invariance at finite spin density.
Here $\beta(\tau)\equiv 1/T(\tau)$ with $T$ co-moving temperature and we used:
\begin{equation}\label{def: Omega}
    \SP(\tau)\equiv \frac{\Omega(\tau)}{T(\tau)}\;\;,
\end{equation}
where $\Omega$ is the proper \emph{spin potential} which, in analogy with the chemical potential $\mu$ has dimensions of an energy in natural units. Our naming of the reduced spin potential $\SP=\Omega/T$ is done in analogy with the reduced chemical potential $\zeta=\mu/T$.
The above density operator can be written as:
\begin{equation}\label{eq:density operator split}
    \wrho_{\rm LE}(\tau)=\frac{1}{Z_{\rm LE}(\tau)}\exp\left[-\frac{\widehat{\Pi}_\Omega(\tau)}{T(\tau)}
    +\zeta(\tau)\wQ\right]\;,\
\end{equation}
where $\wQ$ is the charge operator, $\widehat{\Pi}_\Omega$ is the effective spin potential-dependent Hamiltonian: 
\begin{equation}\label{Pi tau generico}
   \begin{split}
         \wPi_\Omega(\tau)&\equiv \tau\int\di^2{\rm x}_{\rm T}\di\eta\left[\wT^{\tau\tau}-\Omega(\tau)\wS^{\tau,xy}\right]\;,\\
        \wQ&\equiv\tau\int\di^2{\rm x}_{\rm T}\di\eta \,\wj^\tau\;.
   \end{split}
\end{equation}
An analogous expression can be written for the true non equilibrium operator, simply by substituting $\tau$ with the equilibration proper time $\tau_0$ in eq.~\eqref{eq:density operator split}.

By construction the local equilibrium state in eq.~\eqref{eq:density operator split} is invariant under translation in the transverse direction and in the longitudinal one:
\begin{equation}\label{invariant rho}
    \e^{\ii\widehat{\bf P}_{\rm T}\cdot{\bf x}_{\rm T}}\,\wrho_{\rm LE}(\tau)\e^{-\ii\widehat{\bf P}_{\rm T}\cdot{\bf x}_{\rm T}}=\wrho_{\rm LE}(\tau)\;,\qquad\e^{\ii\widehat{{\rm K}}_z\eta}\,\wrho_{\rm LE}(\tau)\e^{-\ii\widehat{{\rm K}}_z\eta}=\wrho_{\rm LE}(\tau)\;,
\end{equation}
it is not, however, independent of the proper time $\tau$.
We remark that splitting the operator \eqref{Pi tau generico} in two, with one integral of $\wT^{00}$ and a second one on $\wS^{0,12}$ is futile in the case of Milne spacetime. Indeed, since $\hat{\tau}$ is not a Killing vector, $T^{\mu\nu}\hat{\tau}_\nu=T^{\mu0}$ is not a true vector density, and the Gauss theorem doesn't imply that the Hamiltonian is hypersurface independent, see also the discussion in refs. \cite{Rindori:2021quq}. However if we had chosen a pseudogauge such that $\nabla_\mu S^{\mu,\nu\rho}=0$, then $ \tau\int\di^2{\rm x}_{\rm T}\di\eta \wS^{\tau,xy}$ would define a conserved global charge, as $\hat{x}$ and $\hat{y}$ are Killing vectors. 

The goal of this work is to compute exactly expectation values for Dirac fermions with the density operator \eqref{eq:density operator split}. We will do this for the Belinfante and the canonical pseudogauges. While the debate on which spin tensor shall be used in the context of QGP physics is still ongoing, we choose these two pseudogauges for their simplicity. There are also theoretical arguments that seem to favour these choices. For instance, an (almost) pseudogauge invariant density operator can be defined, whose result coincide with those of the Belinfante pseudogauge \cite{Becattini:2025twu}, whereas the canonical pseudogauge provides the only known spin tensor which obeys $SO(3)$ algebra for Dirac fields \cite{Dey:2023hft}, and can emerge effectively from interacting theories \cite{Buzzegoli:2024mra}. 

The invariance of the statistical operator under a given symmetry group fixes the form of the expectation values computed with it \cite{Rindori:2021quq}. The density operator in eq.~\eqref{eq:density operator split} is invariant under longitudinal boost and parity transformations\footnote{The reduced spin potential component $\SP_{xy}$ is the component of a pseudovector. Indeed in equilibrium $\SP_{xy}\mapsto\omega_z$, the $z$-component of the angular velocity.} and, using the fact that the canonical energy momentum tensor of a Dirac field is not necessarily symmetric, and that the canonical spin tensor is dual to the axial current, expectation values \eqref{eq: local state} must have the following tensor forms:
\begin{equation}\label{eq: expectation values expansion}
    \begin{split}
        T^{\mu\nu}(x)&=\mathcal{E}(\tau)\hat{\tau}^\mu \hat{\tau}^\nu+\mathcal{P}_{\rm T}(\tau)\left(\hat{x}^\mu\hat{x}^\nu+\hat{y}^\mu\hat{y}^\nu\right)\\
        &+  \mathcal{P}_{\rm L}(\tau)\hat{\eta}^\mu\hat{\eta}^\nu+\frac{\mathcal{T}(\tau)}{2}\left(\hat{y}^\mu\hat{x}^\nu-\hat{x}^\mu\hat{y}^\nu\right)\;,\\
    j^\mu\left(x\right)&=\mathcal{Q}(\tau)\hat{\tau}^\mu\;,\qquad S^{\mu,\lambda\nu}(x)=\epsilon^{\mu\lambda\nu\rho}\mathcal{S}(\tau)\hat{\eta}_\rho\;.
    \end{split}
\end{equation}
Note that, compared to~\cite{Rindori:2021quq}, we have the additional appearance of a \emph{spin-orbit torque density} $\mathcal{T}(\tau)$, which is due to the fact that the energy momentum tensor is not assumed to be symmetric under exchange of indices: $T^{\mu\nu}\neq T^{\nu\mu}$.

Inverting relations \eqref{eq: expectation values expansion} one can express the thermodynamic functions in terms of the expectation value of the components of the stress-energy tensor, four-current and spin tensor:
\begin{equation}\label{def: E PL PT}
    \begin{split}
        \mathcal{E}(\tau)&=\langle\wT^{\tau\tau}\rangle\;,\quad \mathcal{P}_{\rm L}(\tau)=\tau^{2}\langle\wT^{\eta\eta}\rangle\;,\quad\mathcal{P}_{\rm T}(\tau)=\frac{\langle\wT^{xx}\rangle+\langle\wT^{yy}\rangle}{2}\;,\;\\
        \mathcal{T}(\tau)&=\langle\wT^{yx}\rangle-\langle\wT^{xy}\rangle\;,\;\quad \mathcal{S}(\tau)=\langle\wS^{\tau,xy}\rangle\;.
    \end{split}
\end{equation}
Assuming that the expectation values  \eqref{def: E PL PT} are properly normalized and that they satisfy the same conservation equations \eqref{eq: continuity}, we obtain:
\begin{equation*}
    \begin{split}
        \nabla_\mu\langle\wT^{\mu\nu}\rangle&=\frac{1}{\sqrt{-g}}\partial_\mu\left(\sqrt{-g}\langle\wT^{\mu\nu}\rangle\right)+\Gamma^\nu_{\mu\lambda}\langle\wT^{\mu\lambda}\rangle=\frac{1}{\tau}\frac{\partial}{\partial\tau}\left(\tau\mathcal{E}(\tau)\right)+\tau\frac{\mathcal{P}_{\rm L}(\tau)}{\tau^2}=0\;,\\
        \nabla_\mu\langle\wj^\mu\rangle&=\frac{1}{\sqrt{-g}}\partial_\mu\left(\sqrt{-g}\langle\wj^{\mu}\rangle\right)=\frac{1}{\tau}\frac{\partial}{\partial \tau}\left(\tau\mathcal{Q}(\tau)\right)=0\;,\\
        \nabla_\mu\langle\wS^{\mu,\lambda\nu}\rangle&=\frac{1}{\sqrt{-g}}\partial_\mu\left(\sqrt{-g}\wS^{\mu,\lambda\nu}\right)+\Gamma^{\lambda}_{\mu\rho}\wS^{\mu,\rho\nu}+\Gamma^{\nu}_{\mu\rho}\wS^{\mu,\lambda\rho}=\frac{1}{\tau}\frac{\partial}{\partial\tau}\left(\tau\mathcal{S}(\tau)\right)=\mathcal{T}(\tau)\;,
    \end{split}
\end{equation*}
which yields a set of differential equations relating the different components of the expectation values of the conserved densities and the spin tensor:
\begin{subequations}\label{eq: equation ob}
    \begin{align}
        \dot{\mathcal{E}}(\tau)+\frac{\mathcal{E}(\tau)+\mathcal{P}_{\rm L}(\tau)}{\tau}&=0\;,\\
        \dot{\mathcal{Q}}(\tau)+\frac{\mathcal{Q}(\tau)}{\tau}&=0\;,\\
        \dot{\mathcal{S}}(\tau)+\frac{\mathcal{S}(\tau)}{\tau}-\mathcal{T}(\tau)&=0\;.
    \end{align}
\end{subequations}
The interpretation of $\mathcal{T}$ as a torque density is motivated by the fact that the above relation is the equivalent, in Milne coordinates, to the equation relating the change of angular momentum $\bm{L}$ to external torque $\bm{\tau}$ in classical mechanics: $\dot{\bm{L}}={\bm\tau}$.
Other than the expectation value of the conserved densities we will be interested in computing another phenomenologically important observable, that is the spin density matrix $\Theta_{rs}$~\cite{Becattini:2020sww}:
\begin{equation}\label{eq: def spin density matrix}
    \Theta_{rs}(p) = 
    \frac{\langle\widehat{a}^\dagger_s(p)\widehat{a}_{r}(p)\rangle}
    {\sum_t\langle\widehat{a}^\dagger_t(p)\widehat{a}_{t}(p)\rangle},
\end{equation}
with the goal of obtaining exact expressions for the spin polarization of Dirac fermions at finite spin density.

From the spin density matrix~\eqref{eq: def spin density matrix} one extracts the mean spin of the particle, which in the particle's rest frame reads~\cite{Becattini:2020sww}:
\begin{equation}\label{eq: polarizz}
    {\bf S}^*=\tr\left(\Theta(p)\boldsymbol{\sigma}\right)\;.
\end{equation}
The \emph{polarization vector} is defined as $\mathbf{P}={\bf S}^*/S$, $S$ being the particle's spin, and is thus normalized to unity.

An exact solution of Eq.~\eqref{eq: polarizz} provides a valuable benchmark for testing linear-response formulas and for clarifying the role of symmetries in the generation of spin polarization. Indeed, a naïve expectation based on linear-response theory would suggest that, since the thermal shear is finite and the reduced chemical potential is not constant in our setup (see Eq.~\eqref{eq: varpi e xi}), a non-vanishing polarization should arise even in the Belinfante pseudogauge. In the following, we demonstrate that this is not the case: symmetries play a crucial role, and the mere presence of thermal shear or not constant chemical potential are not sufficient to generate a net polarization.


\section{Free Dirac field in Milne coordinates}\label{Sec: Dirac Milne}
In the previous section, we have shown the prominent role of Milne coordinates in the analytical treatment of boost invariant fluids. To calculate expectation values, it is therefore convenient to use the Dirac modes of Milne spacetime. 

To determine them, one needs to solve the Dirac equation in Milne coordinates, which reads:
\begin{equation}\label{Dirac Eq Milne}
 \left[\ii\gamma^{\underline{0}}\,\partial_\tau+\ii\boldsymbol{\gamma}_{\rm T}\cdot\boldsymbol{\nabla}_{\rm T} + \frac{\ii\gamma^{\underline{3}}}{\tau}\left(\partial_\eta-\frac{\gamma^\uthree\gamma^\uzero}{2}\right)-m\right]\psi\left(\tau,{\bf x}_{\rm T},\eta\right)=0\;.
\end{equation}
The vector ${\boldsymbol{\gamma}}_{\rm T}\equiv \left(\gamma^\uone,\gamma^\utwo\right)$ contains the gamma matrices associated with the transverse directions whereas  the gradient in the transverse direction is denoted by ${\boldsymbol{\nabla}}_{\rm T}=\left(\partial_x,\partial_y\right)$. Details about Milne coordinates and the derivation of the Dirac equation are reported in appendix \ref{app:dirac eq}.

As mentioned earlier, the metric \eqref{Milne Chart} is manifestly non-static, as it depends explicitly on the time coordinate~$\tau$. Consequently, the spacetime does not admit a global time-like Killing vector field. In general, non-static metrics are associated with gravitational particle production, since it is impossible to define a unique set of field modes with well-defined positive and negative frequencies throughout the entire spacetime~\cite{Fulling_1989}.
However, the metric \eqref{Milne Chart} merely represents flat Minkowski spacetime expressed in a different coordinate chart. Therefore, any apparent particle production in this context must be a coordinate artifact rather than a genuine physical effect~\cite{Padmanabhan:1990fm}. Indeed, it has been shown rigorously~\cite{Arcuri:1994kd} that the free Klein--Gordon equation in Milne coordinates admits a complete set of solutions that can be expressed as linear combinations of Minkowski plane waves of definite frequency sign. This demonstrates that the scalar field can be expanded in terms of Minkowski modes without any mixing between positive- and negative-frequency components. As a result, no real particle creation occurs.
While this conclusion was explicitly established for scalar fields, it does not depend on the spin of the field. The same reasoning applies to fields of arbitrary spin, including the Dirac field. 
To prove this, we show that we can build a solution of the Dirac equation in Milne coordinates in terms of the positive-frequency modes in Cartesian coordinates.

We will follow the procedure outlined in \cite{Akkelin:2018hpu} for scalar fields, adapting it to the case of the Dirac field. First we define:
\begin{equation}\label{new field}
    \psi(\tau,{\bf x}_{\rm T},\eta)\equiv e^{\ii\eta\Sigma^{\uzero\uthree}}\widetilde{\psi}(\tau,{\bf x}_{\rm T},\eta)\;,
\end{equation}
where the operator $\ii\Sigma^{\uzero\uthree}=(1/2)\gamma^\uthree\gamma^\uzero$ is the generator of the boosts in the $z$-direction in the spinorial representation. In terms of \eqref{new field}, equation \eqref{Dirac Eq Milne} reads:
\begin{equation*}
    \left(\ii\gamma^{\underline{0}}e^{\ii\eta\Sigma^{\uzero\uthree}}\partial_\tau+\ii\boldsymbol{\gamma}_{\rm T}\cdot\boldsymbol{\nabla}_{\rm T}e^{\ii\eta\Sigma^{\uzero\uthree}}\ + \frac{\ii\gamma^{\underline{3}}}{\tau}e^{\ii\eta\Sigma^{\uzero\uthree}}\partial_\eta-m e^{\ii\eta\Sigma^{\uzero\uthree}}\right)\widetilde{\psi}(\tau,{\bf x}_{\rm T},\eta)=0\;,
\end{equation*}
where the $\eta$-derivative acting on the exponential factor cancels the spin connection term.

Using the algebra of the Dirac gamma matrices it's easy to prove that:
\begin{equation*}
    e^{-\ii\eta\Sigma^{\uzero\uthree}}\gamma^\uzero e^{\ii\eta\Sigma^{\uzero\uthree}}=\cosh\eta\,\gamma^\uzero-\sinh\eta\,\gamma^\uthree\;, \qquad
    e^{-\ii\eta\Sigma^{\uzero\uthree}}\gamma^\uthree e^{\ii\eta\Sigma^{\uzero\uthree}}=\cosh\eta\,\gamma^\uthree-\sinh\eta\,\gamma^\uzero\;,
\end{equation*}
while $[\Sigma^{03},\bm{\gamma}_T]=0$. Thus, the equation for $\tilde{\psi}$ reduces to:
\begin{equation}\label{eqDir2}
    \begin{split}
        &\left[\gamma^{\underline{0}}\left(\cosh\eta\, \ii\partial_\tau-\frac{\sinh\eta}{\tau}\ii\partial_\eta\right)+\ii\boldsymbol{\gamma}_{\rm T}\cdot\boldsymbol{\nabla}_{\rm T}\right.\\
        &\left.+ \gamma^{\underline{3}}\left(-\sinh\eta\, \ii\partial_\tau+\frac{\cosh\eta}{\tau}\ii\partial_\eta\right)-m\right]\widetilde{\psi}\left(\tau,{\bf x}_{\rm T},\eta\right)=0\;.
    \end{split}
\end{equation}
We look for solutions $\tilde{\psi}$ which are expressible in terms of the Minkowski plane wave modes. To this end, it is useful to express the four-momentum in terms of the \emph{transverse mass} $m_{\rm T}=(m^2+{ p}^2_{\rm T})^{1/2}$ and the \emph{momentum-space rapidity} $\vartheta$ as: 
\begin{equation*}
    p^\mu=\left(m_{\rm T}\cosh\vartheta,p_x,p_y,m_{\rm T}\sinh\vartheta\right)\;.
\end{equation*}
In terms of Milne coordinates, $m_{\rm T}$, and $\vartheta$, 
and the positive frequency plane waves of the Minkowski mode read:
\begin{equation*}
    \widetilde{\psi}^{(+)}\sim e^{-ip\cdot x}u(p) =  e^{-\ii\tau m_{\rm T}\cosh(\vartheta-\eta)+\ii{\bf p}_{\rm T}\cdot{\bf x}_{\rm T}}u(p)\;,
\end{equation*}
where $u(p)$ is a complex bi-spinor.
Plugging this ansatz in eq. \eqref{eqDir2} and using the addition formulae of the hyperbolic functions, one finds:
\begin{equation*}
    \Big[\gamma^\uzero m_{\rm T}\cosh\vartheta-{\boldsymbol{\gamma}}_{\rm T}\cdot{\bf p}_{\rm T}+\gamma^\uthree(-m_{\rm T}\sinh\vartheta)-m\Big]u(p)=(\slashed{p}-m)u(p)\;.
\end{equation*}
The above is just the standard Dirac equation in momentum space, which is solved by the usual Minkwowski free-spinor. For massive particles, one has:
\begin{equation}\label{eq:massive spinor}
    u_r(p)=\frac{m+\slashed{p}}{\sqrt{2(m+\varepsilon)}}\xi_r\;,
\end{equation}
where $\xi_r$ is the standard eigen-spinor of $\gamma^\uthree$:
\begin{equation}\label{def: chi pm}
    \xi_r=\begin{pmatrix}
        \chi_r\\
        \chi_r
    \end{pmatrix}\;,\qquad \chi_+=\begin{pmatrix}
        1\\
        0
    \end{pmatrix}\,,\quad\chi_-=\begin{pmatrix}
        0\\
        1
    \end{pmatrix}\;.
\end{equation}
Therefore, recalling the definition~\eqref{new field}, the positive energy solution of the Dirac equation \eqref{Dirac Eq Milne} can be written as:
\begin{equation*}
    \psi^+(\tau,{\bf x}_{\rm T},\eta)=\frac{e^{\ii\eta\Sigma^{\uzero\uthree}}}{(2\pi)^{3/2}}\sum_{r=\pm}\int\frac{\di^2{\bf p}_{\rm T}\di p_z}{\sqrt{2\varepsilon(p)}}e^{-\ii\tau m_{\rm T}\cosh(\vartheta-\eta)+\ii{\bf p}_{\rm T}\cdot{\bf x}_{\rm T}}\widehat{a}_r(p)u_r(p)\;.
\end{equation*}
Following ref. \cite{Akkelin:2018hpu,Akkelin:2020cfs, Rindori:2021quq}, the creation and annihilation operators of a Milne particle, which will be denoted with capital letters $\widehat{A},\ \widehat{A}^\dagger$, can be related with the Minkowski one as:
\begin{equation}\label{eq: Mink a in terms of A}
    \begin{split}
        \widehat{a}_r(p)&=\frac{1}{\sqrt{2\pi m_{\rm T}\cosh\vartheta}}\int^{+\infty}_{-\infty}\di\mu\, e^{\ii\mu\vartheta}\widehat{A}_r({\bf p}_{\rm T},\mu)\;,
        \\
        \widehat{a}^\dagger_r(p)&=\frac{1}{\sqrt{2\pi m_{\rm T}\cosh\vartheta}}\int^{+\infty}_{-\infty}\di\mu\, e^{-\ii\mu\vartheta}\widehat{A}^\dagger_r({\bf p}_{\rm T},\mu)\;.
    \end{split}
\end{equation}
where $\mu$ is a new variable introduced as the conjugate of momentum-space rapidity. Inverting the above relations,
\begin{equation}\label{eq: A a in terms of Mink}
    \begin{split}
        \widehat{A}_r(\bf{p}_{\rm T},\mu)&=\int^{+\infty}_{-\infty}\di\vartheta\sqrt{\frac{m_{\rm T} \cosh\vartheta}{2\pi}}\, e^{-\ii\mu\vartheta}\widehat{a}_r(p)\;,
        \\
        \widehat{A}^\dagger_r({\bf{p}}_{\rm T},\mu)&=\int^{+\infty}_{-\infty}\di\vartheta\sqrt{\frac{m_{\rm T} \cosh\vartheta}{2\pi}}\, e^{\ii\mu\vartheta}\widehat{a}^\dagger_r(p)\;,
    \end{split}
\end{equation}
one can immediately show that the  Milne operators $\widehat{A}_r$ and $\widehat{B}^\dagger_r$ satisfy the canonical anti-commutation relations:
\begin{equation}\label{canonical anticommutator}
    \begin{split}
        \left\{\widehat{A}_r({\bf p}_{\rm T},\mu),\widehat{A}^\dagger_{r'}({\bf p}_{\rm T}',\mu')\right\}&=\left\{\widehat{B}_r({\bf p}_{\rm T},\mu),\widehat{B}^\dagger_{r'}({\bf p}_{\rm T}',\mu')\right\}\\
        &=\delta^2\left({\bf p}_{\rm T}-{\bf p}'_{\rm T}\right)\delta\left(\mu-\mu'\right)\delta_{rr'}\;,
    \end{split}
\end{equation}
all other anticommutators vanishing.

In terms of the operator $\widehat{A}_r$ and after a change of variable $p_z\mapsto m_{\rm T}\sinh\vartheta$ one obtains:
\begin{equation*}
    \psi^{+}(\tau,{\bf x}_{\rm T},\eta)=\frac{e^{\ii\eta\Sigma^{\uzero\uthree}}}{\sqrt{2}(2\pi)^{2}}\sum_{r=\pm}\int\di^2{\bf p}_{\rm T}\,\di \mu \,\di\vartheta\, e^{-\ii\tau m_{\rm T}\cosh(\vartheta-\eta)+\ii{\bf p}_{\rm T}\cdot{\bf x}_{\rm T}}e^{\ii\mu\vartheta}\widehat{A}_r({\bf p}_{\rm T},\mu)u_r(p)\;.
\end{equation*}
In order to deal with the overall exponential factor, the spinor $u_r(p)$ can be written as the longitudinal boost with rapidity $\vartheta$ of a spinor with purely transverse momentum $p^\mu=\left(m_{\rm T},{\bf p}_{\rm T},0\right)$. Denoting such a spinor as  $u_r({\bf p}_{\rm T})$, this amounts to the transformation:
\begin{equation*}
    u_r(p)=e^{-\ii\vartheta\Sigma^{\uzero\uthree}}u_r({\bf p}_{\rm T})\;.
\end{equation*}
Hence, translating the integration variable $\vartheta\mapsto\vartheta+\eta$, we get:
\begin{align*}
    \psi^+\left(\tau,{\bf x}_{\rm T},\eta\right)&=\frac{1}{\sqrt{2}(2\pi)^{2}}\sum_{r=\pm}\int\di^2{\bf p}_{\rm T}\,\di \mu \, e^{\ii{\bf p}_{\rm T}\cdot{\bf x}_{\rm T}+\ii\mu\eta}\\
    &\times\left(\int\di\vartheta\, e^{-\ii\tau m_{\rm T}\cosh\vartheta-\vartheta\left(-\ii\mu+\ii\Sigma^{\uzero\uthree}\right)}\right)\widehat{A}_r({\bf p}_{\rm T},\mu)u_r({\bf p}_{\rm T})\;.
\end{align*}

The same procedure can be applied to the negative energy modes, so that the solution of the Dirac equation in Milne coordinates, eq.~\eqref{Dirac Eq Milne}, is:
\begin{align*}
    &\psi\left(\tau,{\bf x}_{\rm T},\eta\right)=\\
    &\frac{1}{\sqrt{2}(2\pi)^{2}}\sum_{r=\pm}\int\di^2{\bf p}_{\rm T}\,\di \mu \, \Big[e^{\ii{\bf p}_{\rm T}\cdot{\bf x}_{\rm T}+\ii\mu\eta}\left(\int\di\vartheta\, e^{-\ii\tau m_{\rm T}\cosh\vartheta-\vartheta\left(-\ii\mu+\ii\Sigma^{\uzero\uthree}\right)}\right)\widehat{A}_r({\bf p}_{\rm T},\mu)u_r({\bf p}_{\rm T})\\
    &\qquad\left.+e^{-\ii{\bf p}_{\rm T}\cdot{\bf x}_{\rm T}-\ii\mu\eta}\int \di \vartheta \left(e^{\ii\tau m_{\rm T}\cosh\vartheta-\vartheta\left(\ii\mu+\ii\Sigma^{\uzero\uthree}\right)}\right)\widehat{B}^\dagger_r({\bf p}_{\rm T},\mu)v_r({\bf p}_{\rm T})\right],
\end{align*}
where $v$ is the standard Minkowski $v$-spinor related with $u_r$ by charge conjugation \cite{Peskin:1995ev}: $v_r=-\ii\gamma^\utwo (u_r)^*\equiv \mathcal{C} (u_r)^*$ .

Between parentheses, one recognizes the integral representation of the Hankel functions \cite{gradshteyn2007}:
\begin{subequations}\label{def: hankels}
    \begin{align}
    {\rm H}^{(1)}_{\nu}(z)&=\frac{e^{-\ii\pi\nu/2}}{\ii\pi}\int \di \vartheta e^{\ii z\cosh \vartheta-\vartheta\nu}\;,\\
    {\rm H}^{(2)}_{\nu}(z)&=\frac{e^{\ii\pi\nu/2}}{-\ii\pi}\int_{-\infty}^\infty \di\vartheta e^{-\ii z\cosh\vartheta-\vartheta\nu}\;,
\end{align}
\end{subequations}
with $z\equiv m_{\rm T}\tau$. In contrast to the case of the scalar field, where $\nu$ was a purely imaginary number, the order $\nu$ of Hankel functions is now a matrix valued complex index:
\begin{equation*}
    \nu\equiv\frac{1}{2}\gamma^{\uthree}\gamma^\uzero-\ii\mu\I\;,\qquad \bar{\nu}=\frac{1}{2}\gamma^{\uthree}\gamma^\uzero+\ii\mu\I\;.
\end{equation*}
In the Weyl basis of the $\gamma$ matrices, which we will use throughout this work, $\gamma^\uthree\gamma^\uzero/2=\mathrm{diag}(1,-1,-1,1)$, so it is clear that the occurrence of a real part in the index of the Hankel function, is directly related to the spin of the Dirac field. In appendix \ref{app:hankel}, we report some useful identities for Hankel functions, both of scalar and matrix-valued orders.

Introducing the following short-hand notation:
\begin{equation}\label{eq: accorciabro}
    \begin{split}
        \fp&=\left({\bf p}_{\rm T},\mu\right)\,,\qquad\di^3\fp=\di\mu\di^2{\bf p}_{\rm T}\;,\\
        \fr&=\left({\bf x}_{\rm T},\eta\right)\;,\qquad \di^3\fr=\di\eta\di^2{\bf x}_{\rm T}
    \end{split}
\end{equation}
 the general solution of \eqref{Dirac Eq Milne} can finally be written as:
\begin{equation}\label{eq: field final expansion}
    \psi\left(\tau,{\bf x}_{\rm T},\eta\right)=\frac{1}{\left(2\pi\right)^{3/2}}\sum_{r=\pm}\int\di^3\fp\left[e^{\ii\fp\cdot\fr}U_r(\tau,\fp)\widehat{A}_r(\fp)+e^{-\ii\fp\cdot\fr}V_r(\tau,\fp)\widehat{B}^\dagger_r(\fp)\right]\;,
\end{equation}
where $U_r$ and $V_r$ are the proper Milne four-spinors defined as:
\begin{subequations}\label{eq: Milne spinors}
    \begin{align}
        U_r\left(\tau,\fp\right)&=-\frac{\ii\sqrt{\pi}}{2} e^{-\ii\pi\nu/2}{\rm H}^{(2)}_\nu(m_{\rm T}\tau)\,u_r({\bf p}_{\rm T})\;,\\
        V_r\left(\tau,\fp\right)&=\frac{\ii\sqrt{\pi}}{2} e^{\ii\pi\bar{\nu}/2}{\rm H}^{(1)}_{\bar{\nu}}(m_{\rm T}\tau)\, v_r({\bf p}_{\rm T})\;,
    \end{align}
\end{subequations}
where the spinor $u_r({\bf p}_{\rm T})$ is the purely-transverse Minkwoski spinor:
\begin{equation}\label{def: Orthogonal four-spinor}
    u_r({\bf p}_{\rm T})=\frac{m+m_{\rm T}\gamma^\uzero+{\bf p}_{\rm T}\cdot\boldsymbol{\gamma}_{\rm T}}{\sqrt{2\left(m_{\rm T}+m\right)}}\xi_r
    \;,
\end{equation}
and the dependence on $\mu$ is hidden in the matrix order $\nu$. 

Our solutions are also valid for massless spinors, with the simple exchange of the massive Minkowski $u({\bf p}_{\rm T})$ in eq.~\eqref{def: Orthogonal four-spinor} for its massless analogue. 
They have been constructed explicitly using Minkowski modes, and indeed, using the asymptotic expression of the Hankel function for large arguments \eqref{app: hank asymptotic grande z}, it is easy to prove that the above spinors reduce to the usual Minkowski free-spinors in the limit $m_{\rm T}\tau\mapsto\infty$ (modulo a complex overall phase). Hence \eqref{eq: field final expansion} coincides with the Cartesian field expansion for large $m_{\rm T}\tau$. 

In the rest of this work we will focus on massive particles.
From the standard normalization of spinors, it follows that the Milne spinors \eqref{eq: Milne spinors} satisfy the following orthogonality relations:
\begin{equation}\label{ortho conditions}
    \begin{split}
        U^\dagger_r\left(\tau,\fp\right)U_s\left(\tau,\fp\right)&=V^\dagger_r\left(\tau,\fp\right)V_s\left(\tau,\fp\right)=\frac{\delta_{rs}}{\tau}\;,\\
        \overline{U}_r\left(\tau,\fp\right)V_s\left(\tau,\fp\right)&=\overline{V}_r\left(\tau,\fp\right)U_s\left(\tau,\fp\right)=0\;,\\
        U^\dagger_r\left(\tau,\fp\right)V_s\left(\tau,-\fp\right)&=V^\dagger_r\left(\tau,\fp\right)U_s\left(\tau,-\fp\right)=0\;,
    \end{split}
\end{equation}
which are derived in appendix \ref{app: spinors}, together with many other useful identities to be used later on\footnote{Note that, from eqs.~\eqref{eq: Milne spinors}, for $\fp\mapsto-\fp$ the variable $\mu$ changes sign, and thus $\nu\mapsto\bar{\nu}$ in the Hankel functions.}. The above contractions implies that the spinor basis associated with the operators $\widehat{A},\ \widehat{A}^\dagger$ is indeed orthogonal with respect to the Dirac inner product defined as:
\begin{equation}\label{app hank: Dirac inner product}
    \left(\psi_1,\psi_2\right)_{\rm Dirac}\equiv\int_{\Sigma}\di\Sigma_\mu \opsi_1(x)\gamma^\mu(x)\psi_2(x)\;,
\end{equation}
where $\psi_1,\psi_2$ are two solutions of the Dirac equation \eqref{Dirac Eq Milne} and $\Sigma $ is an arbitrary space-like hypersurface under which choice \eqref{app hank: Dirac inner product} is independent. 
From \eqref{ortho conditions} the Dirac inner product \eqref{app hank: Dirac inner product} implies that:
\begin{equation*}
    \begin{split}
        \left(\psi^{(\pm)}_r\left(\tau,\fp\right),\psi^{(\pm)}_{r'}\left(\tau,\fp'\right)\right)_{\rm Dirac}&=\delta_{rs}\,\delta^{3}\left(\fp-\fp'\right)\,,\\
        \left(\psi^{(\pm)}_r\left(\tau,\fp\right),\psi^{(\mp)}_{r'}\left(\tau,\fp'\right)\right)_{\rm Dirac}&=0\,,
    \end{split}
\end{equation*}
which is equivalent to require that the creation-annihilation operators $\widehat{A},\widehat{B}^\dagger$ satisfy the usual anti-commutation relations \eqref{canonical anticommutator}.
 
Eq. \eqref{eq: field final expansion} is the solution of the Dirac equation in Milne spacetime we were looking for. Note that, according to \eqref{eq: Mink a in terms of A} the Minkowski-particle number in a given state is directly computed in terms of the Milne modes:
\begin{equation}\label{particle number a e A}
    \begin{split}
        \langle\widehat{a}^\dagger_r(p)\widehat{a}_{s}(p')\rangle&=\frac{1}{2\pi m_{\rm T}\sqrt{\cosh\vartheta\cosh\vartheta'}}\\
        &\times\int^{+\infty}_{-\infty}\di\mu\int^{+\infty}_{-\infty}\di\mu'\e^{-\ii\left(\vartheta\mu+\vartheta'\mu'\right)}\langle\widehat{A}^\dagger_r({\bf p}_{\rm T},\mu)\widehat{A}_s({\bf p}'_{\rm T},\mu')\rangle\;.
    \end{split}
\end{equation}
In particular, on the Minkowski vacuum state, one has:
\begin{equation*}
    0=\widehat{a}_r(p)\left|0_{\rm mink}\right\rangle=\frac{1}{\sqrt{2\pi m_{\rm T}\cosh\vartheta}}\int^{+\infty}_{-\infty}\di\mu\, e^{\ii\mu\vartheta}\widehat{A}_r({\bf p}_{\rm T},\mu)\left|0_{\rm mink}\right\rangle\implies \widehat{A}_r(\fp)\left|0_{\rm mink}\right\rangle=0\;,
\end{equation*}
hence the Milne modes \eqref{eq: A a in terms of Mink} annihilate  the Minkowski vacuum  and no spontaneous particle production occurs, as expected.

\subsection{The Zubarev operator of a free Dirac field in a boost invariant background}\label{sec:zubarev longitudinal}
Using the Milne modes of Dirac equation, we can work out explicitly the non-equilibrium density operator, and in particular its exponent $\widehat{\Upsilon}$ given by eq.~\eqref{Pi tau generico}. In this section, we will consider the specific cases of the \emph{canonical} and the \emph{Belinfante} pseudo-gauge. 

In the canonical pseudo-gauge the stress-energy tensor and the spin tensor are directly obtained from the Lagrangian with the Noether theorem.
Their expression for a free Dirac field in curvilinear coordinates \eqref{Milne Chart} are simply obtained from the Cartesian ones replacing partial with covariant derivatives, and promoting the gamma matrices to space-time dependent ones:
\begin{subequations}\label{eq: canonical tensors}
    \begin{align}
    \wT^{\mu\nu}_{\rm C}(x)&=\frac{\ii}{2}\opsi(x)\gamma^\mu(x)D^\nu\psi(x)-\frac{\ii}{2}D^\nu\opsi(x)\gamma^\mu(x)\psi(x),\label{subeq: canonical T}\\
    \wS^{\mu,\nu\lambda}_{\rm C}(x)&=\frac{1}{2}\epsilon^{\rho\mu\nu\lambda}(x)\opsi(x)\gamma_\rho(x)\gamma^5\psi(x)\;,\label{subeq: canonical S}\\
    \wj^\mu(x)&=\opsi(x)\gamma^\mu(x)\psi(x)\;\label{subeq: canonical j},
    \end{align}
\end{subequations}
where the covariant derivative $D_\mu$ is defined as eq. \eqref{Def: CoDev spinor} while $\epsilon^{\mu\nu\lambda\rho}$ is the totally anti-symmetric Levi-Civita pseudo-tensor in curvilinear coordinates. The fifth gamma matrix $\gamma^5=\ii\gamma^\uzero\gamma^\uone\gamma^\utwo\gamma^\uthree$ is independent of the coordinate system. Note that, for the non-interacting Dirac field, the canonical spin tensor is totally anti-symmetric. 

For the Belinfante pseudogauge, operators can be directly read from the canonical ones. Indeed, for the free Dirac field the energy-momentum in the Belinfante pseudogauge is simply:
\begin{equation}\label{eq: Belinfante Semt}
    \wT^{\mu\nu}_{\rm B}(x)=\frac{1}{2}\left(\wT^{\mu\nu}_{\rm C}(x)+\wT^{\nu\mu}_{\rm C}(x)\right)\,,
\end{equation}
while $\wS^{\mu,\lambda\nu}_{\rm B}=0$. 

In order to compute the operators in eq.~\eqref{Pi tau generico}, we only need the components $\wT^{\tau\tau}$, $\wS^{\tau,xy}$, and $\widehat{j}^\tau$. Since $\wT_{\rm C}^{\tau\tau}=\wT_{\rm B}^{\tau\tau}$ and $\wS_{\rm B}=0$, we will omit the subscript unless necessary, and simply write:
\begin{equation*}
   \begin{split}
        \wT^{\tau\tau}(x)&=\frac{\ii}{2}\psi^\dagger(x)\dot{\psi}(x)-\frac{\ii}{2}\dot{\psi}^\dagger(x)\psi(x)\;,\\
        \wS^{\tau,xy}(x)&=\frac{1}{2}\psi^\dagger(x)\gamma^{\uzero}\gamma^{\uthree}\gamma^{5}\psi(x)=\frac{\ii}{2}\psi^\dagger(x)\gamma^{\uone}\gamma^{\utwo}\psi(x)\;,\\
        \wj^\tau(x)&=\psi^\dagger(x)\psi(x)\;,
   \end{split}
\end{equation*}
where we used that, in curvilinear coordinates $\epsilon^{3012}=-1/\tau$ and that $\gamma^3(x)=\gamma^\uthree/\tau$.
With these, the effective hamiltonian and charge operator in eq.~\eqref{Pi tau generico} reduce:
\begin{equation}\label{eq:Pi1}
    \begin{split}
        \widehat{\Pi}_\Omega&=\tau\int\di^3\fr\;\left\{{\rm Im}\left[\dot{\psi}^\dagger\left(\tau,\fr\right)\psi\left(\tau,\fr\right)\right]-\ii\Omega(\tau)\psi^\dagger\left(\tau,\fr\right)\gamma^{\uone}\gamma^{\utwo}\psi\left(\tau,\fr\right)\right\}\;,\\
    \widehat{Q}&=\tau\int\di^3\fr\,\left[\psi^\dagger\left(\tau,\fr\right)\psi\left(\tau,\fr\right)\right]\;.
    \end{split}
\end{equation}
 The above expressions hold in the canonical pseudogauge, whereas the Belinfante pseudogauge can be immediately recovered setting $\Omega=0$ in $\wPi_\Omega$. 
 
The operator \eqref{eq:Pi1} can be expressed in a compact matrix form introducing the following notation:
\begin{equation}\label{def: NG spinor}
    \Phi^T(\fp)\equiv\begin{pmatrix}
        \widehat{A}_+(\fp),
        \widehat{A}_-(\fp),
        \widehat{B}^\dagger_+(-\fp),
        \widehat{B}^\dagger_-(-\fp)\;
    \end{pmatrix}.
\end{equation}
Then, using the field expansion eq.~\eqref{eq: field final expansion} and carrying out the Milne spinor products as shown in appendix \ref{app: computation and asymptotics}, the above operators read:
\begin{equation}\label{eq: Pi matrix}
    \begin{split}
        \widehat{\Pi}_\Omega(\tau)&\equiv\int\di^3\fp\,\tau\,\Phi^\dagger(\fp)\mathcal{H}_{\rm tot}(\tau,\fp)\Phi(\fp)\;,\\
        \widehat{Q}(\tau)&=\int\di^3\fp\,\tau\,\Phi^\dagger(\fp)\Phi(\fp)\;,
    \end{split}
\end{equation}
where we introduced the notation $\mathcal{H}_{\rm tot}= \mathcal{H}- \tau\Omega \mathcal{S}/2$ and the matrices $\mathcal{H}$ and $\mathcal{S}$, which have dimensions of energy, are given by:
\begin{equation}\label{def: matrices}
    \begin{split}
        \mathcal{H}&=m_{\rm T}^2\tau\begin{pmatrix}
        \mathfrak{h}(\tau,\mathfrak{p})\,\I_{2\times2}&{\rm j}^*(\tau,\mathfrak{p})\,\sigma^x\\
        {\rm j}(\tau,\mathfrak{p})\,\sigma^x&-\mathfrak{h}(\tau,\mathfrak{p})\,\I_{2\times2}
    \end{pmatrix}\;,\\
    \quad \mathcal{S}&=\begin{pmatrix}
        \frac{m}{m_{\rm T}\tau}\sigma^z-{\rm s}(\tau,\mathfrak{p})\left(p^x\sigma^x+p^y\sigma^y\right)&{\rm t}^*(\tau,\mathfrak{p})\left( \ii p^x\I_{2\times2}+ p^y\sigma^z\right)\\
        {\rm t}(\tau,\mathfrak{p})\left(p^y\sigma^z-\ii p^x\I_{2\times 2}\right)&-\frac{m}{m_{\rm T}\tau}\sigma^z+{\rm s}(\tau,\mathfrak{p})\left(p^x\sigma^x-p^y\sigma^y\right)
    \end{pmatrix}\;.
    \end{split}
\end{equation}
The dimensionless functions ${\rm h}(\tau,\mathfrak{p})$, ${\rm j}(\tau,\mathfrak{p})$, ${\rm s}(\tau,\mathfrak{p})$, and ${\rm t}(\tau,\mathfrak{p})$ appear when computing Milne spinor products, and are defined as:
\begin{subequations}\label{def: functions}
    \begin{align}
    {\rm h}(\tau,\mathfrak{p})&\equiv \frac{\pi e^{-\pi\mu}}{4}\left[\hnku_{-\frac{1}{2}+\ii\mu}(z)\frac{\di\hnkd_{-\frac{1}{2}-\ii\mu}(z)}{\di z}+\hnku_{\frac{1}{2}+\ii\mu}(z)\frac{\di\hnkd_{\frac{1}{2}-\ii\mu}(z)}{\di z}\right]\;,\\
    \mathfrak{h}(\tau,\mathfrak{p})&\equiv-{\rm Im}[{\rm h}(\tau,\mathfrak{p})]\;\label{def: frak h}\\
        {\rm j}(\tau,\mathfrak{p})&\equiv-\frac{\pi }{4}\left[ \hnkd_{\frac{1}{2}-\ii\mu}(z)\frac{\di\hnkd_{\frac{1}{2}+\ii\mu}(z)}{\di z}+\hnkd_{-\frac{1}{2}-\ii\mu}(z)\frac{\di\hnkd_{-\frac{1}{2}+\ii\mu}(z)}{\di z}\right]\;,\\
        {\rm s}(\tau,\mathfrak{p})&\equiv\frac{\pi e^{-\pi\mu}}{4}\left[\left|\hnku_{\frac{1}{2}+\ii\mu}(z)\right|^2-\left|\hnku_{-\frac{1}{2}+\ii \mu}(z)\right|^2\right]\;,\\
        {\rm t}(\tau,\mathfrak{p})&\equiv-\frac{\pi}{2}\hnkd_{\frac{1}{2}-\ii\mu}(z)\hnkd_{\frac{1}{2}+\ii\mu}(z)\;,
        \end{align}
\end{subequations}
with $z=m_T\tau$.
The general method to deal with Milne spinor products, as well as other auxiliary functions and all spinor products used in this work are reported in appendix \ref{app: spinors}.

It is important to mention that the functions in eq.~\eqref{def: functions} are not independent. On the contrary, they are related through identities stemming from the Wronskian and other notable properties of Hankel functions (see appendix \ref{app:hankel}).
For instance, we have:
\begin{equation}\label{eq: relazione h j}
   \mathfrak{h}^2(\tau,\mathfrak{p})+\left|{\rm j}(\tau,\mathfrak{p})\right|^2= \frac{(m_{\rm T}^2+\mu^2/\tau^2)^2}{m^4_{\rm T}\,\tau^2}\equiv\frac{\varepsilon^2(\tau,\fp)}{ m^4_{\rm T}\,\tau^2}\;,
\end{equation}
where we introduced the notation $\varepsilon(\tau)=(m_T^2+\mu^2/\tau^2)^{1/2}$, which denotes the energy of a free particle of mass $m$ expressed in Milne coordinates.
This relation is similar to the one relating $K_\mathcal{E}$ and $\Lambda_\mathcal{E}$ in ref. \cite{Rindori:2021quq}.  Other two important identities we can derive are (see appendix \ref{app: spinors}):
\begin{subequations}\label{eq: relazione s e t E hs e jt}
    \begin{align}
        {\rm s}^2(\tau,\mathfrak{p})+\left|{\rm t}(\tau,\mathfrak{p})\right|^2&=\frac{1}{m^2_{\rm T}\tau^2}\;,\\
        {\mathfrak{h} }(\tau,\mathfrak{p}){\rm s}(\tau,\mathfrak{p})+{\rm Im}\left({\rm j}(\tau,\mathfrak{p}){\rm t}^*(\tau,\mathfrak{p})\right)&=-\frac{\mu}{m^3_{\rm T}\tau^3}\;.\label{eq: relazione s e t E hs e jt b}
    \end{align}
\end{subequations}

Eq. \eqref{eq: Pi matrix} makes it apparent that the density operator is non-diagonal in the basis of Milne modes $\widehat{A}_r$ and $\widehat{B}^\dagger_r$. The next step in order to compute exact expectation values is to diagonalize the density operator transforming the Milne modes to new modes, $\widehat{\alpha}_r$ and $\widehat{\beta}^\dagger_s$, with a Bogoliubov transformation. Namely, we will look for a transformation $\mathcal{U}$ such that $\mathcal{U}^\dagger \mathcal{H}_{\rm tot} \mathcal{U} $ is a diagonal matrix, and such that the new modes:
\begin{equation}\label{def: diagonalization}
    \Psi(\fp)\equiv\begin{pmatrix}
       \widehat{\alpha}_+(\fp)\\
       \widehat{\alpha}_-(\fp)\\
       \widehat{\beta}^\dagger_+(\fp)\\
       \widehat{\beta}^\dagger_-(\fp)\end{pmatrix}=\mathcal{U}^\dagger\Phi(\fp)\;,
\end{equation}
continue obeying the canonical anticommutation relations:
\begin{equation}\label{eq: anticomm alpha}
    \Big\{\widehat{\alpha}_s(\fp),\widehat{\alpha}^\dagger_{r}(\fp')\Big\}=\left\{\widehat{\beta}_s(\fp),\widehat{\beta}^\dagger_{r}(\fp')\right\}=\delta_{sr}\delta^3\left(\fp-\fp'\right)\;.
\end{equation}

Before performing the explicit diagonalization, we end the section by remarking that, in our case, the unitarity of $\mathcal{U}$ alone ensures the anticommutation relations. Indeed, assuming a block structure:
\begin{equation}
    \mathcal{U}=\begin{pmatrix}
        u &v\\
        w & z
    \end{pmatrix}\;,
\end{equation}
one can see that the anticommutation rules of $\widehat{A}$ and $\widehat{B}$ imply canonical anticommutation for $\widehat{\alpha}$ and $\widehat{\beta}$ only if:
\begin{align*}
    u^\dagger u+w^\dagger w =& u u^\dagger+v v^\dagger =\mathbb{1}\,,\\
    v^\dagger v + z^\dagger z =& w w^\dagger +z z^\dagger= \mathbb{1}\;,\\
    u^\dagger v+ w^\dagger z =& 0\;.
\end{align*}
However, these conditions are the same ensuring $\mathcal{U}^\dagger \mathcal{U} =\mathbb{1}$. Therefore, any unitary transformation that diagonalizes $\mathcal{H}_{\rm tot}$ is a valid Bogoliubov transformation.

\subsection{Expectation values from Milne modes and their renormalization}\label{sec:exp values and renormalization}

For future reference, in this section we express the expectation values of interest in terms of Milne particle-antiparticle operators. This will be done in the canonical pseudogauge, from which the Belinfante expectation values can be easily recovered using eq. \eqref{eq: Belinfante Semt} and setting $\Omega=0$. In particular, denoting with a ${\rm C}$ or ${\rm B}$ superscript respectively the canonical and Belinfante expectation values, one has: 
\begin{equation}\label{eq: belinfante substitutions}
    \begin{split}
        \mathcal{E}^{({\rm B})}&=\left.\mathcal{E}^{({\rm C})}\right|_{\Omega=0}\;,\qquad
        \mathcal{P}_{\rm T}^{({\rm B})}=\left.\mathcal{P}_{\rm T}^{({\rm C})}\right|_{\Omega=0}\;,\\
        \mathcal{P}_{\rm L}^{({\rm B})}&=\left.\mathcal{P}_{\rm L}^{({\rm C})}\right|_{\Omega=0}\;,\qquad
        \!\mathcal{T}^{({\rm B})}=\mathcal{S}^{({\rm B})}=0\;.
    \end{split}
\end{equation} 
These expectation values are obtained by plugging the field expansion eq.~\eqref{eq: field final expansion} inside the stress-energy tensor operator eq.~\eqref{subeq: canonical T}, calculating the contractions between the Milne spinors \eqref{eq: Milne spinors}, and computing the expectation values of the combinations of Milne creation and annihilation operators. 

Note that the field operator in Milne coordinate transform under translations in the transverse and rapidity directions as:
\begin{align*}
    \e^{\ii\widehat{\bf P}_{\rm T}\cdot{\bf y}_{\rm T}}\,\psi\left(\tau,{\bf x}_{\rm T},\eta\right)\e^{-\ii\widehat{\bf P}_{\rm T}\cdot{\bf y}_{\rm T}}&=\psi\left(\tau,{\bf x}_{\rm T}+{\bf y}_{\rm T},\eta\right)\;,\\
    \e^{\ii\widehat{{\rm K}}_z\varrho}\,\psi\left(\tau,{\bf x}_{\rm T},\eta\right)\e^{-\ii\widehat{{\rm K}}_z\varrho}&=\psi\left(\tau,{\bf x}_{\rm T},\eta+\varrho\right)\;.
\end{align*}
which implies:
\begin{subequations}
    \begin{align*}
        \e^{\ii\widehat{\bf P}\cdot\bf{y}_{\rm T}+\ii\widehat{{\rm K}}_z\varrho}\widehat{A}_r(\fp)\e^{-\ii\widehat{\bf P}\cdot\bf{y}_{\rm T}-\ii\widehat{{\rm K}}_z\varrho}&=\e^{-\ii\fr\cdot\fp}\widehat{A}_r(\fp)\;,\\
        \e^{\ii\widehat{\bf P}\cdot\bf{y}_{\rm T}+\ii\widehat{{\rm K}}_z\varrho}\widehat{A}^\dagger_r(\fp)\e^{-\ii\widehat{\bf P}\cdot\bf{y}_{\rm T}-\ii\widehat{{\rm K}}_z\varrho}&=\e^{\ii\fr\cdot\fp}\widehat{A}^\dagger_r(\fp)\;.
    \end{align*}
\end{subequations}

Using the above equations and the invariance of the statistical operator under translations in the transverse and longitudinal directions, eq. \eqref{invariant rho}, the expectation values of Milne creation and annihilation operators can be factorised as follows:
\begin{equation}\label{eq: tev AB}
    \begin{split}
        \langle\widehat{A}^\dagger_r(\fp)\widehat{A}_{r'}(\fp ')\rangle_{\rm LE}&\equiv\mathcal{A}_{rr'}(\tau,\fp)\delta^3(\fp-\fp')\,,\\
        \langle\widehat{B}_r(\fp)\widehat{B}^\dagger_{r'}(\fp ')\rangle_{\rm LE}&\equiv\overline{\mathcal{B}}_{rr'}(\tau,\fp)\delta^3(\fp-\fp')\equiv\left(\delta_{rr'}-\mathcal{B}_{r'r}\left(\tau,\fp\right)\right)\delta^3\left(\fp-\fp'\right)\;,\\
        \langle\widehat{A}^\dagger_r(\fp)\widehat{B}^\dagger_{r'}(\fp ')\rangle_{\rm LE}&\equiv\mathcal{C}_{rr'}(\tau,\fp)\delta^3(\fp+\fp')\;,\\
        \langle\widehat{B}_r(\fp)\widehat{A}_{r'}(\fp ')\rangle_{\rm LE}&\equiv\mathcal{C}^*_{r'r}(\tau,\fp')\delta^3(\fp+\fp')=-\mathcal{C}^*_{r'r}(\tau,\fp)\delta^3(\fp+\fp')\;,
    \end{split}
\end{equation}
where in the second line we used eq.~\eqref{canonical anticommutator} and implicitly defined $\langle\widehat{B}^\dagger_{r'}(\fp ')\widehat{B}_r(\fp)\rangle_{\rm LE}=\mathcal{B}_{r'r}(\tau,\fp)$. These expectation values are in general non-diagonal in the spin indices $r,r'$, whereas the $\tau$ dependence is inherited from the local equilibrium operator $\wrho_{\rm LE}$.  The relation $\mathcal{C}(-\fp)=-\mathcal{C}(\fp)$ follows from parity:
\begin{equation*}
  \begin{split}
      &\mathcal{C}_{rr'}(\tau,-\fp)\delta^3(\fp+\fp') =\langle\widehat{A}^\dagger_r(-\fp)\widehat{B}^\dagger_{r'}(-\fp ')\rangle_{\rm LE}\\
      &=\eta_A\eta_B\langle \widehat{\rm P}\widehat{A}^\dagger_r(\fp)\widehat{\rm P}^2\widehat{B}^\dagger_{r'}(\fp ')\widehat{\rm P}\rangle_{\rm LE}=-\mathcal{C}_{rr'}(\tau,\fp)\delta^3(\fp+\fp')\;, 
  \end{split}
\end{equation*}
where we also used the invariance of the density operator under parity $\widehat{\rm P}$, and $\eta_A\eta_B=-1$ follows from the transformation rules of the Dirac field under parity (see e.g. \cite{Peskin:1995ev}).

Before setting to the explicit calculation of Bogolyubov transformations and the diagonalization of the density operator, we remark that the two-point functions in eq. \eqref{eq: tev AB} yield divergent thermodynamic expectation values, unless properly renormalized. For a free theory, renormalization can be done simply by removing the vacuum contribution from expectation values, however, for field theories in non-stationary backgrounds, the notion of vacuum state is particularly delicate\footnote{This is a well known issue for quantum field theory in curved space-time (see for instance \cite{Fulling_1989}). For truly curved background indeed the definition of a vacuum is, to a large extent, arbitrary. This is not the case for the Milne space-time in which the Minkowski vacuum is still naturally defined and, as we have shown, it is actually the state annihilated by the Milne modes.}.

Even if the Milne space-time is just a re-parametrization of the Minkowski spacetime,
in ref. \cite{Rindori:2021quq} it was explicitly shown that renormalizing expectation values by subtracting the static Minkowski vacuum is insufficient, as expectation values computed this way are still divergent.
Two alternative methods were proposed: the first consist in subtracting the \emph{time-dependent} Milne vacuum $|0_\tau\rangle$, corresponding to the lowest lying state of the Milne Hamiltonian $H(\tau)=\int \di^3\fp \Phi^\dagger \mathcal{H}\Phi$. Such a vacuum is proper-time dependent because, as explained after eq. \eqref{Pi tau generico}, $T^{00}$ is not a proper vector density as $\hat{\tau}$ is not a Killing vector. This implies that expectation values renormalized this way, although finite, won't obey the equations of motion. For instance, one would have:
\begin{equation*}
\nabla_\mu \left(\langle \wT^{\mu\nu}\rangle-\langle0_\tau|\wT^{\mu\nu}|0_\tau\rangle\right)\neq 0.    
\end{equation*}
In what follows, expectation values renormalized like this will be referred to as \emph{local equilibrium expectation values}. 
The true non equilibrium expectation values, however, should be renormalized with respect to the vacuum at a constant proper time, that is the state $|0_{\tau_0}\rangle$, where $\tau_0$ can be interpreted as decoupling time. Assuming an instantaneous freeze-out, the choice of $\tau_0$ as the decoupling time is natural, as $\tau_0$ is the only proper-time in which the system is in local thermodynamic equilibrium and can be described in terms of quasi-free fields. 

Since the limit $T\rightarrow 0$ of the density operator usually selects naturally the vacuum state, this renormalization procedure was implemented in ref. \cite{Rindori:2021quq} (see also \cite{Becattini:2022bia,Becattini:2024vtf}) by computing the $T\rightarrow 0$ limit of some expectation values, interpreting the finite results of such a limit as vacuum contributions, and subtracting such vacuum remainder from the original expectation value. Such a subtraction indeed produces finite expressions for the expectation value of the stress-energy tensor, however it can yield incorrect results when additional Lagrange multipliers, such as the chemical potential, are included in the density operator. For instance, it is known that the pressure of a gas of free massless fermions at zero temperature is $\mathcal{P}\propto \mu^4$, but such a term would be overlooked with this procedure because it would survive to the $T\rightarrow 0$ limit and hence be incorrectly removed in vacuum subtraction.

In our case this problem is easily solved, simply taking \emph{first} the limit of $\zeta,\SP\rightarrow0$ and \emph{only then} the limit of $T\rightarrow0$. This procedure correctly maps the density operator of a boost invariant fluid into the projector on the vacuum state of the Milne hamiltonian:
\begin{equation}\label{def: Vacuum Sub}
    \lim_{T\rightarrow0}\left(\lim_{\zeta\rightarrow0}\lim_{\SP\rightarrow0}\wrho\right)=\lim_{T\rightarrow0}\frac{1}{Z_{\zeta,\SP=0}}\exp\left(-\frac{1}{T(\tau_0)}\int\di^3\mathfrak{r}\; \wT^{00} \right)=|0_{\tau_0}\rangle\langle0_{\tau_0}|\;.
\end{equation}
In other words, to renormalize the expectation values in eqs.\eqref{eq: tev AB}, it suffices to compute, for example:
\begin{equation}
    :\mathcal{A}_{rr'}(\tau_0,\mathfrak{p}):\,\equiv\mathcal{A}_{rr'}(\tau_0,\mathfrak{p})-\lim_{T\rightarrow0}\mathcal{A}_{rr'}(\tau_0,\mathfrak{p})\Big|_{\zeta,\SP=0},
\end{equation}
which defines our normal ordering. From eqs. \eqref{eq: tev AB}, one realises that this limit boils down to, for example, $:\overline{\mathcal{B}}_{rs}(\tau,\fp):=-:\mathcal{B}_{sr}(\tau,\fp):$.

We can now express the expectation values of interest in terms of $\mathcal{A}$, $\mathcal{B}$, and $\mathcal{C}$, to facilitate the calculations after the explicit Bogoliubov transformation are found.
Starting from the energy density and proceeding like in appendix \ref{app: computation and asymptotics}, using the field expansion eq.~\eqref{eq: field final expansion}, the definitions in eqs.~\eqref{eq: tev AB}, and the identities in eqs.~\eqref{app: derivate spinori finale}, we have:
\begin{equation}\label{Energy generale}
    \begin{split}
        \mathcal{E}^{(ren,\ \tau)}_{\rm LE}(\tau)&=\frac{1}{\left(2\pi\right)^3}\sum_{r,r'}\int\di^3\fp\,m^2_{\rm T}\Big\{\delta_{rr'}\,\mathfrak{h}\left(\tau,\mathfrak{p}\right)\left[:\mathcal{A}_{rr'}\left(\tau,\fp\right):+:\mathcal{B}_{r'r}\left(\tau,\fp\right):\right]\\
        &+(\sigma^x)_{rr'}\,\left[\,{\rm j}^*\left(\tau,\mathfrak{p}\right):\mathcal{C}_{rr'}\left(\tau,\fp\right):+{\rm j}\left(\tau,\mathfrak{p}\right):\mathcal{C}^*_{r'r}\left(\tau,\fp\right):\right]\Big\}\;.
    \end{split}
\end{equation}
The longitudinal pressure is computed in a similar fashion. The spinor contractions are computed in appendix \ref{app: spinors}, and one finds:
\begin{equation}\label{Press L generale}
    \begin{split}
    \mathcal{P}^{{\rm L},(ren,\ \tau)}_{\rm LE}(\tau)&=-\frac{1}{(2\pi)^3\,\tau}\sum_{rr'}\int\di^3\fp\,\mu\,m_{\rm T}\Big\{\,{\rm s}\left(\tau,\fp\right)\delta_{rr'}\left[:\mathcal{A}_{rr'}\left(\tau,\fp\right):+:\mathcal{B}_{r'r}\left(\tau,\fp\right):\right]\\
    &+\ii(\sigma^x)_{rr'}\left[{\rm t}\left(\tau,\fp\right):\mathcal{C}^*_{r'r}(\tau,\mathfrak{p}):-{\rm t}^*\left(\tau,\fp\right):\mathcal{C}_{rr'}(\tau,\mathfrak{p}):\right]\Big\}\;.
    \end{split}
\end{equation}
The expression for the transverse pressure is more involved due to the more complex spinor product, which is reported in eq. \eqref{app eq: spinor contractions PT}. To avoid overburdening the text, we don't substitute the explicit spinor products, and simply write:
\begin{equation}\label{Press T generale}
    \begin{split}
        \mathcal{P}^{ {\rm T},(ren,\ \tau)}_{\rm LE}(\tau)&=\frac{1}{2}\frac{1}{\left(2\pi\right)^3}\sum_{r,r'}\int\di^3\fp\, p^x \Big\{:\mathcal{A}_{rr'}(\tau,\fp):U_r^\dagger(\tau,\fp)\gamma^\uzero\gamma^\uone U_{r'}(\tau,\fp)\\
        &+:\mathcal{B}_{r'r}(\tau,\fp):V_r^\dagger(\tau,\fp)\gamma^\uzero\gamma^\uone V_{r'}(\tau,\fp)\\
        &+:\mathcal{C}_{rr'}(\tau,\fp):U_r^\dagger(\tau,\fp)\gamma^\uzero\gamma^\uone V_{r'}(\tau,-\fp)\\
        &+:\mathcal{C}^*_{r'r}(\tau,\fp):V_r^\dagger(\tau,\fp)\gamma^\uzero\gamma^\uone U_{r'}(\tau,-\fp)\Big\}\\
        &\!+\frac{1}{2}\frac{1}{\left(2\pi\right)^3}\sum_{r,r'}\int\di^3\fp\, p^y \Big\{ \ \gamma^\uone\mapsto\gamma^\utwo \ \Big\}\;.
    \end{split}
\end{equation}
Concerning observables that may be non-vanishing only in the canonical pseudogauge, the spin density reads:
\begin{equation}\label{Spin density generale}
    \begin{split}
        \mathcal{S}^{(ren,\ \tau)}_{\rm LE}(\tau)&=\frac{1}{2\left(2\pi\right)^3}\sum_{rr'}\int\di^3\fp\,\Big\{:\mathcal{A}_{rr'}\left(\tau,\fp\right):\left[\frac{ m}{m_{\rm T}\tau}\left(\sigma^z\right)_{rr'}-\,{\rm s}(\tau,\mathfrak{p})\left(p^x\sigma^x+p^y\sigma^y\right)_{rr'}\right]\\
        &-:\mathcal{B}_{r'r}\left(\tau,\fp\right):\left[-\frac{ m}{m_{\rm T}\tau}(\sigma^z)_{rr'}+\,{\rm s}(\tau,\mathfrak{p})\left(p^x\sigma^x-p^y\sigma^y\right)_{rr'}\right]\\
        &+:\mathcal{C}_{rr'}:{\rm t}^*(\tau,\mathfrak{p})\,\left(\ii p^x\I+ p^y\sigma^z\right)_{rr'}+:\mathcal{C}^*_{r'r}:{\rm t}(\tau,\mathfrak{p})\,\left(-\ii p^x\I+ p^y\sigma^z\right)_{rr'}\Big\}\;,
    \end{split}
\end{equation}
while the torque is:
\begin{equation}\label{Spin torque generale}
    \begin{split}
        \mathcal{T}^{(ren,\ \tau)}_{\rm LE}(\tau)&=\frac{1}{\left(2\pi\right)^3}\sum_{r,r'}\int\di^3\fp\, p^x \Big\{:\mathcal{A}_{rr'}(\tau,\fp):U_r^\dagger(\tau,\fp)\gamma^\uzero\gamma^\utwo U_{r'}(\tau,\fp)\\
        &+:\mathcal{B}_{r'r}(\tau,\fp):V_r^\dagger(\tau,\fp)\gamma^\uzero\gamma^\utwo V_{r'}(\tau,\fp)\\
        &+:\mathcal{C}_{rr'}(\tau,\fp):U_r^\dagger(\tau,\fp)\gamma^\uzero\gamma^\utwo V_{r'}(\tau,-\fp)\\
        &+:\mathcal{C}^*_{r'r}(\tau,\fp):V_r^\dagger(\tau,\fp)\gamma^\uzero\gamma^\utwo U_{r'}(\tau,-\fp)\Big\}\\
        &-\frac{1}{\left(2\pi\right)^3}\sum_{r,r'}\int\di^3\fp\, p^y \Big\{ \ \gamma^\utwo\mapsto\gamma^\uone \ \Big\}\;.
    \end{split}
\end{equation}
All the needed contractions of the Milne spinors are computed explicitly in appendix \ref{app: spinors}.

Finally, the spin density matrix of particle states in a boost invariant fluid can be simply expressed using eqs. \eqref{eq: tev AB}, \eqref{particle number a e A}, and the definition \eqref{eq: def spin density matrix}, which yield:
\begin{equation}\label{eq: spin density matrix with A/trA}
    \Theta_{sr}\left(\tau,{\bf p}_{\rm T}\right) = \frac{\int_{-\infty}^\infty \di \mu\;: \mathcal{A}_{rs}(\tau,\mathfrak{p}):}{\int_{-\infty}^\infty \di \mu\; \tr(:\mathcal{A}(\tau,\mathfrak{p}):)}\;.
\end{equation}
Once the explicit expectation values in eqs. \eqref{eq: tev AB} are known, the expressions reported above can be used to calculate thermodynamic quantities.

%


\section{Belinfante Pseudo-gauge }\label{sec: bel}

We start by considering the simpler case of the Belinfante pseudo-gauge, which can be obtained from \eqref{eq: Pi matrix} setting $\Omega=0$. Ignoring the charge operator for the moment, since it is proportional to the identity matrix in agreement with the conservation of the four-current, we have:
\begin{equation}
    \wPi(\tau)=\int\di^3\fp\,\Phi^\dagger(\fp)\,\mathcal{H}\left(\tau,\fp\right)\,\Phi(\fp)\,,
\end{equation}
with $\Phi$ and $\mathcal{H}$ being defined in eqs.~\eqref{def: NG spinor} and \eqref{def: matrices}, respectively. 
We note, by direct computation, that:
\begin{equation}\label{eq: spectrum Ht2}
    \mathcal{H}^2=m^4_{\rm T}\tau^2\left(\mathfrak{h}^2(\tau,\fp)+\left|{\rm j}(\tau,\mathfrak{p})\right|^2\right)\I=\varepsilon^2(\tau,\fp)\,\I\;,
\end{equation}
where we used the relation \eqref{eq: relazione h j} and we recall $\varepsilon(\tau,\fp)=({m_{\rm T}^2+\mu^2/\tau^2})^{1/2}$. We conclude that the matrix $\mathcal{H}$ has two disjoint real eigenvalues, both with double algebraic multiplicity:
\begin{equation*}
    \varepsilon^+(\tau,\fp)=\sqrt{m^2_{\rm T}+\frac{\mu^2}{\tau^2}}\;,\qquad \varepsilon^-(\tau,\fp)=-\sqrt{m^2_{\rm T}+\frac{\mu^2}{\tau^2}}\;,
\end{equation*}
and thus the spectrum is simply: $\left\{\varepsilon^+,\varepsilon^+,\varepsilon^-,\varepsilon^-\right\}$ with eigenvalues \emph{independent} of the spin of the mode.
The  eigenspace of $\mathcal{H}$ is spanned by the following orthonormal vectors:
\begin{equation}\label{def: eigenspace belin}
    \begin{split}
        \left|\varepsilon^+_s\right\rangle&=\frac{1}{\sqrt{2\varepsilon(\tau,\fp)\left(\varepsilon(\tau,\fp)-m^2_{\rm T}\tau\,\mathfrak{h}\left(\tau,\mathfrak{p}\right)\right)}}\begin{pmatrix}
        m^2_{\rm T}\tau\,{\rm j}^*\left(\tau,\mathfrak{p}\right)\,\chi_s\\
        \left(\varepsilon(\tau,\fp)-m^2_{\rm T}\tau\,\mathfrak{h}\left(\tau,\mathfrak{p}\right)\right)\,\chi_{-s}
    \end{pmatrix}\;,\\
    \left|\varepsilon^-_s\right\rangle&=\frac{1}{\sqrt{2\varepsilon(\tau,\fp)\left(\varepsilon(\tau,\fp)+m^2_{\rm T}\tau\,\mathfrak{h}\left(\tau,\mathfrak{p}\right)\right)}}\begin{pmatrix}
        -m^2_{\rm T}\tau\,{\rm j}^*\left(\tau,\mathfrak{p}\right)\,\chi_s\\
        \left(\varepsilon(\tau,\fp)+m^2_{\rm T}\tau\,\mathfrak{h}\left(\tau,\mathfrak{p}\right)\right)\,\chi_{-s}
    \end{pmatrix}\;,
    \end{split}
\end{equation}
where $s=\pm$ and $\chi_s$ are the standard eigenvectors of $\sigma^z$, see eq.~\eqref{def: chi pm}. The matrix $\mathcal{U}$ diagonalizing $\mathcal{H}$ is then obtained directly from eq.~\eqref{def: eigenspace belin}:
\begin{equation}\label{eq: diag matrix Bel}
    \mathcal{U}\left(\tau,\fp\right)=\begin{pmatrix}
        \frac{m^2_{\rm T}\tau\,{\rm j}^*(\tau,\fp)}{\sqrt{2\varepsilon(\tau,\fp)(\varepsilon(\tau,\fp)-m^2_{\rm T}\tau\,\mathfrak{h}(\tau,\fp))}}\,\I_{2\times2}&-\frac{m^2_{\rm T}\tau\,{\rm j}^*(\tau,\fp)}{\sqrt{2\varepsilon(\tau,\fp)(\varepsilon(\tau,\fp)+m^2_{\rm T}\tau\,\mathfrak{h}(\tau,\fp))}}\,\I_{2\times2}\\
        \frac{\varepsilon(\tau,\fp)-m^2_{\rm T}\tau\,\mathfrak{h}(\tau,\fp)}{\sqrt{2\varepsilon(\tau,\fp)(\varepsilon(\tau,\fp)-m^2_{\rm T}\tau\,\mathfrak{h}(\tau,\fp))}}\,\sigma_x& \frac{\varepsilon(\tau,\fp)+m^2_{\rm T}\tau\,\mathfrak{h}(\tau,\fp)}{\sqrt{2\varepsilon(\tau,\fp)(\varepsilon(\tau,\fp)+m^2_{\rm T}\tau\,\mathfrak{h}(\tau,\fp))}}\,\sigma_x
    \end{pmatrix}\;,
\end{equation}
and one has:
\begin{equation*}
    \mathcal{D}(\tau,\fp)=\mathcal{U}^\dagger\left(\tau,\fp\right)\mathcal{H}\left(\tau,\fp\right)\mathcal{U}\left(\tau,\fp\right)\;,\qquad\mathcal{D}(\tau,\fp)=\begin{pmatrix}
        \varepsilon(\tau,\fp)\,\I_{2\times 2}&0\\
        0&-\varepsilon(\tau,\fp)\,\I_{2\times2}
    \end{pmatrix}\;.
\end{equation*}
The diagonalizing set of creation and annihilation operators is found using eq.~\eqref{def: diagonalization}, and they read:%
\begin{equation}\label{eq: Bogolubov}
    \begin{split}
        \widehat{\alpha}_s(\fp)&=\frac{m^2_{\rm T}\tau\,{\rm j}\left(\tau,\mathfrak{p}\right)}{\sqrt{2\varepsilon(\tau)\left(\varepsilon(\tau,\fp)-m^2_{\rm T}\tau\,\mathfrak{h}\left(\tau,\mathfrak{p}\right)\right)}}\,\widehat{A}_s(\fp)\\
        &+\sqrt{\frac{\varepsilon(\tau,\fp)-m^2_{\rm T}\tau\,\mathfrak{h}\left(\tau,\mathfrak{p}\right)}{2\varepsilon(\tau,\fp)}}\,\left(\sigma^x\right)_{ss'}\widehat{B}^\dagger_{s'}(-\fp)\;,\\
        \widehat{\beta}^\dagger_s(\fp)&=-\frac{m^2_{\rm T}\tau\,{\rm j}\left(\tau,\mathfrak{p}\right)}{\sqrt{2\varepsilon(\tau,\fp)\left(\varepsilon(\tau,\fp)+m^2_{\rm T}\tau\,\mathfrak{h}(\tau,\mathfrak{p})\right)}}\,\widehat{A}_s(\fp)\\
        &+\sqrt{\frac{\varepsilon(\tau,\fp)+m^2_{\rm T}\tau\,\mathfrak{h}\left(\tau,\mathfrak{p}\right)}{2\varepsilon(\tau,\fp)}}\,\left(\sigma^x\right)_{ss'}\widehat{B}^\dagger_{s'}(-\fp)\;.
    \end{split}
\end{equation}
Note that since the matrix $\mathcal{U}$ is unitary by construction, $\widehat{\alpha}$ and $\widehat{\beta}$ obey the canonical commutation rules as explained at the end of section \ref{sec:exp values and renormalization}. This can be also checked explicitly, making use of the identity \eqref{eq: relazione h j}.

Furthermore, using the the inverse Bogolubov transformation $\Phi=\mathcal{U}\,\Psi$, and the Milne creation and annihilation operators can be expressed in terms of $\alpha$ and $\beta^\dagger$:
\begin{equation}\label{eq: bogolubov-1}
    \begin{split}
        \widehat{A}_s(\fp)&=\frac{m^2_{\rm T}\tau\,{\rm j}^*\left(\tau,\mathfrak{p}\right)}{\sqrt{2\varepsilon(\tau,\fp)\left(\varepsilon(\tau,\fp)-m^2_{\rm T}\tau\,\mathfrak{h}\left(\tau,\mathfrak{p}\right)\right)}}\,\widehat{\alpha}_s(\fp)\\
        &-\frac{m^2_{\rm T}\tau\,{\rm j}^*\left(\tau,\mathfrak{p}\right)}{\sqrt{2\varepsilon(\tau,\fp)\left(\varepsilon(\tau,\fp)+m^2_{\rm T}\tau\,\mathfrak{h}\left(\tau,\mathfrak{p}\right)\right)}}\,\widehat{\beta}^\dagger_s(\fp)\;,\\
        \widehat{B}^\dagger_s(-\fp)&=\sqrt{\frac{\varepsilon(\tau,\fp)-m^2_{\rm T}\tau\,\mathfrak{h}\left(\tau,\mathfrak{p}\right)}{2\varepsilon(\tau,\fp)}}\,\left(\sigma^x\right)_{ss'}\widehat{\alpha}_{s'}(\fp)\\
        &+\sqrt{\frac{\varepsilon(\tau,\fp)+m^2_{\rm T}\tau\,\mathfrak{h}\left(\tau,\mathfrak{p}\right)}{2\varepsilon(\tau,\fp)}}\left(\sigma^x\right)_{ss'}\widehat{\beta}^\dagger_{s'}(\fp)\;.
    \end{split}
\end{equation}

The $\wPi$ operator can then be expressed in terms of the diagonal set in eq.~\eqref{eq: Bogolubov} and reads:
\begin{equation*}
    \wPi(\tau)=\int\di^3\fp\sum_{s=\pm}\left(\varepsilon(\tau,\fp)\,\widehat{\alpha}^\dagger_s(\fp)\widehat{\alpha}_s(\fp)-\varepsilon(\tau,\fp)\widehat{\beta}_s(\fp)\widehat{\beta}^\dagger_s(\fp)\right)\;.
\end{equation*}
Finally, using the anti-commutation relations \eqref{eq: anticomm alpha}, $\wPi$ can be written:
\begin{equation}\label{eq: Pi diag}
    \wPi(\tau)=\int\di^3\fp\,\varepsilon(\tau,\fp)\sum_{s=\pm}\left(\,\widehat{\alpha}^\dagger_s(\fp)\widehat{\alpha}_s(\fp)+\widehat{\beta}^\dagger_s(\fp)\widehat{\beta}_s(\fp)\right)-E_{0}(\tau)\;,
\end{equation}
where:
\begin{equation}
    E_{0}(\tau)\equiv -\frac{2}{\tau}\delta^3(0)\int\di  p^x\di p^y\di\mu\sqrt{m^2_{\rm T}+\frac{\mu^2}{\tau^2}}\;,
\end{equation}
is the infinite vacuum energy at proper time $\tau$, which can be absorbed in the definition of the partition function.

Since the charge $\wQ$ is proportional to the identity matrix in the Milne basis $\Phi$, the action of the unitary transformation \eqref{eq: Bogolubov} is trivial, and the charge reads:
\begin{equation}
    \widehat{Q}(\tau)=\int\di^3\fp\,\mu(\tau)\sum_{s}\left(\widehat{\alpha}^\dagger_s(\fp)\widehat{\alpha}_s(\fp)-\widehat{\beta}^\dagger_s(\fp)\widehat{\beta}_s(\fp)+\delta^3(0)\right)\;.
\end{equation}
In the diagonal basis, the  operator \eqref{eq: Pi diag} has exactly the canonical form and local-equilibrium thermal expectation values of the quadratic combinations involving the Bogolubov operators $\widehat{\alpha}_r$ and $\widehat{\beta}_r$ are now expressed in terms of the Fermi-Dirac distribution function:
\begin{equation}\label{Fermi Dirac}
    n_{\rm F}\left[\varepsilon(\tau,\fp),\mp\zeta(\tau,\fp)\right]\equiv n^{\mp}_{\rm F}(\tau,\fp)=\frac{1}{\exp\left[\frac{\varepsilon(\tau,\fp)}{T(\tau)}\mp\zeta(\tau)\right]+1}\;.
\end{equation}
Explicitly, one readily finds:
\begin{equation}\label{eq: exp val diago}
    \begin{split}
        \left\langle\widehat{\alpha}^\dagger_s(\fp)\widehat{\alpha}_{s'}(\fp')\right\rangle_{\rm LE}&= \delta_{ss'}\delta^3\left(\fp-\fp'\right)\, n^-_{\rm F}(\tau,\fp)\;,
        \\
        \left\langle\widehat{\beta}^\dagger_s(\fp)\widehat{\beta}_{s'}(\fp')\right\rangle_{\rm LE}&= \delta_{ss'}\delta^3\left(\fp-\fp'\right)\, n_{\rm F}^+\left(\tau,\fp\right)\;,
        \\
        \left\langle\widehat{\beta}_s(\fp)\widehat{\alpha}^\dagger_{s'}(\fp')\right\rangle_{\rm LE}&=\left\langle\widehat{\beta}^\dagger_s(\fp)\widehat{\alpha}^\dagger_{s'}(\fp')\right\rangle_{\rm LE}=\left\langle\widehat{\beta}_s(\fp)\widehat{\alpha}_{s'}(\fp')\right\rangle_{\rm LE}=0\;.
    \end{split}
\end{equation}
Using the above equations, together with the inverse Bogolubov transformation \eqref{eq: bogolubov-1}, one finally finds the local expectation values of the Milne field operators:
\begin{equation}\label{eq: final exp}
    \begin{split}
        \left\langle\widehat{A}^\dagger_s(\fp)\widehat{A}_{s'}(\fp')\right\rangle_{\rm LE}&=\delta_{ss'}\,\delta^3\left(\fp-\fp'\right)\frac{\varepsilon(\tau,\fp)+m^2_{\rm T}\tau\,\mathfrak{h}\left(\tau,\mathfrak{p}\right)}{2\varepsilon(\tau,\fp)}\\
        &\times\left[n^-_{\rm F}(\tau,\fp)+\frac{\varepsilon(\tau,\fp)-m^2_{\rm T}\tau\,\mathfrak{h}\left(\tau,\mathfrak{p}\right)}{\varepsilon(\tau,\fp)+m^2_{\rm T}\tau\,\mathfrak{h}\left(\tau,\mathfrak{p}\right)}\left(1-n^+_{\rm F}(\tau,\fp)\right)\right]\;,\\
        \left\langle\widehat{B}^\dagger_s(\fp)\widehat{B}_{s'}(\fp')\right\rangle_{\rm LE}&=\delta_{ss'}\,\delta^3\left(\fp-\fp'\right)\frac{\varepsilon(\tau,\fp)+m^2_{\rm T}\tau\,\mathfrak{h}\left(\tau,\mathfrak{p}\right)}{2\varepsilon(\tau,\fp)}\\
        &\times\left[n^+_{\rm F}(\tau,\fp)+\frac{\varepsilon(\tau,\fp)-m^2_{\rm T}\tau\,\mathfrak{h}\left(\tau,\mathfrak{p}\right)}{\varepsilon(\tau,\fp)+m^2_{\rm T}\tau\,\mathfrak{h}\left(\tau,\mathfrak{p}\right)}\left(1-n^-_{\rm F}(\tau,\fp)\right)\right]\;,\\
        \left\langle\widehat{A}^\dagger_s(\fp)\widehat{B}^\dagger_{s'}(-\fp')\right\rangle_{\rm LE}&=\frac{m^2_{\rm T}\tau\,{\rm j}\left(\tau,\mathfrak{p}\right)}{2\varepsilon(\tau,\fp)}\left[n^-_{\rm F}(\tau,\fp)+n^+_{\rm F}(\tau,\fp)-1\right]\delta^3\left(\fp-\fp'\right)\left(\sigma^x\right)_{s's}\;,\\
        \left\langle\widehat{B}_s(-\fp)\widehat{A}_{s'}(\fp')\right\rangle_{\rm LE}&=\frac{m^2_{\rm T}\tau\,{\rm j}^*\left(\tau,\mathfrak{p}\right)}{2\varepsilon(\tau,\fp)}\left[n^-_{\rm F}(\tau,\fp)+n^+_{\rm F}(\tau,\fp)-1\right]\delta^3\left(\fp-\fp'\right)\left(\sigma^x\right)_{ss'}\;.
    \end{split}
\end{equation}
Note that the mixed expectation values are odd under $\fp\mapsto -\fp$, which is in agreement with eq.~\eqref{eq: parity prop}, whereas the $\langle \widehat{A}^\dagger \widehat{A}\rangle$ and $\langle \widehat{B}^\dagger \widehat{B}\rangle$ terms are even.
Also, according to the asymptotic expressions (see  \eqref{app: funz tau GRANDE}), in the far future the above expectation values reduce to the one we would obtain in the usual asymptotic Minkowski space-time:
\begin{equation*}
    \begin{split}
        \lim_{\tau\to\infty}\left\langle\widehat{A}^\dagger_s(\fp)\widehat{A}_{s'}(\fp')\right\rangle_{\rm LE}&=n^-_{\rm F}\left(\frac{\varepsilon}{T}\right)\delta^3\left(\fp-\fp'\right)\delta_{ss'}\;,\\
        \lim_{\tau\to\infty}\left\langle\widehat{B}^\dagger_s(\fp)\widehat{B}_{s'}(\fp')\right\rangle_{\rm LE}&=n^+_{\rm F}\left(\frac{\varepsilon}{T}\right)\delta^3\left(\fp-\fp'\right)\delta_{ss'}\;,\\
        \lim_{\tau\to\infty}\left\langle\widehat{A}^\dagger_s(\fp)\widehat{B}^\dagger_{s'}(-\fp')\right\rangle_{\rm LE}&=
        \left\langle\widehat{B}_s(-\fp)\widehat{A}_{s'}(\fp')\right\rangle_{\rm LE}=0\;.
    \end{split}
\end{equation*}

The above expressions are valid for the local equilibrium operator $\wrho_{\rm LE}(\tau)$. 
From the definition \eqref{def: ZUB1} the corresponding non-equilibrium expectation values, that this those computed in the true state of the system, are obtained from \eqref{eq: final exp} setting $\tau=\tau_0$. Finally, in order to remove vacuum contributions, we subtract from the expressions \eqref{eq: final exp} their limit $\mu(\tau_0)\rightarrow 0$ and $T(\tau_0)\rightarrow 0$, as explained at the end of section \ref{sec:exp values and renormalization}. The renormalized result reads:
\begin{equation}\label{eq: final exp NON EQ renormalized}
    \begin{split}
        \left\langle:\widehat{A}^\dagger_s(\fp)\widehat{A}_{s'}(\fp'):\right\rangle&=\delta_{ss'}\,\delta^3\left(\fp-\fp'\right)\frac{\varepsilon(\tau_0,\fp)+m^2_{\rm T}\tau_0\,\mathfrak{h}\left(\tau_0,\fp\right)}{2\varepsilon(\tau_0,\fp)}\\
        &\times \left[n^-_{\rm F}(\tau_0,\fp)-\frac{\varepsilon(\tau_0,\fp)-m^2_{\rm T}\tau_0\,\mathfrak{h}(\tau_0,\fp)}{\varepsilon(\tau_0,\fp)+m^2_{\rm T}\tau_0\,\mathfrak{h}(\tau_0,\fp)}n^+_{\rm F}(\tau_0,\fp)\right]\;,\\
        \left\langle:\widehat{B}^\dagger_s(\fp)\widehat{B}_{s'}(\fp'):\right\rangle&=\delta_{ss'}\,\delta^3\left(\fp-\fp'\right)\frac{\varepsilon(\tau_0,\fp)+m^2_{\rm T}\tau_0\,\mathfrak{h}(\tau_0,\fp)}{2\varepsilon(\tau_0,\fp)}\\
        &\times \left[n^+_{\rm F}(\tau_0,\fp)-\frac{\varepsilon(\tau_0,\fp)-m^2_{\rm T}\tau_0\,\mathfrak{h}(\tau_0,\fp)}{\varepsilon(\tau_0,\fp)+m^2_{\rm T}\tau_0\,\mathfrak{h}(\tau_0,\fp)}n^-_{\rm F}(\tau_0,\fp)\right]\;,\\
        \left\langle:\widehat{A}^\dagger_s(\fp)\widehat{B}^\dagger_{s'}(-\fp):\right\rangle&=\frac{m^2_{\rm T}\tau_0\,{\rm j}(\tau_0,\fp)}{2\varepsilon(\tau_0,\fp)}\left[n^-_{\rm F}(\tau_0,\fp)+n^+_{\rm F}(\tau_0,\fp)\right]\delta^3\left(\fp-\fp'\right)\left(\sigma^x\right)_{s's}\;,\\
        \left\langle:\widehat{B}_s(-\fp)\widehat{A}_{s'}(\fp):\right\rangle&=\frac{m^2_{\rm T}\tau_0\,{\rm j}^*(\tau_0,\fp)}{2\varepsilon(\tau_0,\fp)}\left[n^-_{\rm F}(\tau_0,\fp)+n^+_{\rm F}(\tau_0,\fp)\right]\delta^3\left(\fp-\fp'\right)\left(\sigma^x\right)_{ss'}\;.
    \end{split}
\end{equation}

We conclude this section with an important remark.
The Bogoliubov transformation~\eqref{eq: Bogolubov} is reminiscent of the Unruh effect \cite{Crispino:2007eb} for accelerated observers and of particle production in genuinely curved space-times. However, as discussed in~\cite{Rindori:2021quq}, this analogy can be misleading, as the underlying physics is fundamentally different. Although the four-velocity field we consider has non-vanishing acceleration, the quantization is performed on a connected patch of Minkowski space-time. In particular, the Milne particle operator $\widehat{A}$ is expressed as a linear combination of Cartesian particle operators $\widehat{a}$ without any mixing between particle and antiparticle modes, unlike in the Rindler case, where such mixing arises due to the presence of causally disconnected wedges.
Furthermore, the explicit time dependence of the effective Hamiltonian suggests that its instantaneous ground state at a given proper time $\tau$ does not coincide with that at a later time $\tau_1>\tau$, which may appear analogous to the situation in genuinely time-dependent \cite{Fulling_1989} (e.g.\ expanding) backgrounds. Once again, this similarity can be misleading. In non-static space-times, there is in general no preferred vacuum state, nor a unique asymptotic definition of particle modes; selecting a vacuum at some time typically leads to particle production at later times due to the dynamical background. In contrast, in the present case the time dependence of the effective Hamiltonian is purely a coordinate artifact: independently of $\tau$, the Milne annihilation operators always annihilate the Minkowski vacuum, and no genuine particle production occurs.
\subsection{Expectation values and spin polarization}
From eq.~\eqref{def: E PL PT}, and taking into account that in the Belinfante pseudogauge the spin-torque density vanishes, one can see that the energy momentum tensor depends on three scalar functions:
\begin{equation}\label{def: LTE obs}
    \mathcal{E}_{\rm LE}(\tau)=\langle\wT^{\tau\tau}\rangle_{\rm LE}\;,\quad\mathcal{P}^{\rm T}_{\rm LE}(\tau)=\frac{1}{2}\left(\langle\wT^{xx}\rangle_{\rm LE}+\langle\wT^{yy}\rangle_{\rm LE}\right)\;,\quad \mathcal{P}^{\rm L}_{\rm LE}(\tau)\equiv\tau^{2}\langle\wT^{\eta\eta}\rangle_{\rm LE}\;.
\end{equation}
In ref. \cite{Rindori:2021quq}, it was found that the local equilibrium components for the free scalar field in Milne coordinates reduce to the usual kinetic expressions. We will show that this is also true for Dirac fermions.

We start from the local equilibrium energy density eq. $\mathcal{E}_{\rm LE}$. Plugging eqs. \eqref{eq: field final expansion} and \eqref{eq: final exp} in \eqref{def: LTE obs} we get, after integrating over $\di^3\fp'$ and summing over $r,r'$, and some algebra involving eq. \eqref{eq: relazione h j}: 
\begin{equation}\label{eq: E2}
    \begin{split}
        \mathcal{E}_{\rm LE}(\tau)=\frac{2}{\left(2\pi\right)^3\tau}\int\di p_x\di p_y\di\mu\sqrt{m^2_{\rm T}+\frac{\mu^2}{\tau^2}}\left[n^-_{\rm F}(\varepsilon(\tau))+n^+_{\rm F}(\varepsilon(\tau))-1\right]\;,
    \end{split}
\end{equation}
which coincides with the expression for the energy density of relativistic  fermions given by kinetic theory minus the term associated with the divergent vacuum contribution. This divergence can be cured with the time-dependent vacuum subtraction described at the end of section \ref{sec:exp values and renormalization}. Namely one has:
\begin{equation}\label{E LE norm Bel}
     \begin{split}
         \mathcal{E}^{(ren,\tau)}_{\rm LE}(\tau)&= \mathcal{E}_{\rm LE}(\tau)- \mathcal{E}_{\rm LE}(\tau)\Big|_{\zeta=T=0}\\
         &=\frac{2}{\left(2\pi\right)^3\tau}\int\di p_x\di p_y\di\mu\sqrt{m^2_{\rm T}+\frac{\mu^2}{\tau^2}}\left[n^-_{\rm F}(\varepsilon(\tau))+n^+_{\rm F}(\varepsilon(\tau))\right].
     \end{split}
\end{equation}
For the longitudinal and transverse pressure we proceed analogously. Inserting eq. \eqref{eq: final exp} in eqs. \eqref{Press L generale} and \eqref{Press T generale}, and using eqs.  \eqref{eq: relazione s e t E hs e jt b} and \eqref{eq: relazione hfjz}, we get for the renormalized pressures:
\begin{subequations}
    \begin{align}
        \mathcal{P}^{{\rm L},\,(ren,\ \tau)}_{\rm LE}(\tau)&=\frac{2}{(2\pi)^3\tau}\int\di p_x\di p_y\di\mu\frac{\mu^2}{\tau^2\varepsilon(\tau)}\left[n^-_{\rm F}(\tau)+n^+_{\rm F}(\tau)\right]\;,\label{PLLE Bel}\\
        \mathcal{P}^{{\rm T},\,(ren,\ \tau)}_{\rm LE}(\tau)&=\frac{2}{(2\pi)^3\tau}\int\di p_x\di p_y\di\mu\,\frac{p_{\rm T}^2}{2\varepsilon(\tau)}\left[n^-_{\rm F}(\tau)+n^+_{\rm F}(\tau)\right]\;,        \label{PTLE Bel}
    \end{align}
\end{subequations}
which again are the same we would obtain simply using relativistic kinetic theory.
Hence, the local expectation values of energy and pressure in the Belinfante pseudogauge are equal to the classical ones, with no quantum correction. In particular, no anisotropy is developed between longitudinal and transverse pressure, which was the case also for scalar fields \cite{Rindori:2021quq}, and the longitudinal and transverse pressures are actually the same.

The true non-equilibrium expectation value of the stress-energy tensor can be computed in a similar fashion plugging the expectation values \eqref{eq: final exp NON EQ renormalized}, which already takes into account vacuum subtraction in \eqref{def: Vacuum Sub}.
For the out-of equilibrium energy density we thus find:
\begin{align}\label{E NE BEL}
    \mathcal{E}^{(ren,\ \tau_0)}(\tau)=\frac{2}{\left(2\pi\right)^3\tau_0}&\int\di p_x\di p_y\di\mu\,\frac{m^4_{\rm T}\tau^2_0}{\varepsilon(\tau_0)}\left[\mathfrak{h}\left(\tau,\mathfrak{p}\right)\mathfrak{h}\left(\tau_0,\fp\right)+{\rm Re}\left({\rm j}\left(\tau,\mathfrak{p}\right){\rm j}^*\left(\tau_0,\fp\right)\right)\right]\nonumber\\
    &\times\left[\left(n^-_{\rm F}(\tau_0)+n^+_{\rm F}(\tau_0)\right)\right]\;,
\end{align}
while for the longitudinal and transverse pressure we get:
\begin{subequations}\label{P NE Bel}
    \begin{align}
        \mathcal{P}^{(ren,\ \tau_0)}_{\rm L}(\tau)&=-\frac{2}{(2\pi)^3\tau_0}\int\di p_x\di p_y\di\mu\frac{m^3_{\rm T}\tau^2_0}{\varepsilon(\tau_0,\fp)}\frac{\mu}{\tau}\left[{\rm s}\left(\tau,\fp\right)\mathfrak{h}\left(\tau_0,\fp\right)+{\rm Im}\left[{\rm j}\left(\tau_0,\fp\right){\rm t}^*\left(\tau,\fp\right)\right]\right]\nonumber\\
    &\times\left[\left(n^-_{\rm F}(\tau_0)+n^+_{\rm F}(\tau_0)\right)\right]\;,\\
        \mathcal{P}^{(ren,\ \tau_0)}_{\rm T}(\tau)&=\frac{2}{(2\pi)^3\tau_0}\int\di^3\fp\frac{m^2_{\rm T}\tau^2_0}{\varepsilon(\tau_0,\fp)}\frac{p^2_{\rm T}}{2}\left[2\mathfrak{h}\left(\tau_0,\fp\right)\,{\rm Im}\left[{\rm f}\left(\tau,\fp\right)\right]-{\rm Re}\left[{\rm j}\left(\tau_0,\fp\right){\rm z}\left(\tau,\fp\right)\right]\right]\nonumber\\
    &\times\left[\left(n^-_{\rm F}(\tau_0)+n^+_{\rm F}(\tau_0)\right)\right]\;.
    \end{align}
\end{subequations}
Note that the non-equilibrium expectation values are properly renormalized subtracting the vacuum at the initial (decoupling) time $\tau_0$ and that, as expected, they coincide with the normalized local equilibrium one for $\tau=\tau_0$. 
In general the transverse and longitudinal pressure are different and thus the system may develop an anisotropy.
We properly refer to these quantities computed with the fixed non-equilibrium operator \eqref{def: ZUB1} as \emph{free-streaming} solutions, given that they represent the energy density and pressure of a system of free-particles evolving from an initial local equilibrium state at $\tau=\tau_0$. 

The main conclusion is that the free-streaming energy density and pressures are modified with respect to classical expectations\footnote{Note that the only difference with the local equilibrium expressions \eqref{E LE norm Bel}, \eqref{PLLE Bel} and \eqref{PTLE Bel} lies in the fact that the Fermi-Dirac distribution is evaluated at the decoupling time $\tau_0$.}:
\begin{subequations}\label{class free}
    \begin{align}
        \mathcal{E}_{\rm Class}(\tau)&=\frac{2}{(2\pi)^3\tau}\int\di p_x\di p_y\di\mu\sqrt{m^2_{\rm T}+\frac{\mu^2}{\tau^2}}\left(n^-_{\rm F}(\tau_0)+n^+_{\rm F}(\tau_0)\right)\;,\\
        \mathcal{P}^{\rm L}_{\rm Class}(\tau)&=\frac{2}{(2\pi)^3\tau}\int\di p_x\di p_y\di\mu\frac{\mu^2}{\tau^2\sqrt{m^2_{\rm T}+\frac{\mu^2}{\tau^2}}}\left(n^-_{\rm F}(\tau_0)+n^+_{\rm F}(\tau_0)\right)\;,\\
        \mathcal{P}^{\rm T}_{\rm Class}(\tau)&=\frac{2}{(2\pi)^3\tau}\int\di p_x\di p_y\di\mu\frac{|{\bf p}_{\rm T}|^2}{2\sqrt{m^2_{\rm T}+\frac{\mu^2}{\tau^2}}}\left(n^-_{\rm F}(\tau_0)+n^+_{\rm F}(\tau_0)\right)\;,
    \end{align}
\end{subequations}
due to the initial, non trivial, mixing of Milne modes, in perfect analogy with the scalar field results \cite{Rindori:2021quq}. The discrepancy between quantum field and classical theory predictions originates  from the non-equivalence between the Minkowski vacuum and the lowest-energy state defined with respect to the effective Milne Hamiltonian at the decoupling time (recall the effective Hamiltonian depends on time) and this mismatch induces genuine quantum corrections. Since this applied both for scalar and Dirac field, we conjecture that the only difference between a full quantum free field theory calculation and the classical kinetic theory expectation boils down to a vacuum (renormalization) effect also for higher spin states. 
 Such effects have been estimated for scalar fields within the Wigner-function formalism in Ref.~\cite{Tinti:2023mtv}, and they can become particularly relevant in the vicinity of the decoupling hypersurface. In that regime, deviations from the classical kinetic description may modify the free-streaming properties of the produced particles immediately after freeze-out.  
However, for baryons produced in realistic heavy-ion collisions these corrections are expected to be extremely small.

As an example, consider the $\Lambda$ hyperon and a realistic decoupling time $\tau_0 \simeq 10\,{\rm fm}/c$. In this case, we can consider $m_{\rm T}\tau_0 \gg 1$, in which limit the Hankel functions can be approximated by their large-argument asymptotic expansion (see eq.~\eqref{app: funz tau GRANDE}), yielding, for example, for the energy density:
\begin{equation*}
\begin{split}
    \mathcal{E}^{(ren,\tau_0)}(\tau)
&\approx
\mathcal{E}_{\rm Class}(\tau)
\\
&+
\frac{1}{2(2\pi)^3\tau}\int\di p_x\di p_y\di\mu\frac{\mu^2 m^2_{\rm T}}{\tau^2\varepsilon(\tau,\fp)}\frac{\cos\left[2\left(\varphi(\tau)-\varphi(\tau_0)\right)\right]\left(n^-_{\rm F}(\tau_0)+n^+_{\rm F}(\tau_0)\right)}{\tau^2_0\varepsilon(\tau_0,\fp)}\;,
\end{split}
\end{equation*}
where $\varphi$ is a phase given by $\varphi=m_{\rm T}\tau-{\rm arcsinh}(\mu/(m_{\rm T}\tau))$.
The structure of this correction is similar to what is found in genuinely curved space-times~\cite{Becattini:2024vtf}, where the dimensionless parameter $m_{\rm T}\tau_0$ controls the \emph{adiabaticity} of the dynamics. For $m_{\rm T}\tau_0 \gg 1$, the evolution is adiabatic and the quantum corrections are suppressed by inverse powers of this parameter, so that the result reduces to the classical kinetic-theory expression up to rapidly oscillating and parametrically small terms.  
In the present case the space-time is flat, and there is no genuine gravitational curvature. The adiabaticity parameter therefore does not measure curvature effects, but rather quantifies how close the Milne mode functions are to the standard Minkowski plane waves. When $m_{\rm T}\tau_0 \gg 1$, the two modes effectively coincide, and the quantum field--theoretical evolution becomes indistinguishable from classical free streaming.

To conclude this section, we consider the spin polarization vector. 
Using Eq.~\eqref{eq: spin density matrix with A/trA}, and noticing that 
$\langle \widehat{A}_r^\dagger(p)\widehat{A}_s(p)\rangle \propto \delta_{rs}$,
it immediately follows that the spin density matrix corresponds to an isotropic spin distribution:
\begin{equation}\label{eq: spin matrix BEL}
    \Theta_{rs}(p)=\frac{1}{2}\,\delta_{rs}\;,
\end{equation}
therefore, the spin polarization vector of particles decoupling from the fluid vanishes identically.

This result implies that, in a boost-invariant system, the presence of non-vanishing thermal shear and gradients of the chemical potential is not sufficient to generate a net polarization. We emphasize that this is an exact statement, and therefore independent of the magnitude of such gradients, which may in principle be arbitrarily large.
Furthermore, various expressions for the spin polarization vector of spinor fields at local thermodynamic equilibrium have been proposed in the literature, notably in refs.~\cite{Becattini:2021suc,Liu:2021uhn}. In particular, the formula derived in ref.~\cite{Becattini:2021suc} is based on linear response theory and involves some geometric approximations of the decoupling hypersurface (see also \cite{Palermo:2025imv}).  Because of that, the formula derived in \cite{Becattini:2021suc} allows a non-vanishing shear-induced polarization\footnote{In practice, numerical evidence shows that the formulae of ref.\cite{Becattini:2021suc} and \cite{Liu:2021uhn} agree at midrapidity, so the effect must be very small, see ref. \cite{Alzhrani:2022dpi}.} in a boost-invariant flow, which our result clearly rules out. Recently, a more general formula has been derived in ref.~\cite{Sheng:2025cjk}, which doesn't rely on any geometric approximation of the decoupling hypersurface and thus improves the result of ref.~\cite{Becattini:2021suc}. The shear-induced polarization and the spin hall term computed using ref.~\cite{Sheng:2025cjk}, as well as those of ref.~\cite{Liu:2021uhn} vanish at first order in gradients, which is consistent with the exact result derived here.

This result, while seemingly simple, highlights how strongly symmetries constrain the emergence of polarization. Under the assumption of longitudinal boost invariance—an excellent approximation at very high collision energies—a non-vanishing spin polarization must therefore arise from mechanisms beyond thermodynamic gradients alone, such as interactions or the presence of a finite spin potential, which we will investigate in the following section.



\section{Canonical pseudo-gauge}\label{sec: can}
We now come to the study of the Dirac field in the canonical pseudogauge with a finite spin potential. This case is particularly relevant in the context of spin hydrodynamics, and will allow clarifying aspects such as the thermodynamic properties of such a theory, as well as provide the exact spin polarization ensuing from a fluid with finite spin potential.

In this case, the effective hamiltonian reads, using eq. \eqref{def: matrices}:
\begin{equation}
\begin{split}
    \wPi_\Omega(\tau)=&\int\di^3\fp\,\Phi^\dagger(\fp)\left[\mathcal{H}\left(\tau,\fp\right)-\frac{\tau\Omega}{2}\mathcal{S}\left(\tau,\fp\right)\right]\Phi(\fp)= \int\di^3\fp\,\Phi^\dagger(\fp)\mathcal{H}_{tot}\Phi(\fp)\;.
\end{split}
\end{equation}
with  $\mathcal{H}_{\rm tot}=\mathcal{H}-\tau\Omega\mathcal{S}/2$:
\begin{align}\label{Htot}
    &\mathcal{H}_{\rm tot}=\\
    &\!\!\!{ \begin{pmatrix}
        m_{\rm T}^2\tau\mathfrak{h}\,\I_{2\times2}-\frac{\Omega m}{2m_{\rm T}}\sigma^z+\frac{\tau\Omega\,{\rm s}}{2}\left(p^x\sigma^x-p^y\sigma^y\right)
        &m_{\rm T}^2\tau{\rm j}^*\,\sigma^x-\frac{\tau\Omega\,{\rm t}}{2}\left( \ii p^x\I_{2\times2}+ p^y\sigma^z\right)\\ 
        m_{\rm T}^2\tau{\rm j}\,\sigma^x-\frac{\tau\Omega\,{\rm t}^*}{2}\left(p^y\sigma^z-\ii p^x\I_{2\times 2}\right)
        &-m_{\rm T}^2\tau\mathfrak{h}\,\I_{2\times2}+\frac{\Omega m}{2m_{\rm T}}\sigma^z-\frac{\tau\Omega\,{\rm s}}{2}\left(p^x\sigma^x-p^y\sigma^y\right)
    \end{pmatrix}}\;,\nonumber
\end{align}
where the dependences on $\tau$ and $\fp$ of the functions \eqref{def: functions} have been omitted for the sake of conciseness.

As for the Belinfante pseudo-gauge,  our first task is to diagonalize $\wPi_\Omega$ in order to compute thermal expectation values. In this case, the analytic diagonalization is much harder than for $\Omega=0$, as the total Hamiltonian density $\mathcal{H}_{\rm tot}$ is a dense $4\times4$ matrix involving non-trivial combinations of Hankel functions.

Before solving the eigenvalue problem for the full effective hamiltonian, is therefore instructive to study two limiting cases: the case of purely longitudinal momentum, i.e. $p_x=p_y=0$ and of very large spin potential $\Omega$. In these two cases we will confine ourselves to the study of the spin density matrix.

\subsection{Spin density matrix for purely longitudinal particles}

Starting from the latter case, we see that for longitudinal momentum $\mathcal{H}^{L}_{tot}$ (where the superscript specifies that we are considering $p_{\rm T}=0$) boils down to $\mathcal{H}$ in eq. \eqref{def: matrices}, plus a diagonal spin dependent correction, proportional to $\gamma^\uthree\gamma^\uzero$. An explicit calculation shows that $\mathcal{H}$ and $\gamma^\uthree\gamma^\uzero$ commute, so it suffices to diagonalize $\mathcal{H}$ to diagonalize $\mathcal{H}_{\rm tot}^{L}$. This means that the eigenvectors of $\mathcal{H}_{\rm tot}^L$ have been already found, and are those in eq. \eqref{def: eigenspace belin}. The diagonalization proceeds exactly as in the previous section and we find:
\begin{equation*}
    \mathcal{D}\left(\tau,\mu\right)=\mathcal{U}^\dagger\left(\tau,\mu\right)\,\mathcal{H}^L_{\rm tot}\left(\tau,\mu\right)\,\mathcal{U}\left(\tau,\mu\right)=
    \begin{pmatrix}
        m_{\rm L}\left(\tau,\mu\right)-\frac{\sigma^z}{2}\Omega(\tau)&0\\
        0 &-m_{\rm L}\left(\tau,\mu\right)-\frac{\sigma^z}{2}\Omega(\tau)
    \end{pmatrix}\;,
\end{equation*}
with the Bogoliubov transformation given by eq. \eqref{eq: diag matrix Bel} and with $m_{\rm L}(\mu)$ being the \emph{longitudinal mass}:
\begin{equation*}
    m_{\rm L}(\mu)\equiv\sqrt{m^2+\frac{\mu^2}{\tau^2}}\;.
\end{equation*}
The number of Milne modes $\langle A^\dagger_r(\fp) A_s(\fp)\rangle$ can then be read off \eqref{eq: final exp}, such that, recalling $\SP=\beta\Omega$, the spin density matrix is:
\begin{equation}\label{p_T=0}
    \begin{split}
        \Theta_{rs}(\SP)&=\frac{1}{2}\Bigg[\delta_{rs}+ \sigma^z_{rs}\frac{\sinh(\SP/2)}{\tau m^2}\\
        &\times\frac{\int^{\infty}_0\di\mu \left[\cosh(\beta m_{\rm L}(\mu))+\cosh(\SP/2)\right]^{-1}\;}{\int^{\infty}_0\di\mu \;\frac{\mathfrak{h}(\tau,\mu)}{m_{\rm L}(\mu)}\left[e^{-\beta m_{\rm L}(\mu)}+\cosh(\SP/2)\right]\left[\cosh(\beta m_{\rm L}(\mu))+\cosh(\SP/2)\right]^{-1}\;}\Bigg]\;,
    \end{split}
\end{equation}
where  we used that the integral over $\mu$ is even.
This is the first exact result for quantum particles with a finite spin potential. We notice that the formula resembles the one obtained in global equilibrium \cite{Palermo:2023cup}, with the vorticity replaced by the spin potential, and the additional presence of other terms acting as integral weights. A full comparison is impossible, and we don't generally expect the two results to coincide once the reduced spin potential $\SP$ is set equal to the thermal vorticity $\varpi$: as we have commented early on, in a boost invariant flow there is no thermal vorticity, so the reduced spin potential can never relax to equilibrium without breaking the symmetries assumed during our calculations.

From \eqref{p_T=0} it is immediate to compute the spin polarization vector in the fermion rest frame through formula \eqref{eq: polarizz} and using that $P_i=S^i/S$:
\begin{equation}\label{Pz Long}
    \begin{split}
        P_z(T,\Omega)&=\frac{\sinh\left(\frac{\Omega}{2T}\right)}{\tau m^2}\\
        &\times\frac{\int^{\infty}_0\di\mu \left[\cosh\left( \frac{m_{\rm L}(\mu)}{T}\right)+\cosh\left(\frac{\Omega}{2T}\right)\right]^{-1}\;}{ \int^{\infty}_0\di\mu \;\frac{\mathfrak{h}(\tau,\mu)}{m_{\rm L}(\mu)}\left[e^{-\frac{m_{\rm L}(\mu)}{T}}+\cosh\left(\frac{\Omega}{2T}\right)\right]\left[\cosh\left(\frac {m_{\rm L}(\mu)}{T}\right)+\cosh\left(\frac{\Omega}{2T}\right)\right]^{-1}\;}\;,
    \end{split}
\end{equation}
with the transverse component vanishing $P_x=P_y=0$ and where we expressed the explicit dependence on the thermodynamic fields $T$ and $\Omega$. In the limit of very large spin potential $\Omega/T\gg1$ this formula saturates to one whereas in the opposite limit it tends toward zero with a linear dependence (see section \ref{Sec: Analysis}).

\subsection{Spin density matrix in the limit of large spin potential}
We now consider a second limiting case in which the contribution of the energy term in the exponent is parametrically small compared with the spin–potential term. Concretely, for the modes that dominate the trace, we assume
\begin{equation}\label{Grande SP}
    \beta(\tau)\,\omega(\tau,\fp) \ll |\SP(\tau)|\, |\lambda_p|\; ,
\end{equation}
so that the factor $\beta \widehat{\mathcal H}$ in the exponent can be neglected with respect to the spin potential contribution. In this regime, the effective operator $\widehat{\Pi}_\Omega$ reduces to the pure spin contribution,
\begin{equation*}
    \frac{\widehat{\Pi}_\Omega(\tau)}{T(\tau)}\simeq -\,\SP(\tau)\,\widehat{\mathcal S}\;,
\end{equation*}
and the statistical operator becomes dominated by the constraint on the spin density, while the matching to the mean energy–momentum tensor is effectively relaxed. Equivalently, one may view this situation as an auxiliary ensemble in which only the spin-tensor constraint in Eq.~\eqref{eq: local state} is enforced. Note that the limit \eqref{Grande SP} requires the eigenvalue of the spin part to be bigger than the thermal one, which is controlled by $m_{\rm T}$ as we have explicitly computed in the Belinfante case. Hence the approximation is only valid for:
\begin{equation*}
    \beta m_{\rm T}\ll\SP\implies p_{\rm T}\ll\Omega\; {\rm and}\; m\ll\Omega \;,
\end{equation*}
 and is expected to break down at large transverse momenta or for very heavy particles. 

For the canonical Dirac spin operator, $\widehat{\mathcal S}$ is the generator of spatial rotations in the spin-$1/2$ representation; therefore its eigenvalues are $\lambda=\pm 1/2$. In the single-particle Dirac space, and in a basis where $\widehat{\mathcal S}$ is diagonal (e.g. spin quantized along the $z$-axis for both particle and antiparticle sectors), this implies:
\begin{equation}
    D_{\mathcal S}
    = \mathrm{diag}\!\left(
        \frac{1}{2},-\frac{1}{2},-\frac{1}{2},\frac{1}{2}
      \right),
\end{equation}
The diagonalizing matrix is:
\begin{align}
    {\mathcal{U}=
    \begin{pmatrix}
        -\ii \tau  {\rm t}\, {\rm s} \sqrt{\frac{m_{\rm T}^2-m_{\rm T} m}{2\,|{\rm t}|^2}} 
        & -\ii \tau  \,{\rm t}^*\, {\rm s} \sqrt{\frac{m_{\rm T}^2+m_{\rm T} m}{2\,|{\rm t}|^2}} 
        & \frac{\tau  (p_y+\ii p_x)\, {\rm t}^* \sqrt{m_{\rm T}}}{\sqrt{2(m_{\rm T}-m)}}
        & -\frac{\tau  (p_y+\ii p_x)\, {\rm t}^*\sqrt{m_{\rm T}}}{\sqrt{2(m_{\rm T}+m)}}\\
        \frac{(p_y-\ii p_x)\, {\rm t}^*}{\sqrt{2(m_{\rm T}^2-m_{\rm T} m)|{\rm t}(\tau ,\fp )|^2}} 
        & \frac{(-p_y+\ii p_x)\, {\rm t}^*}{\sqrt{2(m_{\rm T}^2+m_{\rm T} m)|{\rm t}(\tau ,\fp )|^2}}
        & 0 & 0 \\
        0 & 0 
        & \sqrt{\frac{m_{\rm T}-m}{2m_{\rm T}}} & \sqrt{\frac{m_{\rm T}+m}{2m_{\rm T}}} \\
        \tau \sqrt{\frac{p_{\rm T}^2\,|{\rm t}|^2 m_{\rm T}}{2(m+m_{\rm T})}} 
        & \tau \sqrt{\frac{p_{\rm T}^2\,|{\rm t}|^2 m_{\rm T}}{2(m-m_{\rm T})}}
        &\frac{\tau  (p_x-\ii p_y)\, {\rm s}\sqrt{m_{\rm T}}}{\sqrt{2(m_{\rm T}-m)}}
        & \frac{\tau  (-p_x+\ii p_y)\, {\rm s} \sqrt{m_{\rm T}}}{\sqrt{2(m_{\rm T}+m)}}
    \end{pmatrix}}.
\end{align}
In this case, we find for the spin density matrix:
\begin{equation}\label{high Omega}
    \Theta^\mathcal{S}_{rs}\left(p_{\rm T},\SP\right)=\frac{\delta_{rs}}{2}+\frac{\left(\sigma^z\right)_{rs}}{2}\,\frac{m}{m_{\rm T}}\tanh\left(\frac{\SP}{4}\right),
\end{equation}
so again, particles can only be longitudinally polarized. We should remark that, since we have neglected the term involving the temperature, the renormalization procedure described in section \ref{sec:exp values and renormalization} cannot be properly applied. The result reported above has been obtained without vacuum subtraction.

The spin polarization vector can be obtained immediately and reads:
\begin{equation}\label{Pz grande SP}
    P_z\left( p_{\rm T};\SP\right)=\frac{m}{\sqrt{m^2+p^2_{\rm T}}}\tanh\left(\frac{\SP}{4 }\right)\;.
\end{equation}
For finite transverse momentum $p_{\rm T}\lesssim\Omega$ one can clearly see, comparing eq.~\eqref{Pz grande SP} with eq.~\eqref{Pz Long} that in the limit of large spin potential $\Omega\gg T$, the polarization does not saturate at 1, but always at lower value:
\begin{equation*}
    P_z\left(p_{\rm T};\SP\right)\mapsto\frac{m}{\sqrt{m^2+p^2_{\rm T}}}<1\;,\qquad{\rm for}\; \SP\gg1\;,
\end{equation*}
implying that the maximum polarization is obtained for purely longitudinal particles.

Note that, except for the possible implicit dependence on $\tau$ through $\SP=\Omega/T$, the spin density matrix and thus the spin polarization vector are independent of time in the limit of very large spin potential.

\subsection{Exact diagonalization}

Finally we come to the diagonalization of the operator $\mathcal{H}_{\rm tot}$.
In order to analytically diagonalize it, we proceed in analogy with the Belinfante case, employing $\mathcal{H}_{\rm tot}^2$ instead of $\mathcal{H}_{\rm tot}$, as it turns out to be easier. One has:
\begin{align}\label{HT2}
        &\mathcal{H}^2_{\rm tot}=\\
        &{
        \begin{pmatrix}
        \frac{\Omega^2}{4} +\varepsilon^2+m m_{\rm T}\Omega\tau\mathfrak{h} &(p_x-\ii p_y)\frac{\Omega\mu}{m_T\tau} &0 & m m_{\rm T}\Omega\tau \,{\rm j}^*\\
        (p_x+\ii p_y)\frac{\Omega\mu}{m_{\rm T}\tau} &\frac{\Omega^2}{4} +\varepsilon^2-m m_{\rm T}\Omega\tau\mathfrak{h} & -m m_{\rm T}\Omega\tau \,{\rm j}^* &0\\
        0 & -m m_{\rm T}\Omega\tau \,{\rm j} &\frac{\Omega^2}{4} +\varepsilon^2+m m_{\rm T}\Omega\tau\mathfrak{h}  & (p_x+\ii p_y)\frac{\Omega\mu}{m_{\rm T}\tau}\\
         m m_{\rm T}\Omega\tau \,{\rm j} &0  &(p_x-\ii p_y)\frac{\Omega\mu}{m_{\rm T}\tau} &\frac{\Omega^2}{4} +\varepsilon^2-m m_{\rm T}\Omega\tau\mathfrak{h} 
    \end{pmatrix}
    }\;,\nonumber
\end{align}
where again $\varepsilon(\tau,\fp)=(m_{\rm T}^2+\mu^2/\tau^2)^{1/2}$ is the single particle energy, and we omitted the dependence on $\tau$ and $\fp$ of the various functions for the sake of brevity. The matrix \eqref{HT2} has 2 disjoint eigenvalues:
\begin{equation}\label{H2 eigenvalue}
    \omega^2_{s}(\tau,\fp)\equiv\varepsilon^2(\tau,\fp)+\frac{\Omega^2(\tau)}{4}+s|\Omega(\tau)|m_{\rm L}(\mu)\;;\quad s=\pm1\;.
\end{equation}
Comparing eq. \eqref{H2 eigenvalue} with the spectrum of the matrix $\mathcal{H}^2$, eq.~\eqref{eq: spectrum Ht2}, we observe that $\omega_{\pm}^2$ reduces to $\varepsilon^2$ in the limit of vanishing spin potential $\Omega\mapsto0$. Moreover the presence of a non-vanishing spin potential resolves the degeneracy in spin due to the linear term in $|\Omega|$. Note that the splitting of eigenvalues is larger at at higher $\mu/\tau$, i.e larger $p_z$. From now on, we will assume $\Omega>0$.

Eq. \eqref{H2 eigenvalue} implies that the spectrum of $\mathcal{H}_{\rm tot}$ is composed of $\pm\omega_{\pm}$.
The corresponding eigenvectors $|\pm\omega_\pm\rangle$ can be found from the eigenvectors of $\mathcal{H}_{\text{tot}}^2$ using a projection procedure described explicitly in appendix \ref{app: projection procedure canonical eigenv}, where their analytic form is given. 
The diagonalized Hamiltonian then reads:
\begin{equation*}
    \mathcal{D}\left(\tau,\fp\right)=\mathcal{U}^\dagger\left(\tau,\fp\right)\mathcal{H}_{\rm tot}(\tau,\fp)\,\mathcal{U}(\tau,\fp)=\begin{pmatrix}
        \omega_+(\tau,\fp)&0&0&0\\
        0&\omega_-(\tau,\fp)&0&0\\
        0&0&-\omega_+(\tau,\fp)&0\\
        0&0&0&-\omega_-(\tau,\fp)
    \end{pmatrix}\;,
\end{equation*}
and the diagonalizing matrix $\mathcal{U}$ can be written as:
\begin{equation}\label{diag can gen}
    \mathcal{U}=\begin{pmatrix}
        |\omega_+\rangle,|\omega_-\rangle,|-\omega_+\rangle,|-\omega_-\rangle
    \end{pmatrix}\;,
\end{equation}
with each eigenvector being a column of the matrix. With this transformation, the operator in eq. \eqref{eq: Pi matrix} is written in terms of diagonalizing modes:
\begin{equation}\label{Pi diag can}
    \wPi(\tau)=\int\di^3\fp\sum_{s=\pm}\omega_{s}(\tau,\fp)\left[\widehat{\alpha}^\dagger_s(\fp)\widehat{\alpha}_s(\fp)-\widehat{\beta}_s(\fp)\widehat{\beta}^\dagger_{s}(\fp)\right]\;,
\end{equation}
where we have denoted the spin dependence of the eigenvalues as a subscript:
\begin{equation}\label{exact eigen}
    \omega_{s}(\tau,\fp)=\sqrt{\varepsilon^2(\tau,\fp)+\frac{\Omega^2(\tau)}{4}+s\Omega(\tau)m_{\rm L}(\mu)}\;.
\end{equation}
The diagonal form of eq.\eqref{Pi diag can}, makes it possible to find the exact partition function of the out-of-equilibrium system using standard thermal field theory techniques \cite{Kapusta:2006pm}. Since the trace is invariant under change of basis, summing over states on fixed momentum shells using \eqref{diag can gen}, one finds:
\begin{equation}\label{eq:partition function}
    \log Z = \frac{V}{(2\pi)^3\tau}\int \di^3\fp\sum_{s=\pm}\left[\log\left(1+e^{-\frac{\omega_s}{T}+\zeta}\right)+\log\left(1+e^{-\frac{\omega_s}{T}-\zeta}\right)\right]\:,
\end{equation}
where we have reintroduced the reduced chemical potential $\zeta$ and $V$ is a volume factor. 
Furthermore, the expectation values of combinations of $\widehat{\alpha}$ and $\widehat{\beta}$ operators at local thermodynamic equilibrium read:
\begin{equation}\label{eq: exp val diago can}
    \begin{split}
        \left\langle\widehat{\alpha}^\dagger_s(\fp)\widehat{\alpha}_{s'}(\fp')\right\rangle_{\rm LE}&= \delta^3\left(\fp-\fp'\right)\, n^-_{\rm F}\left(\omega_s(\tau,\fp)\right)\delta_{ss'}\;,
        \\
        \left\langle\widehat{\beta}^\dagger_s(\fp)\widehat{\beta}_{s'}(\fp')\right\rangle_{\rm LE}&= \delta^3\left(\fp-\fp'\right)\, n^+_{\rm F}\left(\omega_s(\tau,\fp)\right)\delta_{ss'}\;,
        \\
        \left\langle\widehat{\beta}_s(\fp)\widehat{\alpha}^\dagger_{s'}(\fp')\right\rangle_{\rm LE}&=\left\langle\widehat{\beta}^\dagger_s(\fp)\widehat{\alpha}^\dagger_{s'}(\fp')\right\rangle_{\rm LE}=\left\langle\widehat{\beta}_s(\fp)\widehat{\alpha}_{s'}(\fp')\right\rangle_{\rm LE}=0\;.
    \end{split}
\end{equation}
Finally, using the above results, and inverting eq. \eqref{def: diagonalization}, we get the Milne modes in term of the diagonal ones:
\begin{equation}\label{diagonalizing can}
    \begin{pmatrix}
        \widehat{A}_r(\fp)\\
        \widehat{B}^\dagger_r(-\fp)
    \end{pmatrix}=\mathcal{U}_{rs}\left(\tau,\fp\right)\begin{pmatrix}
        \widehat{\alpha}_s(\fp)\\
        \widehat{\beta}^\dagger_s(\fp)
    \end{pmatrix}=\begin{pmatrix}
        u_{rs}\left(\tau,\fp\right)&v_{rs}\left(\tau,\fp\right)\\
        w_{rs}\left(\tau,\fp\right)&z_{rs}\left(\tau,\fp\right)
    \end{pmatrix}\begin{pmatrix}
        \widehat{\alpha}_s(\fp)\\
        \widehat{\beta}^\dagger_s(\fp)
    \end{pmatrix}\;.
\end{equation}
We are now in the position of computing the expectation values of Milne creation and annihilation operators, corresponding to what was done in eq. \eqref{eq: final exp} for the case of vanishing spin potential. However,
despite being able to analytically solve the eigenvalue problem, the calculations that follow are extremely difficult to carry out analytically, as the expression involved are very large (see appendix \ref{app: projection procedure canonical eigenv}). Therefore, we will not give analytic expressions for energy density, pressure, nor any other of the observables of our interest.  Nonetheless, exact quantities can be computed numerically starting from the analytic expressions obtained.

From eq. \eqref{diagonalizing can}, the local equilibrium expectation values of the Milne fields operator $\widehat{A}$ and $\widehat{B}$ can be obtained. Using the fact that $\langle\alpha^\dagger_r(\fp)\alpha_s(\fp')\rangle$ and $\langle\beta^\dagger_r(\fp)\beta_s(\fp')\rangle$ are symmetric under $r\leftrightarrow s$ and $\fp\leftrightarrow \fp'$\footnote{Note that, by parity, $\langle\widehat{B}^\dagger_r(\fp)\widehat{B}_{r'}(\fp')\rangle=\langle\widehat{B}^\dagger_r(-\fp)\widehat{B}_{r'}(-\fp')\rangle$.}, we find:
\begin{equation}\label{LTE expval Can}
    \begin{split}
        \left\langle\widehat{A}^\dagger_r(\fp)\widehat{A}_{r'}(\fp')\right\rangle_{\rm LE}&=u_{r's'}\left(\tau,\fp'\right)\left\langle\widehat{\alpha}^\dagger_{s'}(\fp')\widehat{\alpha}_{s}(\fp)\right\rangle_{\rm LE}u^\dagger_{sr}\left(\tau,\fp\right)\\
        &\quad+v_{r's'}\left(\tau,\fp'\right)\left\langle\widehat{\beta}_{s'}(\fp')\widehat{\beta}^\dagger_{s}(\fp)\right\rangle_{\rm LE}v^\dagger_{sr}\left(\tau,\fp\right)\;,\\
        \left\langle\widehat{B}^\dagger_r(\fp)\widehat{B}_{r'}(\fp')\right\rangle_{\rm LE}&=w_{rs}\left(\tau,\fp\right)\left\langle\widehat{\alpha}_s(\fp)\widehat{\alpha}^\dagger_{s'}(\fp')\right\rangle_{\rm LE}w^\dagger_{s'r'}\left(\tau,\fp'\right)\\
        &\quad+z_{rs}\left(\tau,\fp\right)\left\langle\widehat{\beta}^\dagger_s(\fp)\widehat{\beta}_{s'}(\fp')\right\rangle_{\rm LE}z^\dagger_{s'r'}\left(\tau,\fp'\right)\;,\\
        \left\langle\widehat{A}^\dagger_r(\fp)\widehat{B}^\dagger_{r'}(-\fp')\right\rangle_{\rm LE}&=w_{r's'}\left(\tau,\fp'\right)\left\langle\widehat{\alpha}^\dagger_{s'}(\fp')\widehat{\alpha}_{s}(\fp)\right\rangle_{\rm LE}u^\dagger_{sr}\left(\tau,\fp\right)\\
        &\quad+z_{r's'}\left(\tau,\fp'\right)\left\langle\widehat{\beta}_{s'}(\fp')\widehat{\beta}^\dagger_{s}(\fp)\right\rangle_{\rm LE}v^\dagger_{sr}\left(\tau,\fp\right)\;,\\
        \left\langle\widehat{B}_r(\fp)\widehat{A}_{r'}(-\fp')\right\rangle_{\rm LE}&=-\left(\left\langle\widehat{A}^\dagger_{r'}(\fp)\widehat{B}^\dagger_{r}(-\fp')\right\rangle_{\rm LE}\right)^*\;.
    \end{split}
\end{equation}
where for the last equality we have used eq. \eqref{eq: tev AB}.
The renormalized expressions at a generic proper time $\tau$ are computed simply by normal ordering the modes $\alpha$ and $\beta$, which yields:
\begin{equation}\label{LTE expval Can2}
    \begin{split}
        \left\langle\widehat{A}^\dagger_r(\fp)\widehat{A}_{r'}(\fp')\right\rangle_{\rm LE}^{ren}&=u_{r's'}\left(\tau,\fp'\right)\left\langle\widehat{\alpha}^\dagger_{s'}(\fp')\widehat{\alpha}_{s}(\fp)\right\rangle_{\rm LE}u^\dagger_{sr}\left(\tau,\fp\right)\\
        &\quad-v_{r's'}\left(\tau,\fp'\right)\left\langle\widehat{\beta}^\dagger_{s}(\fp)\widehat{\beta}_{s'}(\fp')\right\rangle_{\rm LE}v^\dagger_{sr}\left(\tau,\fp\right)\;,\\
        \left\langle\widehat{B}^\dagger_r(\fp)\widehat{B}_{r'}(\fp')\right\rangle_{\rm LE}^{ren}&=w_{rs}\left(\tau,\fp\right)\left\langle\widehat{\alpha}_s(\fp)\widehat{\alpha}^\dagger_{s'}(\fp')\right\rangle_{\rm LE}w^\dagger_{s'r'}\left(\tau,\fp'\right)\\
        &\quad+z_{rs}\left(\tau,\fp\right)\left\langle\widehat{\beta}^\dagger_s(\fp)\widehat{\beta}_{s'}(\fp')\right\rangle_{\rm LE}z^\dagger_{s'r'}\left(\tau,\fp'\right)\;,\\
        \left\langle\widehat{A}^\dagger_r(\fp)\widehat{B}^\dagger_{r'}(-\fp')\right\rangle_{\rm LE}^{ren}&=w_{r's'}\left(\tau,\fp'\right)\left\langle\widehat{\alpha}^\dagger_{s'}(\fp')\widehat{\alpha}_{s}(\fp)\right\rangle_{\rm LE}u^\dagger_{sr}\left(\tau,\fp\right)\\
        &\quad+z_{r's'}\left(\tau,\fp'\right)\left\langle\widehat{\beta}_{s'}(\fp')\widehat{\beta}^\dagger_{s}(\fp)\right\rangle_{\rm LE}v^\dagger_{sr}\left(\tau,\fp\right)\;,\\
        \left\langle\widehat{B}_r(\fp)\widehat{A}_{r'}(-\fp')\right\rangle_{\rm LE}^{ren}&=-\left(\left\langle\widehat{A}^\dagger_{r'}(\fp)\widehat{B}^\dagger_{r}(-\fp')\right\rangle_{\rm LE}\right)^*\;.
    \end{split}
\end{equation}
where $u$, $v$, $w$, and $z$ are computed numerically from the matrix $\mathcal{U}$ in eq.~\eqref{diag can gen}, and eq.\eqref{eq: exp val diago can} must be used. The analysis of these results and the comparison with the case $\SP=0$ is the object of next section.

We conclude by remarking that eq. \eqref{LTE expval Can} corresponds to the expectation values obtained with the local equilibrium operator \eqref{eq:density operator split}. In order to obtain the actual non-equilibrium expectation value is sufficient to compute the Bogolubov coefficients $u,v,w$ and $z$ for $\tau=\tau_0$ equilibrium time, in analogy with the Belinfante case \eqref{eq: final exp NON EQ renormalized}. In both cases, the renormalization is done as described at the end of section \ref{sec: bel}.

\section{Exact expectation values with a finite spin potential}\label{Sec: Analysis}
In this section we evaluate numerically thermal expectation values with finite spin density. The code used to produce the results of this section is freely available at \cite{repo}. To evaluate with arbitrary precision Hankel functions at large values of $\mu$ and of $m_{\rm T}\tau$ we used the Python library \texttt{mpmath} \cite{mpmath}. The numerical evaluation of such functions and their integration remains a relatively expensive task. Furthermore, the $p_{\rm T}$ and $\mu$ domain over which the integrands are non-negligible varies depending on the mass of the particle $m$, the decoupling time $\tau$, the temperature $T$ and the spin potential $\Omega$. Therefore in this section, unless otherwise specified, we have chosen a small value of decoupling time $\tau_0=1$. This allows to keep the integrands peaked in a relatively small domain, such that the Hankel functions can be calculated precisely relatively fast. Results in this sections are also renormalized at $\tau_0=1$, and we will not study their proper time dependence. For this section, we also set $\zeta=0$, and use $m=T=1$ GeV unless otherwise specified.

We will compare the numerical values obtained explicitly from the integration of the expectation values reported in section \ref{sec:exp values and renormalization} with those based on thermodynamic relations. In particular, having computed exactly the partition function in eq. \eqref{eq:partition function}, we use:
\begin{subequations}
\begin{align}\label{eq:therm.rel.}
    \mathcal{P}&=TV^{-1}\log Z\;,\\
    \mathcal{S}&=\frac{\partial \mathcal{P}}{\partial \Omega}\;,\\
    \mathcal{E}&=-\mathcal{P}+\Omega\frac{\partial \mathcal{P}}{\partial \Omega}+T\frac{\partial \mathcal{P}}{\partial T}\;,
\end{align}
\end{subequations}
with $V$ volume of the system.
In our figures, whenever explicit integration is used, curves will be labeled as ``T.E.V'' standing for ``Thermal Expectation Values''. Instead, the results obtained from eqs. \eqref{eq:therm.rel.}, are labelled ``Therm. rel''.
\begin{figure}
    \centering
    \includegraphics[width=0.45\linewidth]{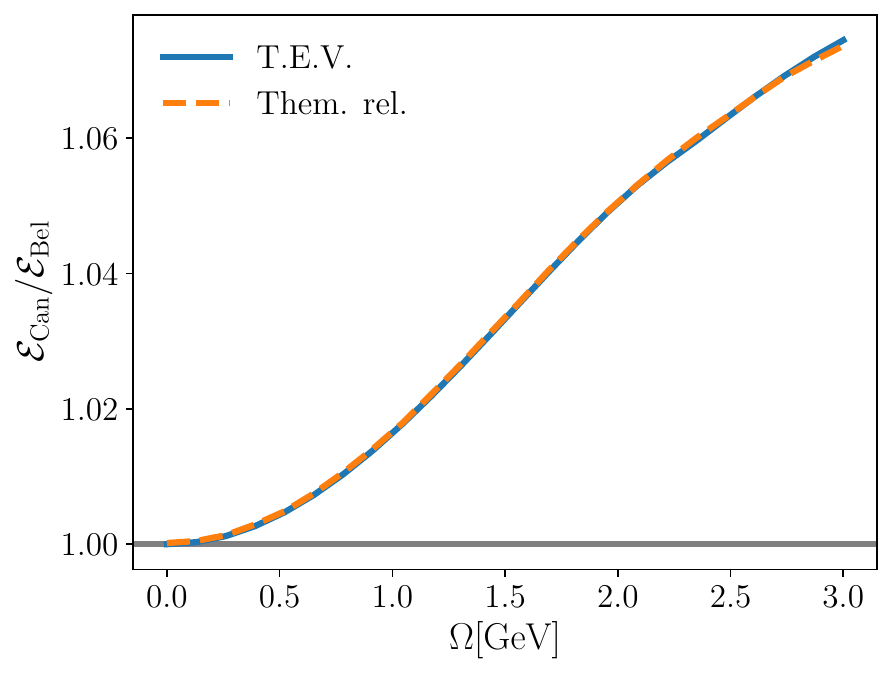}
    \includegraphics[width=0.45\linewidth]{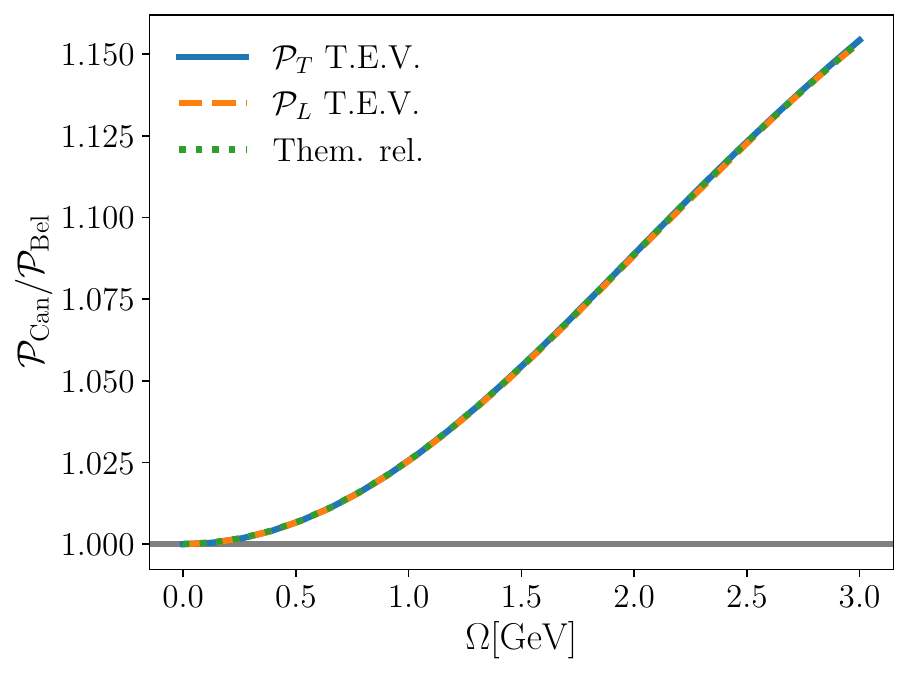}\\    %
    \includegraphics[width=0.45\linewidth]{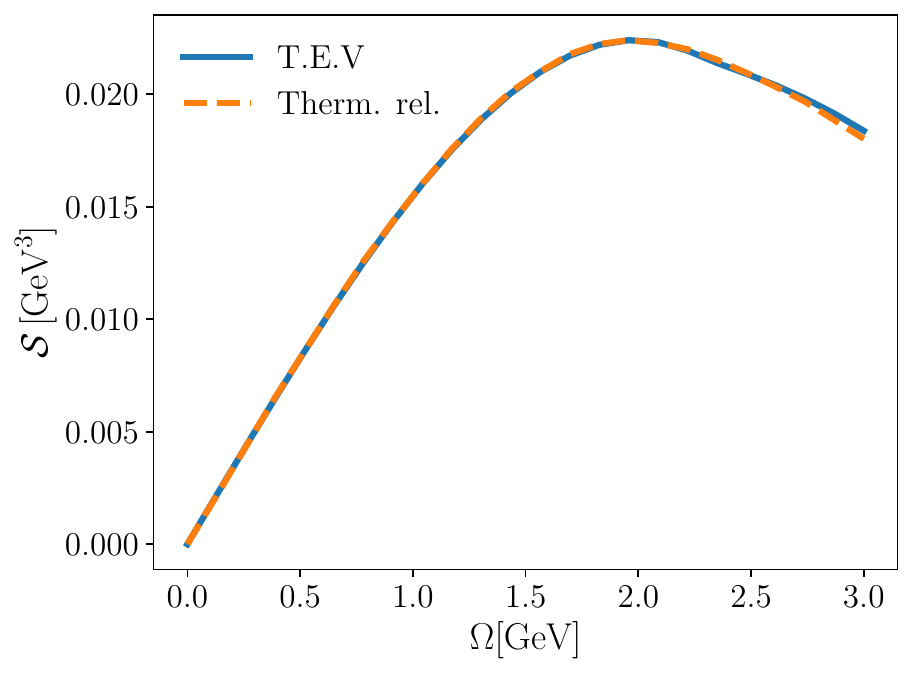}
    \caption{In the top panels we report the non-vanishing components of the energy momentum tensor as a function of spin potential. Top left panel: energy density in the canonical pseudogauge as a function of spin potential normalized to its Belinfante (i.e. $\Omega=0$) value. Top right panel: transverse and longitudinal pressures in the canonical pseudogauge as a function of spin potential normalized to their Belinfante value. A solid horizontal grey line is drawn at the y-axis value of one. Bottom panel: plot of the the spin density as a function of spin potential. Quantities labelled with ``T.E.V'' are computed using the formulae reported in sec. \ref{sec:exp values and renormalization}, whereas the label ``Therm. rel.'' is used for eqs. \eqref{eq:therm.rel.}.}
    \label{fig:Tmunu_and_Spin}
\end{figure}

In figure \ref{fig:Tmunu_and_Spin} we report the components of the energy momentum tensor normalized to their corresponding Belinfante value, as well as the spin tensor density. The spin torque is not reported because, despite the fact that a non-zero value is allowed by symmetry, its thermal expectation value is found to be vanishing numerically, i.e. $\mathcal{T}\sim\mathcal{O}(10^{-15}$ GeV$^4)$.  
We can see that the spin potential induced enhancement in energy density and pressure is at most $6\div15\%$, even at large value of $\Omega=3$GeV. Furthermore, it is interesting to see that the presence of a spin potential doesn't induce any pressure anisotropy: longitudinal and transverse pressure continue to be equal to each other:
\begin{equation*}
    \mathcal{P}_{\rm L}=\mathcal{P}_{\rm T}=\mathcal{P}\;.
\end{equation*}
Finally, we find that in this particular non-equilibrium case the thermodynamic relations connecting the partition function to thermodynamic observable hold very well. The differences between the explicit calculation of thermal expectation values and the result from the partition function agree excellently, with extremely small deviation caused by our numeric implementation.

The validity of thermodynamic relations when a spin potential is included is an active topic of debate in the literature. Even though it appears that on general grounds the relations \eqref{eq:therm.rel.} shouldn't apply \cite{Becattini:2023ouz}, our calculation shows an explicit non-equilibrium system with a finite spin potential where they hold, at least in the canonical pseudogauge.
\begin{figure}
    \centering
    \includegraphics[width=0.45\linewidth]{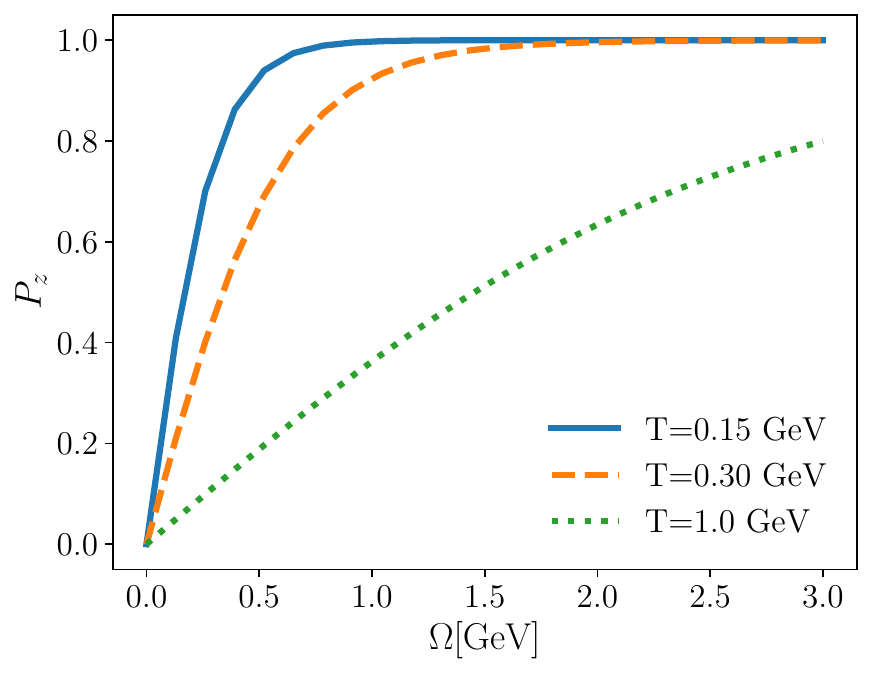}\hfill
    \includegraphics[width=0.45\linewidth]{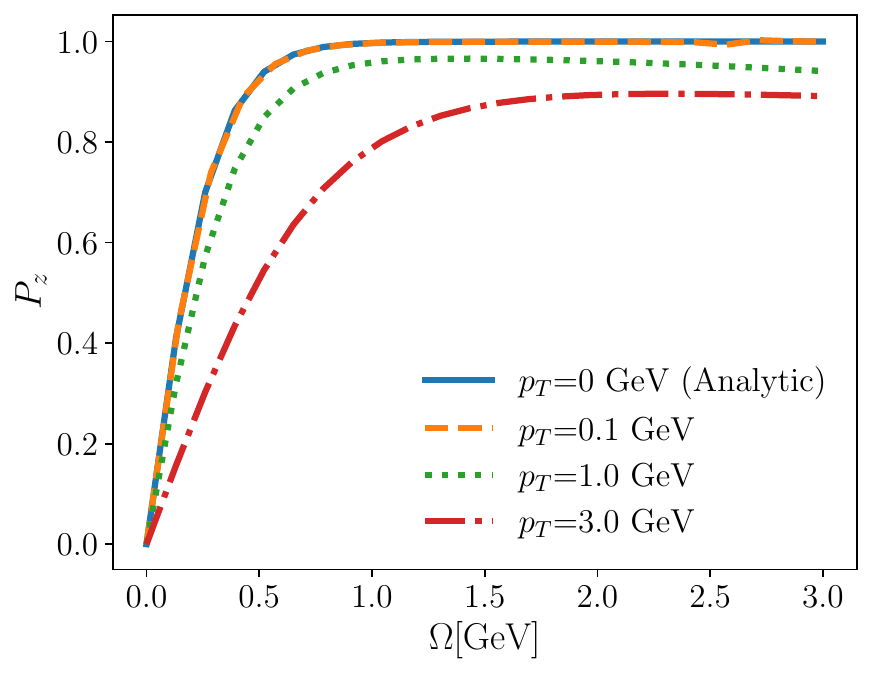}
    \caption{Left panel: longitudinal polarization at zero transverse momentum $P_{z}(p_{\rm T}=0)$, computed from eq.~\eqref{Pz Long}, for $m=1.167\,{\rm GeV}$ and $\tau=10$ fm as a function of $\Omega$ for different values of decoupling temperature. Right panel: longitudinal polarization for different values of transverse momentum, computed with eq.~\eqref{eq: spin density matrix with A/trA} at $T=0.15\,{\rm GeV}$. The curve labelled as \emph{Analytic} has been obtained using the eq.~\eqref{Pz Long}.}
    \label{fig:pz}
\end{figure}

Finally, in figure \ref{fig:pz}, we reported the longitudinal component of the spin polarization vector as a function of the spin potential for different decoupling temperatures. For these plots, we use mass of the $\Lambda$ hyperon, $m=1.167\,{\rm GeV}$, and a realistic decoupling time for heavy-ion collisions, $\tau=10\,{\rm fm}$. This increase in $\tau$ causes the evaluation of the integrands to be significantly more expensive as the integration domain to be considered becomes much larger. However, this is compensated by the fact that in eq. \eqref{eq: spin density matrix with A/trA} we only need to integrate in $\mu$. 

The left panel of fig.~\ref{fig:pz} shows the exact polarization for longitudinally moving $\Lambda$ particles, computed with eq.\eqref{Pz Long}. One can see that polarization increases monotonically with $\SP$, and saturates to $1$ for large values of the spin potential. Lower temperatures lead to larger polarization for fixed $\Omega$. In the right panel, we show polarization for different values of transverse momentum $p_{\rm T}$, using eq. \eqref{eq: spin density matrix with A/trA}. One sees that for small transverse momentum our numerical diagonalization procedure is consistent with the analytic result. As transverse momentum increases, longitudinal polarization decreases. Numerically, we see that the presence of a non-vanishing transverse momentum doesn't lead to non-vanishing transverse components of spin polarization, as we find $P_x=P_y\lesssim10^{-6}$.

\section{Conclusions}\label{sec: Conclusions}
In conclusion, we have computed exact local-equilibrium expectation values for a system of fermions in a boost-invariant configuration with finite spin density. Starting from the Belinfante pseudogauge, we showed that the quantum local-equilibrium stress–energy tensor coincides with the one obtained from classical calculations, in analogy with the scalar case. 
We have then extended the analysis to the canonical pseudogauge, allowing for a finite spin potential. In this case, we computed analytically the exact partition function of the system. This provides, to the best of our knowledge, the first exact analytic relation between thermodynamic functions and the spin potential. We further computed exactly the local-equilibrium stress–energy tensor and the spin tensor, and evaluated numerically the spin potential-induced modifications. This is done both explicitly computing renormalized expectation values and using thermodynamic relations, properly modified to include the presence of spin potential. We found that pressure remains isotropic, and that both energy density and pressure experience a modest enhancement in the presence of the spin potential. Explicit thermal expectation values calculations are consistent with the thermodynamic relations: the system under investigation is therefore an example of an exactly solvable out-of-equilibrium system where these relations hold in the canonical pseudogauge. 

We have performed a detailed study of the exact spin polarization of fermions with a finite spin potential, which is an experimental observable of direct relevance in heavy-ion collisions. We found that in a boost-invariant fluid the spin polarization vector vanishes identically despite the presence of a finite thermal shear tensor and a non-constant chemical potential. Therefore, under the assumptions of local thermodynamic equilibrium, absence of interactions, and boost invariance, the system cannot develop a finite polarization. 
When a finite spin potential is introduced, we have been able to compute analytically the spin polarization of massive particles with $p_{\rm T}=0$ and in the limiting case $\Omega\gg T$. In the general case our analysis shows that lower temperatures increase polarization at a fixed value of spin potential, whereas finite transverse momentum tends to reduce it. Numerically, we see that in our configuration the spin of particles can only point in the $z$-direction. 

A comparison with polarization data suggest that, if present, a spin potential at decoupling should be rather small, of order $\SP\sim10^{-2}$, which is well within the region of applicability of linear response theory. However, our results are completely general and can be applied in any regime of $\SP$, furnishing a valuable benchmark for spin hydrodynamic theory and providing an important toy model to test thermodynamics in the presence of spin.  

\acknowledgments

A.P. is funded by the European Union under grant agreement number 101198140 ``SPINNERET''. Views and
opinions expressed are however those of the author(s) only and do not necessarily reflect those
of the European Union or the granting authority. Neither the European Union nor the granting
authority can be held responsible for them.
D.R. is supported by the Italian 
Ministry of University and Research, project PRIN2022 “Advanced probes of the Quark Gluon Plasma”, Next 
Generation EU, Mission 4 Component 1.

D.R thanks F. Becattini, E. Speranza and E. Grossi for clarifying discussions. A.P. is grateful to M. Chernodub for numerous insights and countless coffees. 


\appendix
\section{Dirac equation in Milne coordinates}\label{app:dirac eq}

The Milne chart is defined in terms of the Cartesian coordinates $t,x,y,z$ by:
\begin{equation}\label{APP: Milne Chart}
    \begin{split}
        t&=\tau\cosh\eta,\quad z=\tau\sinh\eta\;,\\
        \tau&=\sqrt{t^2-z^2},\quad \eta=\frac{1}{2}\ln\frac{t+z}{t-z}\;,
    \end{split}
\end{equation}
where $\tau$ is the proper time and $\eta$ is the space-time rapidity.
In terms of \eqref{APP: Milne Chart} the flat metric, restricted to the future light cone, can be written as:
\begin{equation}\label{APP:Milne metric}
    \di s^2=\di\tau^2-\di x^2-\di y^2-\tau^2\di\eta^2,\quad g_{\mu\nu}=\mbox{diag}\left(1,-1,-1,-\tau^2\right)\;.
\end{equation}
To work with spinors in curvilinear coordinates, we introduce a local tetrad of vierbein $\left\{e_\ua(x)\right\}$ on the space-time point $x$ such that:
\begin{equation}\label{Def: Vierbein}
    \eta_{\ua\ub}=g_{\mu\nu}(x)e^\mu_\ua(x)e^\nu_\ub(x),
\end{equation}
where $\eta_{\ua\ub}$ is the metric in Cartesian coordinates.
A natural basis for the vierbein reads:
\begin{equation}
    e^\uzero=\di\tau,\quad e^\uone=\di x,\quad e^\utwo=\di y,\quad e^\uthree=\tau\di \eta\;, 
\end{equation}
from which we get:
\begin{equation}\label{APP: Vierbein Milne}
    e^\ua_\mu(x)=\mbox{diag}\left(1,1,1,\tau\right),\quad e^\mu_\ua(x)=\mbox{diag}\left(1,1,1,\tau^{-1}\right)\;.
\end{equation}
The Clifford algebra can be expressed in curvilinear coordinates defining the \emph{curvilinear} $\gamma$ matrices:
\begin{equation}
    \gamma^\mu(x)=e^\mu_\ua(x)\gamma^\ua\;,
\end{equation}
which satisfy the following algebra:
\begin{equation}\label{app: Clifford spacetime}
    \left\{\gamma^\mu(x),\gamma^\nu(x)\right\}=2g^{\mu\nu}(x)\;.
\end{equation}
For the Milne chart \eqref{APP: Milne Chart} and using the vierbein eq.~\eqref{APP: Vierbein Milne}, the curvilinear $\gamma$ matrices read:
\begin{equation}\label{APP:Gamma Milne}
    \gamma^0(x)=\gamma^\uzero,\ \gamma^1(x)=\gamma^\uone,\ \gamma^2(x)=\gamma^\utwo,\ \gamma^3(x)=\tau^{-1}\gamma^\uthree.
\end{equation}
where the $\gamma$ matrices with an underlined index are the usual $\gamma$ matrices of Minkowski spacetime and satisfy:
\begin{equation}\label{app: Clifford algebra cartesian}
    \left\{\gamma^{\underline{a}},\gamma^{\underline{a}}\right\}=2\eta^{\underline{a}\underline{b}}\;,
\end{equation}
$\eta^{\underline{a}\underline{b}}=\mathrm{diag}\{1,-1,-1,-1\}$ being the Minkowski metric. For these matrices, we employ the Weyl basis:
\begin{equation}
    \gamma^\uzero=\begin{pmatrix}
        0&\mathbb{1}\\
       \mathbb{1}&0
    \end{pmatrix}\;,\quad \gamma^{\underline{k}}=\begin{pmatrix}
        0&\sigma^k\\
        -\sigma^k&0
    \end{pmatrix}\;,\quad
    \gamma^5 = \begin{pmatrix}
        -\mathbb{1}&0\\
        0&\mathbb{1}
    \end{pmatrix}
\end{equation}
with $\sigma^k$ are Pauli matrices, and we have also introduced the fifth $\gamma$ matrix, $\gamma^5$, which is independent of coordinates.

In terms of the space-time dependent gamma matrices one can then extend the Dirac equation to a curvilinear chart.
In order to do so one must replace the partial derivative with the covariant one defined as:
\begin{equation}\label{Def: CoDev spinor}
    \begin{split}
        \partial_\mu\psi(x)\mapsto D_\mu\psi(x)&\equiv\partial_\mu\psi(x)+\Gamma_\mu(x)\psi(x)\;,\\
        \partial_\mu\overline{\psi}(x)\mapsto D_\mu\overline{\psi}(x)&\equiv\partial_\mu\overline{\psi}(x)-\overline{\psi}(x)\Gamma_\mu(x)\;,
    \end{split}
\end{equation}
so that the Dirac equation reads:
\begin{equation}\label{eq:standard Dirac eq}
    \left(\ii\gamma^\mu(x)D_\mu-m\right)\psi(x)=\left[\ii\gamma^\mu(x)(\partial_\mu+\Gamma_\mu)-m\right]\psi(x)=0\;.
\end{equation}
The operators $\Gamma_\mu(x)$ are the \emph{spin connection coefficients}:
\begin{equation}\label{Def: Spin connection}
    \Gamma_\mu(x)=\frac{1}{4}\omega^{\ \ua\ub}_\mu(x)\gamma_{\ua}\gamma_{\ub}\;,
\end{equation}
and $\omega(x)$ is the \emph{spin connection 1-form}. The spin connection is defined such that $D_\mu e_\ua^\nu=\nabla_\mu e_\ua^\nu+\Gamma_\mu e_\ua^\nu=0$. Using eq. \eqref{APP: Vierbein Milne} and taking into account that the non-vanishing Christoffel symbols are:
\begin{equation}\label{eq: Christoffel}
\Gamma^{\tau}_{\eta\eta}=\tau\;,\qquad\Gamma^\eta_{\eta\tau}=\Gamma^{\eta}_{\tau\eta}=\frac{1}{\tau}\;,  
\end{equation}
the spin connection coefficient turns out to be:
\begin{equation}\label{Result: Spin connection}
    \Gamma_\mu=-\frac{1}{2}\delta^3_\mu\gamma^\uthree\gamma^\uzero\;.
\end{equation}
Finally, using eq.~\eqref{Result: Spin connection} in \eqref{eq:standard Dirac eq}, the Dirac equation reads:
\begin{equation}\label{Eq: Dirac eq CST}
    \left\{\ii\left[\frac{\partial}{\partial\tau}\gamma^\uzero+\gamma^\uone\frac{\partial}{\partial x}+\gamma^\utwo\frac{\partial}{\partial y}+\gamma^\uthree\frac{1}{\tau}\left(\frac{\partial}{\partial\eta}-\frac{1}{2}\gamma^\uthree\gamma^\uzero\right)\right]-m\right\}\psi\left(\tau,{\bf p}_{\rm T},\eta\right) =0\;.
\end{equation}
Note that the equation of motion depends on $\gamma^\uthree\gamma^\uzero/2$ which is related with the boost operator along the $z-$direction:
\begin{equation}
    \Sigma^{\uzero\uthree}=\frac{\ii}{4}\left[\gamma^\uzero,\gamma^\uthree\right]=-\frac{\ii}{2}\gamma^\uthree\gamma^\uzero\;.
\end{equation}
%

\section{Properties of the Hankel functions}\label{app:hankel}
The Hankel functions are defined as linear combinations of the Bessel functions of the first and second kind:
\begin{equation*}
    \hnku_w(z)\equiv J_w(z)+\ii\, Y_w(z)\;, \qquad 
    \hnkd_w(z)\equiv J_w(z)-\ii\, Y_w(z)\;,
\end{equation*}
where $w\in\mathbb{C}$ denotes the \emph{order} of the function. In general, the argument $z$ may also take complex values.  
For real arguments, the Hankel functions admit the following integral representations \cite{gradshteyn2007}:
\begin{equation}\label{app hank: Integral representation}
    \begin{split}
        \hnku_w(z)&=-\frac{\ii}{\pi}\,e^{-\frac{\ii\pi w}{2}}
    \int_{-\infty}^{+\infty}\! \mathrm{d}\vartheta\,
    e^{\ii z\cosh\vartheta-\vartheta w}\;,
    \\
    \hnkd_\nu(z)&=\frac{\ii}{\pi}\,e^{\frac{\ii\pi w}{2}}
    \int_{-\infty}^{+\infty}\! \mathrm{d}\vartheta\,
    e^{-\ii z\cosh\vartheta-\vartheta w}\;,
    \end{split}
\end{equation}
valid in the range $-1<\mathrm{Re}(w)<1$.
The Hankel functions satisfy the standard recursion relations:
\begin{equation}\label{app hank: recursive}
    2z\frac{\mathrm{d} {\rm H}^{(1,2)}_w(z)}{\mathrm{d} z}
    ={\rm H}^{(1,2)}_{w-1}(z)-{\rm H}^{(1,2)}_{w+1}(z)\;,
    \qquad
    2w\, {\rm H}^{(1,2)}_w(z)
    =z\big[{\rm H}^{(1,2)}_{w-1}(z)+{\rm H}^{(1,2)}_{w+1}(z)\big]\;,
\end{equation}
which also imply:
\begin{equation}\label{app hank: derivative relation}
    \begin{split}
        \frac{\di{\rm H}^{(1,2)}_w(z)}{\di z}&={\rm H}^{(1,2)}_{w-1}(z)-\frac{w}{z}{\rm H}^{(1,2)}_w(z)\;,\\
        \frac{\di{\rm H}^{(1,2)}_w(z)}{\di z}&=-{\rm H}^{(1,2)}_{w+1}(z)+\frac{w}{z}{\rm H}^{(1,2)}_w(z)\;.
    \end{split}
\end{equation}
and are normalized through their Wronskian \cite{NIST:DLMF}:
\begin{equation}\label{app hank: Wronskian}
    \mathcal{W}\!\left[\hnku_w(z),\hnkd_w(z)\right]
    =\hnku_{w+1}(z)\hnkd_{w}(z)
    -\hnku_w(z)\hnkd_{w+1}(z)
    =-\frac{4\ii}{\pi z}\;.
\end{equation}
The reflection properties of the Hankel functions are given by
\begin{equation}\label{app hank: reflection}
    {\rm H}^{(1)}_{-w}(z)=e^{\ii\pi w}\,{\rm H}^{(1)}_{w}(z)\;,
    \qquad 
    {\rm H}^{(2)}_{-w}(z)=e^{-\ii\pi w}\,{\rm H}^{(2)}_{w}(z)\;.
\end{equation}
Finally we derive a useful identity valid for the specific case of order $w=1/2-\ii\mu$. Combining Eqs.~\eqref{app hank: recursive} and \eqref{app hank: Wronskian} we get:
\begin{equation}\label{app hank: relazione *}
    \begin{split}
        \hnku_{w^*}(z)\hnkd_{w}(z)+\hnku_{-w}(z)\hnkd_{-w^*}(z)=\ii e^{\pi\mu}\mathcal{W}\left[\hnku_{w-1}(z),\hnkd_{w-1}(z)\right]=\frac{4e^{\pi\mu}}{\pi z}\;,
    \end{split}
\end{equation}
where $ w=\frac{1}{2}-\ii\mu$.
Note that for order $w=1/2-\ii\mu$ the following trivial relations exist:
\begin{equation}\label{app hank: relazioni wpm1}
    w-1=-w^*\;,\qquad-w^*+1=w\,.
\end{equation}
Thanks to the above definition and the reflection relations \eqref{app hank: reflection} it is immediate to prove, for $w=1/2-\ii\mu$, the following useful identity:
\begin{equation}\label{app: combinazione nulla}
    {\rm H}^{(1,2)}_w(z){\rm H}^{(1,2)}_{w^*}(z)+{\rm H}^{(1,2)}_{-w}(z){\rm H}^{(1,2)}_{-w^*}=0\,,
\end{equation}
which, once derived, also implies:
\begin{equation}\label{app: combinazione derivata}
    {\rm H}^{(1,2)}_w(z)\frac{\di{\rm H}^{(1,2)}_{w^*}(z)}{\di z}+{\rm H}^{(1,2)}_{-w}(z)\frac{\di{\rm H}^{(1,2)}_{-w^*}}{\di z}=-{\frac{\di{\rm H}^{(1,2)}_w(z)}{\di z}}{\rm H}^{(1,2)}_{w^*}(z)-\frac{\di{\rm H}^{(1,2)}_{-w}}{\di z}{\rm H}^{(1,2)}_{-w^*}(z)\,.
\end{equation}

We now turn to the asymptotic expansions of the Hankel functions. One particular useful formula is valid in the regime of large arguments $z\gg1$ and non-vanishing order imaginary order $w=1/2-\ii\mu$, with $\mu/z\sim\mathcal{O}(1)$ \cite{Olver:1954wc}:
\begin{equation}\label{app: hank asymptotic grande z mu fix}
    \begin{split}
        \hnku_{w}(z)&=-\ii\sqrt{\frac{2}{\pi}}\frac{\e^{-\pi\mu/2}}{\sqrt{\varrho}}\sqrt{\frac{z}{\varrho+\mu}}\e^{\ii\varphi_+(z)}\;,\\
        \hnkd_w(z)&=\ii\sqrt{\frac{2}{\pi}}\frac{\e^{\pi\mu/2}}{\sqrt{\varrho}}\sqrt{\frac{z}{\varrho+\mu}}\e^{-\ii\varphi_-(z)}\;,
    \end{split}
\end{equation}
where:
\begin{equation}\label{rho e varphi}
    \varrho=\sqrt{z^2+\mu^2}\;,\quad \varphi_\pm=\varrho\pm{\rm arcsinh}\left(\frac{\mu}{z}\right)\;.
\end{equation}
Note that if the order is kept fix in the limit of large argument then $\mu/z\to0$, so $\varrho\to z$ and \eqref{app: hank asymptotic grande z mu fix} reduces to the well known  asymptotic formula \cite{gradshteyn2007,NIST:DLMF}:
\begin{equation}\label{app: hank asymptotic grande z}
    {\rm H}^{(1,2)}_w(z)
    \sim \sqrt{\frac{2}{\pi z}}
    \,e^{\pm\ii\left(z-\frac{\pi w}{2}-\frac{\pi}{4}\right)}\left(1\pm\frac{4w^2-1}{8\ii z}\right)\;,\qquad z\gg1\;.
\end{equation}

The integral representations eqs. \eqref{app hank: Integral representation} can be easily extended to the case where the order is matrix-valued, simply by computing the exponential of the matrix: $w\to \nu^A_B$
\begin{equation}\label{app hank: matrix order}
    \nu^A{}_B
    \equiv
    \frac{1}{2}\left(\gamma^{\uthree}\gamma^{\uzero}\right)^A{}_B
    -\ii\mu\,\delta^A{}_B={\rm diag}\left(\frac{1}{2}-\ii\mu,-\frac{1}{2}-\ii\mu,-\frac{1}{2}-\ii\mu,\frac{1}{2}-\ii\mu\right)\;,
\end{equation}
with $A,B$ denoting spinor indices. 
The Hankel functions thus become matrices in spinor space, defined as
\begin{equation*}
    \left({\rm H}^{(1,2)}_{\nu }(z)\right)^A{}_B
        = \mp\frac{\ii}{\pi}
        \int_{-\infty}^{+\infty}\!\mathrm{d}\vartheta\,
        e^{\pm\ii z\cosh\vartheta}
        \left(e^{\mp\frac{\ii\pi\nu}{2}-\vartheta\nu}\right)^A{}_B\;.
\end{equation*}
In the Weyl basis, we have $\gamma^\uthree\gamma^\uzero={\rm diag}(1,-1,-1,1)$ hence the Hankel functions of matrix order \eqref{app hank: matrix order} are diagonal and read:
\begin{equation}\label{app: Hankel matrix components}
    {\rm H}^{(1,2)}_\nu(z)
    \equiv\begin{pmatrix}
    {\rm H}^{(1,2)}_{w}(z)&0&0&0\\
    0&{\rm H}^{(1,2)}_{-w^*}(z)&0&0\\
    0&0&{\rm H}^{(1,2)}_{-w^*}(z)&0\\
    0&0&0&{\rm H}^{(1,2)}_{w}(z)
    \end{pmatrix}\;,
\end{equation}
where we have reintroduced the notation $w=1/2-\ii\mu$ for the scalar order.
Since the matrix index depends on the Hermitian generator $\ii\Sigma^{\uzero\uthree}$, it follows from the Clifford algebra that
\begin{equation}\label{eq: H with gamma perp}
    \big[{\rm H}^{(1,2)}_\nu(z),\,\boldsymbol{\gamma}_{\rm T}\big]=0\;.
\end{equation}
Using the anticommutation relations, one can also prove the following relations
\begin{equation}\label{eq: H with gamma03}
    {\rm H}^{(1,2)}_\nu(z)\,\gamma^{\uzero,\uthree}
    =\gamma^{\uzero,\uthree}\,
    {\rm H}^{(1,2)}_{\underline{\nu}}(z)\;,
\end{equation}
where we introduced an underlined index $\underline{nu}$ to signal the change of sign in the real part of the matrix-valued order; an overline instead denotes the complex conjugate:
\begin{equation}
    \underline{\nu}=-\frac{1}{2}\gamma^\uthree\gamma^\uzero-\ii\mu\,\I\;,
    \qquad
    \overline{\nu}=\frac{1}{2}\gamma^\uthree\gamma^\uzero+\ii\mu\,\I\;.
\end{equation}
Since $\gamma^\uthree\gamma^\uzero$ is Hermitian and the argument $z$ is real, one further has:
\begin{equation}\label{app hank: dagger relations}
    {\rm H}^{(2)}_\nu(z)^\dagger=\hnku_{\bar{\nu}}(z)\;,
    \qquad
    \hnku_{\bar{\nu}}(z)^\dagger=\hnkd_\nu(z)\;,
\end{equation}
Finally, it follows that all the above properties extend naturally to the case of a matrix-valued order.  
In particular, using Eq.~\eqref{app hank: relazione *}  together with \eqref{app: Hankel matrix components} yields the following identity:
\begin{equation}\label{app hankel: wronskian matrix}
    \hnku_{\bar{\nu}}(z)\hnkd_{\nu}(z)
    +\hnku_{-\nu}(z)\hnkd_{\underline{\nu}}(z)
    =\frac{4}{\pi}\frac{e^{-\pi\mu}}{z}\,\I\;,
\end{equation}
where $\I$ is the identity in the spinor space.

\section{Milne spinor identities}\label{app: spinors}
In this appendix we report the most important results concerning the various products between Milne spinors \eqref{eq: Milne spinors}.

We start proving the orthogonality conditions \eqref{ortho conditions}.
Using the definitions \eqref{eq: Milne spinors}, we calculate:
\begin{equation}\label{eq:spinor U prod}
    U^\dagger_r\left(\tau,\fp\right)U_{r'}\left(\tau,\fp\right)=\frac{\pi e^{-\pi\mu}}{4}u^\dagger_r({\bf p}_{\rm T})\hnku_{\bar \nu}(z)\hnkd_\nu(z)u_{r'}({\bf p}_{\rm T})\;,
\end{equation}
with $z=m_{\rm T}\tau$, and we recall that the purely transverse Minkowski spinor is given by:
\begin{equation*}
    u_r({\bf p}_{\rm T})=\frac{m+m_{\rm T}\gamma^\uzero+{\bf p}_{\rm T}\cdot\boldsymbol{\gamma}_{\rm T}}{\sqrt{2(m_T+m)}}\; \xi_r\;.
\end{equation*}

We present here the general technique used to compute products such as \eqref{eq:spinor U prod}, involving transverse spinors and matrix-valued Hankel functions.
As a first step, we project the product of matrix-valued Hankel function along $\I\pm\gamma^\uthree\gamma^\uzero$, which can be easily done in the Weyl basis due to the fact that $\gamma^\uthree\gamma^\uzero$ is diagonal. In this case we have, using \eqref{app: Hankel matrix components}:
\begin{align}\label{app: espanione I gamma3gamma0}
        \hnku_{\bar \nu}(z)\hnkd_\nu(z)&={\rm diag}\left[\hnku_{w^*}(z)\hnkd_w(z),\hnku_{-w}(z)\hnkd_{-w^*}(z),\hnku_{-w}(z)\hnkd_{-w^*}(z),\hnku_{w^*}(z)\hnkd_w(z)\right]\nonumber\\
        &=\hnku_{w^*}(z)\hnkd_w(z)\frac{\I+\gamma^\uthree\gamma^\uzero}{2}+\hnku_{-w}(z)\hnkd_{-w^*}(z)\frac{\I-\gamma^\uthree\gamma^\uzero}{2}\,.
\end{align}
where $w=1/2-\ii\mu$.
With this expansion the spinor contraction reduces to:
\begin{equation*}
    \begin{split}
        U^\dagger_r\left(\tau,\fp\right)U_{r'}\left(\tau,\fp\right)=\frac{\pi e^{-\pi\mu}}{8}&\Big[\left(\hnku_{w^*}(z)\hnkd_w(z)+\hnku_{-w}(z)\hnkd_{-w^*}(z)\right)u^\dagger_r({\bf p}_{\rm T})u_{r'}({\bf p}_{\rm T})\\
        &+\left(\hnku_{w^*}(z)\hnkd_w(z)-\hnku_{-w}(z)\hnkd_{-w^*}(z)\right)u^\dagger_r({\bf p}_{\rm T})\gamma^\uthree\gamma^\uzero u_{r'}({\bf p}_{\rm T})\Big]\;.
    \end{split}
\end{equation*}
We have now reduced the problem to the calculation of contractions of standard Minkowski spinors with purely transverse momentum. The first term is just $u^\dagger_r({\bf p}_{\rm T}) u_{r'}({\bf p}_{\rm T})=2m_{\rm T} \delta_{rr'}$, whereas the second term is vanishing, $u^\dagger_r({\bf p}_{\rm T})\gamma^\uthree\gamma^\uzero u_{r'}({\bf p}_{\rm T})=0$, as we are considering purely transverse momenta (i.e. $p_z=0$).  Noticing that the combination of Hankel functions in the first term is proportional to the Wronskian (see eq. \eqref{app hank: relazione *}) we obtain:
\begin{equation}\label{eq: UdaggerU=delta}
    U^\dagger_r\left(\tau,\fp\right)U_{r'}\left(\tau,\fp\right)=\frac{\delta_{rr'}}{\tau}\;.
\end{equation}
The relation for the $V_r$ spinors is immediately obtained using charge conjugation transformations. Indeed, being $v_r=-\ii\gamma^\utwo u^*_r$ \cite{Peskin:1995ev}, we have:
\begin{equation}\label{eq: VdaggerV=delta}
    V_r^\dagger\left(\tau,\fp\right)V_{r'}\left(\tau,\fp\right)=U^T_r\left(\tau,\fp\right){\gamma^\utwo}^\dagger\gamma^\utwo U^*_{{r'}}\left(\tau,\fp\right)=\left[U^\dagger_{r}\left(\tau,\fp\right)U_{r'}\left(\tau,\fp\right)\right]^*=\frac{\delta_{rr'}}{\tau}\;,
\end{equation}
where we used the anti-hermiticiy of $\gamma^{\underline{1},\underline{2}}$.

The same method used in \eqref{app: espanione I gamma3gamma0} can be used to compute the mixed terms. For instance we have:
\begin{equation*}
    V^\dagger_r\left(\tau,\fp\right)U_{r'}\left(\tau,-\fp\right)=-\frac{\pi}{4}v_r^\dagger({\bf p}_{\rm T})e^{-\ii\pi\gamma^\uthree\gamma^\uzero/2}\hnkd_\nu(z)\hnkd_{\bar{\nu}}(z)u_{r'}(-{\bf p}_{\rm T})\;.
\end{equation*}
A direct matrix calculation reveals:
\begin{align*}
    e^{-\ii\pi\gamma^\uthree\gamma^\uzero/2}&\hnkd_\nu(z)\hnkd_{\bar{\nu}}(z)=\\
    &=\ii\,\text{diag}\left(
    -\hnkd_{w}(z)\hnkd_{w^*}(z),\hnkd_{-w^*}(z)\hnkd_{-w}(z),\hnkd_{-w^*}(z)\hnkd_{-w}(z),
    -\hnkd_{w^*}(z)\hnkd_{w}(z) 
    \right)
    \\
    &=\ii\left(\hnkd_{-w^*}(z)\hnkd_{-w}(z)\frac{\I-\gamma^\uthree\gamma^\uzero}{2}-\hnkd_{w}(z)\hnkd_{w^*}(z)\frac{\I+\gamma^\uthree\gamma^\uzero}{2}\right)\,.
\end{align*}
Using the fact that $v_r^\dagger({\bf p}_{\rm T})u_{r'}(-{\bf p}_{\rm T})=0$, we are left with
\begin{align*}
    V^\dagger_r\left(\tau,\fp\right)U_{r'}\left(\tau,-\fp\right)=\frac{\ii\pi}{8}\left(\hnkd_{w}(z)\hnkd_{w^*}(z)+\hnkd_{-w^*}(z)\hnkd_{-w}(z)\right)\; \left[v^\dagger_r({\bf p}_{\rm T})\gamma^\uthree\gamma^\uzero u_{r'}\left(-{\bf p}_{\rm T}\right)\right]\,,
\end{align*}
which also vanishes thanks to the identity \eqref{app: combinazione nulla}.
Therefore we have shown:
\begin{equation}\label{eq:VdaggerU=0}
    \begin{split}
        V^\dagger_r\left(\tau,\fp\right)U_{r'}\left(\tau,-\fp\right) &= 0\;,\\
    U^\dagger_r\left(\tau,-\fp\right)V_{r'}\left(\tau,\fp\right) &=0\;,
    \end{split}
\end{equation}
where the second line follows from the first by conjugation. Combining \eqref{eq: UdaggerU=delta}, \eqref{eq: VdaggerV=delta} and \eqref{eq:VdaggerU=0} one has:
\begin{equation}\label{app: orthogonal}
    \begin{split}
         U^\dagger_r\left(\tau,\fp\right)U_s\left(\tau,\fp\right)&=V^\dagger_r\left(\tau,\fp\right)V_s\left(\tau,\fp\right)=\frac{\delta_{rs}}{\tau}\;,\\
        \overline{U}_r\left(\tau,\fp\right)V_s\left(\tau,\fp\right)&=\overline{V}_r\left(\tau,\fp\right)U_s\left(\tau,\fp\right)=0\;,\\
        U^\dagger_r\left(\tau,\fp\right)V_s\left(\tau,-\fp\right)&=V^\dagger_r\left(\tau,\fp\right)U_s\left(\tau,-\fp\right)=0\;,
    \end{split}
\end{equation}
Many more spinor contractions are needed for the purposes of our paper. All of them can be computed using the method outlined above, but not all of them always reduce to the Wronskian of Hankel functions. In order to avoid a much too long appendix, we don't report the full derivation here, but just state the final results.
In order to express in a compact way each contraction we introduce two sets of adimensional auxiliary functions.
 The functions in eq. \eqref{def: functions}, already defined in the main text which appear in the computation of the exponent of the density operator eq. \eqref{def: matrices}, and the functions:
  \begin{subequations}\label{def: app functions2}
     \begin{align}
         {\rm f}(\tau,\fp)&=\frac{\pi}{4}e^{-\pi\mu}\hnku_{-\frac{1}{2}+\ii\mu}(z)\hnkd_{\frac{1}{2}-\ii\mu}(z)\;,\\
         {\rm w}(\tau,\fp)&=\frac{\pi}{4}\left(\hnku_{-\frac{1}{2}+\ii\mu}(z)\hnku_{\frac{1}{2}-\ii\mu}(z)+\hnku_{\frac{1}{2}+\ii\mu}(z)\hnku_{-\frac{1}{2}-\ii\mu}(z)\right)\;,\\
         {\rm z}(\tau,\fp)&=\frac{\pi}{4}\left(\hnku_{\frac{1}{2}+\ii\mu}(z)\hnku_{-\frac{1}{2}-\ii\mu}(z)-\hnku_{-\frac{1}{2}+\ii\mu}(z)\hnku_{\frac{1}{2}-\ii\mu}(z)\right)\;,
     \end{align}
 \end{subequations}
 which instead appear in the computation of the expectation value of the stress-energy tensor's components.

Not all the functions \eqref{def: functions} and \eqref{def: app functions2} are real or even under $\fp\mapsto-\fp$. However making use of the reflection properties of the Hankel functions \eqref{app hank: reflection} and the relations \eqref{app hank: relazioni wpm1} we can prove that the following properties under parity must hold: 
\begin{equation}\label{eq: parity prop}
    \begin{split}
        {\rm h}(\tau,-\mathfrak{p})&={\rm h}(\tau,\mathfrak{p})\;, \qquad\qquad\qquad
        {\rm j}(\tau,-\mathfrak{p})=-{\rm j}(\tau,\mathfrak{p})\;,\\
        {\rm s}(\tau,-\mathfrak{p})&=-{\rm s}(\tau,\mathfrak{p})\;,\qquad\qquad\qquad
        \hspace{-7pt}{\rm t}(\tau,-\mathfrak{p})={\rm t}(\tau,\mathfrak{p})\;,\\
        {\rm Re}[{\rm f}(\tau,-\mathfrak{p})]&=-{\rm Re}[{\rm f}(\tau,\mathfrak{p})]\;,\qquad
        {\rm Im}[{\rm f}(\tau,-\mathfrak{p})]={\rm Im}[{\rm f}(\tau,\mathfrak{p})]\;,\\
        {\rm w}(\tau,-\mathfrak{p})&={\rm w}(\tau,\mathfrak{p})\;,\qquad\qquad\qquad
        \hspace{-3pt}{\rm z}(\tau,-\mathfrak{p})=-{\rm z}(\tau,\mathfrak{p})\;,        
    \end{split}
\end{equation}

Following the same steps outlined in the computation of the contraction \eqref{eq: UdaggerU=delta}, \eqref{eq: VdaggerV=delta} and \eqref{eq:VdaggerU=0} we find that the following contraction between the spinors and their time derivatives:
\begin{equation}\label{app: derivate spinori finale}
    \begin{split}
        U^\dagger_r\left(\tau,\fp\right)\dot{U}_{r'}\left(\tau,\fp\right) &=m^2_{\rm T}{\rm h}(\tau,\mathfrak{p})\,\delta_{rr'}\;,\\
        V^\dagger_r\left(\tau,\fp\right)\dot{V}_{r'}\left(\tau,\fp\right) &=m^2_{\rm T}{\rm h}^*(\tau,\mathfrak{p})\,\delta_{rr'}\;,\\
        V^\dagger_r\left(\tau,\fp\right)\dot{U}_{r'}\left(\tau,-\fp\right)&=\ii m^2_{\rm T}\,{\rm j}(\tau,\mathfrak{p})\,\left(\sigma^x\right)_{rr'}\;,\\
        U^\dagger_r\left(\tau,\fp\right)\dot{V}_{r'}\left(\tau,-\fp\right)&=-\ii m^2_{\rm T}\,{\rm j}^*(\tau,\mathfrak{p})\,\left(\sigma^x\right)_{rr'}\;.
    \end{split}
\end{equation}
The above are used for the computation of the energy density in the Belinfante and Canonical pseudogauges.

For the canonical spin tensor, the following contraction are needed:
\begin{equation}\label{app: contrazioni gamma12}
    \begin{split}
        U^\dagger_r\left(\tau,\fp\right)\gamma^\uone\gamma^\utwo U_{r'}\left(\tau,\fp\right) &=-\frac{\ii m}{m_{\rm T}\tau}\left(\sigma^z\right)_{rr'}+\ii\,{\rm s}(\tau,\mathfrak{p})\left(p^x\sigma^x+p^y\sigma^y\right)_{rr'}\\
    V^\dagger_r\left(\tau,\fp\right)\gamma^\uone\gamma^\utwo V_{r'}\left(\tau,\fp\right)&=\frac{\ii m}{m_{\rm T}\tau}(\sigma^z)_{rr'}-\ii\,{\rm s}(\tau,\mathfrak{p})\left(p^x\sigma^x-p^y\sigma^y\right)_{rr'}\;,\\
    V^\dagger_r\left(\tau,\fp\right)\gamma^\uone\gamma^\utwo U_{r'}\left(\tau,-\fp\right)&={\rm t}(\tau,\mathfrak{p})\,\left(p^x\I+\ii p^y\sigma^z\right)_{rr'}\;\\
    U^\dagger_r\left(\tau,\fp\right)\gamma^\uone\gamma^\utwo V_{r'}\left(\tau,-\fp\right)&={\rm t}^*(\tau,\mathfrak{p})\,\left(p^x\I-\ii p^y\sigma^z\right)_{rr'}\;.
    \end{split}
\end{equation}
From the expression \eqref{app: derivate spinori finale} and \eqref{app: contrazioni gamma12} one can obtain the compact expression for the effective hamiltonian \eqref{eq: Pi matrix}.

If one is also interested in the computation of the expectation value of the stress-energy tensor, other contractions are needed. To derive eq.~\eqref{Press L generale}, one needs the contractions:
\begin{equation}\label{app eq: spinor contractions PL}
    \begin{split}
        U^\dagger_r\left(\tau,\fp\right)\gamma^\uzero\gamma^\uthree U_{r'}\left(\tau,\fp'\right)&=-m_{\rm T}\,\mathrm{s}\left(\tau,\fp\right)\,\delta_{rr'}\;,\\
        V^\dagger_r\left(\tau,\fp\right)\gamma^\uzero\gamma^\uthree V_{r'}\left(\tau,\fp\right)&=-m_{\rm T}\,\mathrm{s}\left(\tau,\fp\right)\,\delta_{rr'}\;,\\
        V^\dagger_r\left(\tau,\fp\right)\gamma^\uzero\gamma^\uthree U_{r'}\left(\tau,-\fp\right)&=-\ii m_{\rm T}\,{\rm t}(\tau,\mathfrak{p})\,(\sigma^x)_{rr'}\;,\\
        U^\dagger_r\left(\tau,\fp\right)\gamma^\uzero\gamma^\uthree V_{r'}\left(\tau,-\fp\right)&=\ii m_{\rm T}\,{\rm t}^*(\tau,\mathfrak{p})\left(\sigma^x\right)_{rr'}\;.
    \end{split}
\end{equation}
Instead, for the transverse pressure and the spin-torque, eqs.~\eqref{Press T generale} and~\eqref{Spin torque generale}, the needed contractions are:
\begin{align}\label{app eq: spinor contractions PT}
        U^\dagger_r(\fp)\gamma^\uzero\gamma^\uone U_{r'}(\fp)&=2p_x{\rm Im}\left[{\rm f}(\tau,\mathfrak{p})\right]\delta_{rr'}\nonumber\\
        &+2{\rm Re}\left[{\rm f}(\tau,\mathfrak{p})\right]\,\left(\frac{m^2+p_y^2+m m_{\rm T}}{m+m_{\rm T}}\sigma^y+\frac{p_x p_y}{m+m_{\rm T}}\sigma^x\right)_{rr'}\;,\nonumber\\
        U^\dagger_r(\fp)\gamma^\uzero\gamma^\uone V_{r'}(-\fp)&=-p_x{\rm z}(\tau,\mathfrak{p})\sigma^x_{rr'}\\
        &+{\rm w}(\tau,\mathfrak{p})\,\left(\frac{m^2+p_y^2+m m_{\rm T}}{m+m_{\rm T}}\sigma^z+\ii\frac{p_x p_y}{m+m_{\rm T}}\mathbb{1}\right)_{rr'}\;,\nonumber\\
        V^\dagger_r(\fp)\gamma^\uzero\gamma^\uone V_{r'}(\fp)&=\left(U^\dagger_r(\fp)\gamma^\uzero\gamma^\uone U_{r'}(\fp)\right)^*\;,\quad V^\dagger_r(\fp)\gamma^\uzero\gamma^\uone U_{r'}(-\fp)=\left(U^\dagger_r(\fp)\gamma^\uzero\gamma^\uone V_{r'}(-\fp)\right)^*\;,\nonumber
\end{align}
and
\begin{align}
        U^\dagger_r(\fp)\gamma^\uzero\gamma^\utwo U_{r'}(\fp)&=2p_y{\rm Im}\left[{\rm f}(\tau,\mathfrak{p}))\right]\delta_{rr'}\nonumber\\
        &-2{\rm Re}\left[{\rm f}(\tau,\mathfrak{p})\right]\,\left(\frac{m^2+p_x^2+m m_{\rm T}}{m+m_{\rm T}}\sigma^x+\frac{p_x p_y}{m+m_{\rm T}}\sigma^y\right)_{rr'}\;,\nonumber\\
    U^\dagger_r(\fp)\gamma^\uzero\gamma^\utwo V_{r'}(-\fp)&=-p_y{\rm z}(\tau,\mathfrak{p})\sigma^x_{rr'}\\
    &-\;{\rm w}(\tau,\mathfrak{p})\,\left(\ii\frac{m^2+p_x^2+m m_{\rm T}}{m+m_{\rm T}}\mathbb{1}+\frac{p_x p_y}{m+m_{\rm T}}\sigma^z\right)_{rr'}\;,\nonumber\\
        V_r^\dagger(\fp)\gamma^\uzero\gamma^\utwo V_{r'}(\fp)&=\left(U^\dagger_r(\fp)\gamma^\uzero\gamma^\utwo U_{r'}(\fp)\right)^*\;,\quad V^\dagger_r(\fp)\gamma^\uzero\gamma^\utwo U_{r'}(-\fp)=\left(U^\dagger_r(\fp)\gamma^\uzero\gamma^\utwo V_{r'}(-\fp)\right)^*\;.\nonumber
\end{align}
We remark that the functions defined in eqs.~\eqref{def: functions} and \eqref{def: app functions2} are not independent. Indeed some relations between them can be found, which are ultimately connected to the Wronskian and the reflection property of the Hankel functions. In particular, we find that these relations hold:
\begin{subequations}\label{eq: relazione s e t E hs e jt app}
    \begin{align}
    \mathfrak{h}^2(\tau,\fp)+\left|{\rm j}(\tau,\mathfrak{p})\right|^2&=\frac{\varepsilon^2(\tau,\fp)}{ m^4_{\rm T}\,\tau^2}\;,\label{eq: relazione hj}\\
    {\rm s}^2(z)+\left|{\rm t}(\tau,\mathfrak{p})\right|^2&=\frac{1}{m^2_{\rm T}\tau^2}\;,\label{eq: relazione st}\\
    {\mathfrak{h} }\left(\tau,\fp\right){\rm s}\left(\tau,\fp\right)+{\rm Im}\left({\rm j}(\tau,\mathfrak{p}){\rm t}^*(\tau,\mathfrak{p})\right)&=-\frac{\mu}{m^3_{\rm T}\tau^3}\;,\label{eq: relazione hsjt}\\
    2\mathfrak{h}\left(\tau,\fp\right){\rm Im}\left({\rm f}\left(\tau,\fp\right)\right)-{\rm Re}\left({\rm j}\left(\tau,\fp\right){\rm z}\left(\tau,\fp\right)\right)&=\frac{1}{m^2_{\rm T}\tau^2}\;, \label{eq: relazione hfjz}
    \end{align}
\end{subequations}
where $\varepsilon^2(\tau,\fp) = m_{\rm T}^2+\mu^2/\tau^2$ and we defined:
\begin{equation}
    \begin{split}
        \mathfrak{h}(\tau,\mathfrak{p})&\equiv\frac{\pi e^{-\pi\mu}}{4}{\rm Im}\left\{\left[\hnkd_{-w^*}(z)\frac{\di\hnku_{-w}(z)}{\di z}+\hnkd_{w}(z)\frac{\di\hnku_{w^*}(z)}{\di z}\right]\right\}\\
        &={\rm Im}({\rm h}^*(\tau,\fp))=-{\rm Im}({\rm h}(\tau,\fp))\;.
    \end{split}
\end{equation}
The proof of these relations relies on the properties of the Hankel functions described in appendix \ref{app:hankel}, notably the eqs.~\eqref{app hank: derivative relation} \eqref{app hank: Wronskian}, and \eqref{app hank: reflection}. Eq.~\eqref{app hank: derivative relation} must be used in order to write the derivatives of Hankel functions in terms of Hankel functions with the real part of their order equal to $\pm 1/2$. 

We report, as an example, the proof of the relation \eqref{eq: relazione st} linking ${\rm s}(\tau,\fp)$ and ${\rm t}(\tau,\fp)$.
Using the reflection property, eq. \eqref{app hank: reflection}, the Wronskian eq. \eqref{app hank: Wronskian} reads:
\begin{equation}\label{app eq: wroknskian simplified}
    \begin{split}
         \frac{4e^{\pi\mu}}{\pi z}&=\hnku_{w^*}(z)\hnkd_w(z)+\hnku_{-w}(z)\hnkd_{-w^*}(z)\\
         &=(\hnku_{w}(z))^*\hnkd_w(z)+e^{\ii\pi w}\hnku_{w}(z) e^{-\ii\pi w^*}\hnkd_{w^*}(z)\\
         &=\left|\hnkd_w(z)\right|^2+e^{2\pi\mu}\left|\hnku_w(z)\right|^2\\
         &\equiv A^2+e^{2\pi\mu}B^2\;,
    \end{split}
\end{equation}
where we defined:
\begin{equation}
    A^2\equiv\left|\hnkd_w(z)\right|^2\;,\qquad B^2\equiv\left|\hnku_{w}(z)\right|^2\;.
\end{equation}
Applying similar manipulations to the function ${\rm s}$, one gets:
\begin{equation}\label{app eq: sw simplified}
    \begin{split}
        {\rm s}(\tau,\mathfrak{p})&=\frac{\pi e^{-\pi\mu}}{4}\left(\left|\hnku_{w^*}(z)\right|^2-\left|\hnku_{-w}(z)\right|^2\right)\\
        &=\frac{\pi e^{-\pi\mu}}{4}\left(\left|\hnku_{w^*}(z)\right|^2-e^{2\pi\mu}\left|\hnku_{w}(z)\right|^2\right)\\
        &=\frac{\pi e^{-\pi\mu}}{4}\left(A^2-e^{2\pi\mu}B^2\right)\;.
    \end{split}
\end{equation}
Equations \eqref{app eq: sw simplified} and \eqref{app eq: wroknskian simplified} can be solved with respect to $A^2$ and $B^2$, yielding:
\begin{equation*}
    \begin{split}
        A^2&=\frac{2e^{\pi\mu}}{\pi}\left(\frac{1}{z}+{\rm s}(\tau,\mathfrak{p})\right)\;,\\
         B^2&=\frac{2e^{-\pi\mu}}{\pi}\left(\frac{1}{z}-{\rm s}(\tau,\mathfrak{p})\right)\;.
    \end{split}
\end{equation*}
From the definition of ${\rm t}(\tau,\fp)$, one has:
\begin{equation*}
    \left|{\rm t}(\tau,\mathfrak{p})\right|^2=\frac{\pi^2}{4}A^2B^2=\frac{1}{z^2}-{\rm s}^2(\tau,\fp)\;,
\end{equation*}
which finally implies
\begin{equation}
    {\rm s}^2(\tau,\fp)+\left|{\rm t}(\tau,\mathfrak{p})\right|^2=\frac{1}{z^2}=\frac{1}{m_{\rm T}^2\tau^2}\;.
\end{equation}

We conclude this appendix providing the large argument asymptotic expansion of the special functions \eqref{def: functions} at late times $z=m_{\rm T}\tau\gg1$. Being $\mu=\tau p_z$ in general $\mu/z\sim\mathcal{O}(1)$, plugging the asymptotic formulae \eqref{app: hank asymptotic grande z mu fix} in \eqref{def: functions} we get:
\begin{equation}\label{app: funz tau GRANDE}
    \begin{split}
        \mathfrak{h}\left(\tau,\mathfrak{p}\right)&=\frac{\varepsilon(\tau,\fp)}{\tau m^2_{\rm T}}+\mathcal{O}\left(\frac{1}{m^3_{\rm T}\tau^3}\right)\;,\quad {\rm j}\left(\tau,\mathfrak{p}\right)=\frac{\mu}{2\tau^3m_{\rm T}\varepsilon^2(\tau,\fp)}\e^{-2\ii\varphi}+\mathcal{O}\left(\frac{1}{m^3_{\rm T}\tau^3}\right)\;;\\
        {\rm s}\left(\tau,\fp\right)&=-\frac{\mu}{\tau^2m_{\rm T}\varepsilon(\tau,\fp)}+\mathcal{O}\left(\frac{1}{m^3_{\rm T}\tau^3}\right)\;,\quad {\rm t}\left(\tau,\fp\right)=-\frac{\e^{-2\ii\varphi}}{\tau\varepsilon(\tau,\fp)}+\mathcal{O}\left(\frac{1}{m^3_{\rm T}\tau^3}\right)\;,
    \end{split}
\end{equation}
where we used that for $z=m_{\rm T}\tau$ the $\varrho$ \eqref{rho e varphi} in the expansion \eqref{app: hank asymptotic grande z mu fix} reduces to $\varrho=\tau\varepsilon$ with $\varepsilon$ single particle energy.

\section{Computation of the effective Hamiltonian }\label{app: computation and asymptotics}

In this appendix we report the explicit computations of the operators $\widehat{\Pi}(\tau)$ and $\wQ(\tau)$ in terms of the Milne modes $\widehat{A}$ and $\widehat{B}$. This serves as an explicit example of the calculations of sections \ref{sec:zubarev longitudinal} and \ref{sec:exp values and renormalization}.

Using the compact notation: $\mathfrak{p}\equiv\left({\bf p}_{\rm T},\mu\right)$ and $\mathfrak{r}\equiv\left({\bf x}_{\rm T},\eta\right)$ such that $\di^3\fp\equiv\di^2{\rm p}_{\rm T}\di \mu$ and  $\di^3\fr=\di^2{\rm x}_{\rm T}\di\eta$ the Dirac field and its $\tau$-derivative can be expanded in Milne modes as follows:
\begin{equation}\label{app: field expansions}
    \begin{split}
        \psi\left(\tau,\fr\right)&=\frac{1}{\left(2\pi\right)^{3/2}}\sum_r\int\di^3\fp\left[e^{\ii\fp\cdot \fr}U_r\left(\tau,\fp\right)\widehat{A}_r(\fp)+e^{-\ii\fp\cdot\fr}V_r\left(\tau,\fp\right)\widehat{B}^\dagger_r(\fp)\right]\;,\\
        \dot{\psi}\left(\tau,\fr\right)&=\frac{1}{\left(2\pi\right)^{3/2}}\sum_r\int\di^3\fp\left[e^{\ii\fp\cdot \fr}\dot{U}_r\left(\tau,\fp\right)\widehat{A}_r(\fp)+e^{-\ii\fp\cdot\fr}\dot{V}_r\left(\tau,\fp\right)\widehat{B}^\dagger_r(\fp)\right]\;.
    \end{split}
\end{equation}
We start computing the current contribution. Being the current conserved, its integral over a space-like hypersurface is constant in time and thus the operator must be diagonal in the Milne set.
Plugging the field expansions eqs.~\eqref{app: field expansions} and integrating over one of the momenta $\di^3\fp'$ we obtain:
\begin{equation*}
    \begin{split}
        \int\di^3\fr\,\psi^\dagger(\tau,\fr)\psi(\tau,\fr)=\sum_{r,r'}\int&\di^3\fp\Big[U^\dagger_r(\tau,\fp)U_{r'}(\tau,\fp)\widehat{A}^\dagger_r(\fp)\widehat{A}_{r'}(\fp)+V^\dagger_r(\fp)V_{r'}(\fp)\widehat{B}_r(\fp)\widehat{B}^\dagger_{r'}(\fp)\\
        &+U^\dagger_r(\fp)V_{r'}(-\fp)\widehat{A}^\dagger_r(\fp)\widehat{B}^\dagger_{r'}(-\fp)+V^\dagger_r(\fp)U_{r'}(-\fp)\widehat{B}_r(\fp)\widehat{A}_{r'}(-\fp)\Big]\;.
    \end{split}
\end{equation*}
Taking advantage of the orthogonality conditions \eqref{ortho conditions}
the current term reduces to:
\begin{equation}
    \int\di^3\fr\,\psi^\dagger(\tau,\fr)\psi(\tau,\fr)=\sum_{r}\int\di^3\fp\left(\widehat{A}^\dagger_r(\fp)\widehat{A}_r(\fp)+\widehat{B}_r(\fp)\widehat{B}^\dagger_r(\fp)\right)\;,
\end{equation}
which is diagonal in the $\widehat{A}_r,\widehat{B}_r$ basis. Using the anticommutation relations \eqref{canonical anticommutator} one can write:
\begin{equation*}
    \widehat{B}_r(\fp)\widehat{B}^\dagger_r(\fp)=-\widehat{B}^\dagger_r(\fp)\widehat{B}_r(\fp)+\delta^3(0)\;,
\end{equation*}
so that the charge operator can be equally written as:
\begin{equation*}
    \widehat{Q}=\sum_{r}\int\di^3\fp\left(\widehat{A}^\dagger_r(\fp)\widehat{A}_r(\fp)-\widehat{B}_r(\fp)\widehat{B}^\dagger_r(\fp)\right)+2\delta^3(0)\int\di^3\fp\;,
\end{equation*}
which shows that particle and anti-particles contribute with opposite sign as expected. The $\delta^3(0)$ term is divergent, and is a typical term appearing in the quadratic operators build in terms of the fields; it is a multiplicative constant that is removed by the partition function in the density operator.

We come now to the computation of the stress-energy tensor term. Similarly to the charge operator, using eqs.~\eqref{app: field expansions} we get:
\begin{equation*}
    \begin{split}
        \int\di^3\fr\,\dot{\psi}^\dagger\left(\tau,\fr\right)\psi\left(\tau,\fr\right)=\sum_{r,r'}\int&\di^3\fp\Big[\dot{U}^\dagger_r(\tau,\fp)U_{r'}(\tau,\fp)\widehat{A}^\dagger_r(\fp)\widehat{A}_{r'}(\fp)+\dot{V}^\dagger_r(\fp)V_{r'}(\fp)\widehat{B}_r(\fp)\widehat{B}^\dagger_{r'}(\fp)\\
        &+\dot{U}^\dagger_r(\fp)V_{r'}(-\fp)\widehat{A}^\dagger_r(\fp)\widehat{B}^\dagger_{r'}(-\fp)+\dot{V}^\dagger_r(\fp)U_{r'}(-\fp)\widehat{B}_r(\fp)\widehat{A}_{r'}(-\fp)\Big]\;.
    \end{split}
\end{equation*}
From the definition of the stress-energy tensor, eq. \eqref{eq: Belinfante Semt}:
\begin{equation*}
    \wT^{00}(x)=\frac{\ii}{2}\psi^\dagger(x)\dot{\psi}(x)-\frac{\ii}{2}\dot{\psi}^\dagger(x)\psi(x)\;,
\end{equation*}
we have:
\begin{equation*}
    \begin{split}
        \frac{\ii}{2}\int\di^3\fr\, &\left[\psi^\dagger\left(\tau,\fr\right)\dot{\psi}\left(\tau,\fr\right)-\dot{\psi}^\dagger\left(\tau,\fr\right)\psi\left(\tau,\fr\right)\right]=\\
        &\frac{\ii}{2}\sum_{rr'}\int\di^3\fp\Big[\widehat{A}^\dagger_r(\fp)\widehat{A}_{r'}(\fp)\left(U^\dagger_r\left(\tau,\fp\right)\dot{U}_{r'}\left(\tau,\fp\right)-\dot{U}^\dagger_r\left(\tau,\fp\right)U_{r'}\left(\tau,\fp\right)\right)\\
        &+\widehat{A}^\dagger_r(\fp)\widehat{B}^\dagger_{r'}\left(-\fp\right)\left(U^\dagger_r(\fp)\dot{V}_{r'}\left(-\fp\right)-\dot{U}^\dagger_r\left(\tau,\fp\right)V_{r'}\left(\tau,-\fp\right)\right)\\
        &+\widehat{B}_r(\fp)\widehat{A}_{r'}\left(-\fp\right)\left(V^\dagger_r\left(\tau,-\fp\right)\dot{U}_{r'}\left(\tau,\fp\right)-\dot{V}^\dagger_r\left(\tau,-\fp\right)U_{r'}\left(\tau,\fp\right)\right)\\
        &+\widehat{B}_r(\fp)\widehat{B}^\dagger_{r'}(\fp)\left(V^\dagger_r\left(\tau,\fp\right)V_{r'}\left(\tau,\fp\right)-\dot{V}^\dagger_r\left(\tau,\fp\right)V_{r'}\left(\tau,\fp\right)\right)\Big]\;,
    \end{split}
\end{equation*}
 and using the relations \eqref{app: derivate spinori finale} and the definition \eqref{def: frak h} we get:
\begin{equation}\label{app: semt term}
    \begin{split}
       \int\di^3\fr\,\wT^{00}\left(\tau,\fr\right)=\sum_{rr'}\int\fp^3&\,m^2_{\rm T}\Big\{\left(\widehat{A}^\dagger_r(\fp)\widehat{A}_{r'}(\fp)-\widehat{B}_r(\fp)\widehat{B}^\dagger_{r'}(\fp)\right)\mathfrak{h}(\tau,\mathfrak{p})\delta_{rr'}\\
        &+\left[\widehat{A}^\dagger_r(\fp)\widehat{B}^\dagger_{r'}\left(-\fp\right){\rm j}^*(\tau,\mathfrak{p})\left(\,\sigma^x\right)_{rr'}+{\rm h.c}\right]\Big\}\,.
    \end{split}
\end{equation}

We now consider the final contribution to the statistical operator due to the spin tensor, which reads:
\begin{equation*}
    \wS^{\tau xy}(x)=\frac{\ii}{2}\psi^\dagger(x)\gamma^\uone\gamma^\utwo\psi(x)\;.
\end{equation*}
Using again the expansion in eqs.~\eqref{app: field expansions} we get:
\begin{equation*}
    \begin{split}
        \frac{\ii}{2}\int\di^3\fr\psi^\dagger\left(\tau,\fr\right)\gamma^\uone\gamma^\utwo\psi\left(\tau,\fr\right)=&\frac{\ii}{2}\sum_{rr'}\int\di^3\fp\Big[\widehat{A}^\dagger_r(\fp)\widehat{A}_{r'}(\fp)U^\dagger_r\left(\tau,\fp\right)\gamma^\uone\gamma^\utwo U_{r'}\left(\tau,\fp\right)\\
        &+\widehat{A}^\dagger_r(\fp)B^\dagger_{r'}(-\fp)U^\dagger_r\left(\tau,\fp\right)\gamma^\uone\gamma^\utwo V_{r'}\left(\tau,-\fp\right)\\
        &+\widehat{B}_r(-\fp)\widehat{A}_{r'}(\fp)V^\dagger_r\left(\tau,\fp\right)\gamma^\uone\gamma^\utwo U_{r'}\left(\tau,-\fp\right)\\
        &+\widehat{B}_r(-\fp)\widehat{B}^\dagger_{r'}(-\fp)V^\dagger_r\left(\tau,-\fp\right)\gamma^\uone\gamma^\utwo V_{r'}\left(\tau,-\fp\right)\Big]\;.
    \end{split}
\end{equation*}
Plugging in the above result the eq. \eqref{app: contrazioni gamma12}, the expression turns out to be:
\begin{equation*}
    \begin{split}
        &\frac{\ii}{2}\int\di^3\fr\psi^\dagger\left(\tau,\fr\right)\gamma^\uone\gamma^\utwo\psi\left(\tau,\fr\right)=\frac{1}{2}\sum_{rr'}\int\di^3\fp\Big[\\
        &\times \widehat{A}^\dagger_r(\fp)\widehat{A}_{r'}(\fp)\left(\frac{m}{m_{\rm T}\tau}(\sigma^z)_{rr'}-{\rm s}(\tau,\mathfrak{p})\left(p_x\sigma^x+p_y\sigma^y\right)_{rr'}\right)\\
        &+\widehat{A}^\dagger_r(\fp)B^\dagger_{r'}(-\fp)\left(\ii p_x\I+ p_y\sigma^z\right)_{rr'}{\rm t}^*(\tau,\mathfrak{p})+\widehat{B}_r(-\fp)\widehat{A}_{r'}(\fp)\left(\ii p_x\I- p_y\sigma^z\right)_{rr'}{\rm t}(\tau,\mathfrak{p})\\
        &+\widehat{B}_r(-\fp)\widehat{B}^\dagger_{r'}(-\fp)\left(-\frac{ m}{m_{\rm T}\tau}(\sigma^z)_{rr'}+{\rm s}(z)\left(p_x\sigma^x-p_y\sigma^y\right)_{rr'}\right)\Big]\;.
    \end{split}
\end{equation*}
The operators $\wPi_\Omega$ and $\wQ$ can finally be written in a compact way introducing the following vector notation for the Milne field operators:
\begin{equation}
    \Phi^T(\fp)\equiv\begin{pmatrix}
        \widehat{A}_+(\fp),
        \widehat{A}_-(\fp),
        \widehat{B}^\dagger_+(-\fp),
        \widehat{B}^\dagger_-(-\fp)\;,
    \end{pmatrix}
\end{equation}
and defining the following matrices:
\begin{equation}
    \begin{split}
        \mathcal{H}&=m^2_{\rm T}\tau\begin{pmatrix}
        \mathfrak{h}(\tau,\mathfrak{p})\I_{2\times2}&{\rm j}^*(\tau,\mathfrak{p})\sigma_x\\
        {\rm j}(\tau,\mathfrak{p})\sigma_x&-\mathfrak{h}(\tau,\mathfrak{p})\I_{2\times2}
    \end{pmatrix}\;,\\
    \quad \mathcal{S}&=\begin{pmatrix}
        \frac{m}{m_{\rm T}\tau}\sigma_z-{\rm s}(\tau,\mathfrak{p})\left(p^x\sigma_x+p^y\sigma_y\right)&{\rm t}^*(\tau,\mathfrak{p})\left( \ii p^x\I_{2\times2}+ p^y\sigma_z\right)\\
        {\rm t}(\tau,\mathfrak{p})\left(p^y\sigma_z-\ii p^x\I_{2\times 2}\right)&-\frac{m}{m_{\rm T}\tau}\sigma_z+{\rm s}(\tau,\mathfrak{p})\left(p^x\sigma_x-p^y\sigma_y\right)
    \end{pmatrix}\;,
    \end{split}
\end{equation}
where $\Omega(\tau)\equiv T\SP(\tau)$. We obtain:
\begin{equation}
    \begin{split}
        \widehat{\Pi}_\Omega(\tau)&=\int\di^3\fp\,\Phi^\dagger(\fp)\Big[\mathcal{H}\left(\tau,\fp\right)-\frac{\tau\Omega(\tau)}{2}\mathcal{S}\left(\tau,\fp\right)\Big]\Phi(\fp)\;,\\
        \widehat{Q}(\tau)&=\int\di^3\fp\,\tau\,\Phi^\dagger(\fp)\Phi(\fp)\;,
    \end{split}
\end{equation}
which are eqs. \eqref{eq: Pi matrix} \eqref{def: matrices} in the main text.
Note that the matrices $\mathcal{H}$ and $\mathcal{S}$ have dimensions of energy in natural units.

\section{Bogoliubov transformation in the Canonical pseudogauge}\label{app: projection procedure canonical eigenv}
We provide here a procedure to compute analytically the eigenvectors of the effective hamiltonian in the canonical pseudogauge. This procedure builds upon the diagonalization of $\mathcal{H}_{\rm tot}^2$, and we obtain the eigenvectors of $\mathcal{H}_{\rm tot}$ with appropriate projectors.

The eigenvalues and eigenvectors of $\mathcal{H}_{\rm tot}^2$ can be found explicitly: the spectrum is given by $\mathcal{\omega}_{\pm}^2$ in eq. \eqref{H2 eigenvalue}, whereas the eigenvectors are spanned by:
\begin{align*}
    w_1^+ &= \left(\frac{\mathfrak{h}(\tau,\fp)-\mathfrak{s}\frac{m_{\rm L}}{ m_{\rm T} m\tau}}{ {\rm j}(\tau,\fp)},\frac{\mu  (p_x+\ii
   p_y)}{m m_{\rm T}^2 \tau ^2 {\rm j}(\tau,\fp)},0,1\right)\;,\\
   w_2^+ &= \left(-\frac{\mu  (p_x-\ii
   p_y)}{m m_{\rm T}^2 \tau ^2 {\rm j}(\tau,\fp)},\frac{\mathfrak{h}(\tau,\fp)+\mathfrak{s}\frac{m_{\rm L}}{ m_{\rm T} m\tau}}{ {\rm j}(\tau,\fp)},1,0\right)\;,\\
    w_1^- &= \left(\frac{\mathfrak{h}(\tau,\fp)+\mathfrak{s}\frac{m_{\rm L}}{ m_{\rm T} m\tau}}{ {\rm j}(\tau,\fp)},\frac{\mu  (p_x+\ii
   p_y)}{m m_{\rm T}^2 \tau ^2 {\rm j}(\tau,\fp)},0,1\right)\;,\\
   w_2^- &= \left(-\frac{\mu  (p_x-\ii
   p_y)}{m m_{\rm T}^2 \tau ^2 {\rm j}(\tau,\fp)},\frac{\mathfrak{h}(\tau,\fp)-\mathfrak{s}\frac{m_{\rm L}}{ m_{\rm T} m\tau}}{ {\rm j}(\tau,\fp)},1,0\right)\;,
\end{align*}
where $\mathfrak{s}=\mathrm{sign}(\Omega)$. These eigenvectors are not orthogonal due to the degeneracy of the eigenvalues $\omega_+^2$ and $\omega_-^2$. In this form they are also not normalized. However, neither of these facts thwarts the following procedure.
We remark that the above eigenvectors boil down to those of $\mathcal{H}$ in the case $\bf{p}_{\rm T}=0$. In this singular case the procedure we are about to present cannot be applied, but this is not a problem as the case of longitudinal momentum can be solved analytically as shown in the main text. 

We can now build eigenvectors of $\mathcal{H}$ corresponding to $\pm \omega_{\pm}$. Focusing on $\omega_+$ as an example, one defines:
\begin{equation*}
    v_{\pm} = \frac{1}{2}\left(\mathbb{1}\pm\frac{\mathcal{H}_{tot}}{\mathcal{\omega}_{+}}\right)(aw_1^++bw_2^+)\equiv \Pi_{\pm}(aw_1^++bw_2^+)\;,
\end{equation*}
where $a$ and $b$ are arbitrary coefficients and we have introduced $\Pi_{\pm}$, which is a projector ($\Pi^2_{\pm}=\Pi_{\pm}$) on the eigenspace of $\mathcal{\omega}_{+}$. One realises easily that $v_{\pm}$ is the eigenvector of $\mathcal{H}_{tot}$ with eigenvalue $\pm \omega_+$:
\begin{equation*}
    \begin{split}
        \mathcal{H}_{\rm tot}v_{\pm}&=\frac{1}{2}\left(\mathcal{H}_{\rm tot}\pm\frac{\mathcal{\omega}_{+}^2}{\mathcal{\omega}_{+}}\right)(aw^+_1+bw^+_2)=\pm \frac{\mathcal{\omega}_{+}}{2}\left(\mathbb{1}\pm\frac{\mathcal{H}_{\rm tot}}{\mathcal{\omega}_{+}}\right)(aw^+_1+bw^+_2)=\pm \mathcal{\omega}_{+}v_{\pm}\;.
    \end{split}
\end{equation*}
Note that since $\Pi_+\Pi_-=0$ on the space spanned by $w^+_{12}$, $v_+$ and $v_-$ are automatically orthogonal.
Introducing the projector on the eigenspace spanned by $w^-_{1,2}$:
\begin{equation*}
    \Delta_{\pm}=\frac{1}{2}\left(\mathbb{1}\pm\frac{\mathcal{H}_{tot}}{\omega_-}\right)\;,
\end{equation*}
and denoting the eigenvectors by the ket of the corresponding eigenvalue, one constructs the eigenvectors of $\mathcal{H}_{tot}$ as:
\begin{align*}
    |\omega_+\rangle=\frac{\Pi_{+}w_1^+}{||\Pi_{+}w_1^+||}\;,\qquad |-\omega_+\rangle=\frac{\Pi_{-}w_2^+}{||\Pi_{-}w_2^+||}\;,\\
    |\omega_-\rangle=\frac{\Delta_{+}w_1^-}{||\Delta_{+}w_1^-||}\;,\qquad
    |-\omega_-\rangle=\frac{\Delta_{-}w_2^-}{||\Delta_{-}w_2^-||}\;.
\end{align*}
These eigenvectors are automatically orthonormal, as they are eigenvectors corresponding to different eigenvalues of a hermitian matrix.

The matrix $\mathcal{U}$ diagonalizing the total Hamiltonian $\mathcal{H}_{tot}$ is then built taking each eigenvector as column:
\begin{equation}
    \mathcal{U}=\begin{pmatrix}
        |\omega_+\rangle, &|\omega_-\rangle, &|-\omega_+\rangle, &|-\omega_-\rangle
    \end{pmatrix}\;,
\end{equation}
and it is unitary by construction.

The explicit expression of each eigenvector is extremely involved.
For the eigenvectors associated with the \emph{positive energy} eigenvalues, i.e $\omega_\pm>0$, we obtain:
\begin{align*}
    |\omega_+\rangle=\mathcal{N}_{++}&\left\{
\frac{m\left( -m^2 + m_{\rm L}^2 + m_{\rm T}^2\right) \mathfrak{s} - m_{\rm L} m_{\rm T}^3 \tau \mathfrak{h} }{2{\rm j}\, m \, m_{\rm T}^2 \mathfrak{s}\, \tau \omega_+}
+ \frac{-m_{\rm L} + m \, m_{\rm T}\, \mathfrak{s} \tau \mathfrak{h}}{2 m \, m_{\rm T} \mathfrak{s} \tau {\rm j}}\right.
\nonumber\\
&+ \frac{\Omega \left( m \, m_{\rm L} - m^2 m_{\rm T} \mathfrak{s} \tau \mathfrak{h} +|{\bf p}_{\rm T}|^2 \mathfrak{s} \mu\, {\rm s} \right)}{4 {\rm j} \,m \, m_{\rm T}^2 \mathfrak{s}\,\tau\omega_+},
\nonumber \\
&\frac{(p_x + \ii p_y) \mu}{2{\rm j} m \, m_{\rm T}^2 \tau^2}
+ \frac{(p_x + \ii p_y) \mu \mathfrak{h}}{2{\rm j} m\,\tau\omega_+ }
\nonumber\\
&+ \frac{(p_x + \ii p_y) \Omega \left[ m_{\rm T}^2 \tau^2 (-m_{\rm L} - m \, m_{\rm T} \mathfrak{s} \tau \mathfrak{h}) {\rm s} - m \mathfrak{s} (\mu + \ii m_{\rm T}^3 \tau^3 {\rm j}^* {\rm t}) \right]}{4  {\rm j}\,m \, m_{\rm T}^3 \mathfrak{s} \tau^2 \omega_+},\nonumber
\\
&\quad \frac{(p_x + \ii p_y)  \mu}{2 m \, \tau \omega_+}
+ \frac{(p_x + \ii p_y) \Omega \left[ -m \, m_{\rm T} \mathfrak{s} \tau {\rm j} {\rm s} - \ii (m_{\rm L} - m \, m_{\rm T} \mathfrak{s} \tau \mathfrak{h}) {\rm t} \right]}{4{\rm j} m \, m_{\rm T} \mathfrak{s} \omega_+},\nonumber
\\
&\quad \left. \frac{1}{2} - \frac{m_{\rm L} m_{\rm T}}{2 m \mathfrak{s} \omega_+} - \frac{\Omega \left( m^2 m_T \tau {\rm j} - \ii|{\bf p}_{\rm T}|^2 \mu\, {\rm t} \right)}{4  {\rm j} m \, m_{\rm T}^2 \tau \omega_+}
\right\}\;,
\end{align*}
and 
\begin{align*}
|\omega_-\rangle=\mathcal{N}_{+-}
&\left\{
\frac{ m\left(-m^2 + m_L^2 + m_T^2\right) \mathfrak{s} + m_{\rm L} m_{\rm T}^3 \tau \mathfrak{h} }{2{\rm j} m \, m_{\rm T}^2 \mathfrak{s} \tau \omega_-}
- \frac{-m_{\rm L} - m \, m_{\rm T} \mathfrak{s} \tau \mathfrak{h}}{2  {\rm j} m \, m_{\rm T} \mathfrak{s} \tau}\right.\nonumber \\
&- \frac{\Omega \left( m \, m_{\rm L} + m^2 m_{\rm T} \mathfrak{s} \tau \mathfrak{h} - |{\bf p}_{\rm T}|^2 \mathfrak{s} \mu {\rm s} \right)}{4 {\rm j} m \, m_{\rm T}^2 \mathfrak{s} \tau \omega_-},\nonumber
\\
&\quad \frac{(p_x + \ii p_y) \mu}{2 {\rm j} m \, m_{\rm T}^2 \tau^2 }
+ \frac{(p_x + \ii p_y) \mu \mathfrak{h}}{2 {\rm j} m \, \tau \omega_- }\nonumber
\\
&+ \frac{(p_x + \ii p_y) \Omega \left( m_{\rm T}^2 \tau^2 (m_{\rm L} - m \, m_{\rm T} \mathfrak{s} \tau \mathfrak{h}) {\rm s} - m \mathfrak{s} (\mu + \ii m_{\rm T}^3 \tau^3  {\rm j}^*  {\rm t}) \right)}{4 {\rm j} m \, m_{\rm T}^3 \mathfrak{s} \tau^2 \omega_- },\nonumber
\\
&\quad \frac{(p_x + \ii p_y) \mu}{2 m \, \tau \omega_-}
+ \frac{(p_x + \ii p_y) \Omega \left[ -m \, m_{\rm T} \mathfrak{s} \tau {\rm j} {\rm s} + \ii (m_L + m \, m_T \mathfrak{s} \tau \mathfrak{h})  {\rm t} \right]}{4 {\rm j} m \, m_{\rm T} \mathfrak{s} \omega_- },\nonumber
\\
&\quad \left. \frac{1}{2} + \frac{m_{\rm L} m_{\rm T}}{2 m \mathfrak{s} \omega_-} - \frac{\Omega \left( m^2 m_{\rm T} \tau {\rm j} - \ii |{\bf p}_{\rm T}|^2 \mu {\rm t} \right)}{4{\rm j} m \, m_{\rm T}^2 \tau \omega_-}
\right\}\;.
\end{align*}
Instead the eigenvectors associated with negative energies, i.e $-\omega_\pm <0$ we have:
 \begin{align*}
    |-\omega_{+}\rangle=\mathcal{N}_{-+}&\left\{
- \frac{(p_x - \ii p_y) \mu}{2{\rm j} m \, m_{\rm T}^2 \tau^2 }
+ \frac{(p_x - \ii p_y) \mu \mathfrak{h}}{2{\rm j} m \, \tau \omega_+ }\right.\nonumber \\
&- \frac{(p_x - \ii p_y) \Omega\left[ m_{\rm T}^2 \tau^2 (m_{\rm L} - m \, m_{\rm T} \mathfrak{s} \tau \mathfrak{h}) {\rm s} - m \mathfrak{s} (\mu + \ii m_{\rm T}^3 \tau^3 {\rm j}^* {\rm t}) \right]}{4{\rm j} m \, m_{\rm T}^3 \mathfrak{s} \tau^2 \omega_+},\nonumber
\\
&\quad \frac{m \left(m^2 - m_{\rm L}^2 - m_{\rm T}^2\right) \mathfrak{s} - m_{\rm L} m_{\rm T}^3 \tau \mathfrak{h}}{2{\rm j} m \, m_{\rm T}^2 \mathfrak{s} \tau \omega_+ }
- \frac{-m_{\rm L} - m \, m_{\rm T} \mathfrak{s} \tau \mathfrak{h}}{2{\rm j} m \, m_{\rm T} \mathfrak{s} \tau }\\
&- \frac{\Omega \left( m \, m_{\rm L} + m^2 m_{\rm T} \mathfrak{s} \tau \mathfrak{h} - |{\bf p}_{\rm T}|^2 \mathfrak{s} \mu {\rm s} \right)}{4{\rm j} m \, m_{\rm T}^2 \mathfrak{s} \tau \omega_+},\nonumber
\\
&\quad \frac{1}{2} - \frac{m_{\rm L} m_{\rm T}}{2 m \mathfrak{s} \omega_+} - \frac{\Omega \left( m^2 m_{\rm T} \tau {\rm j} - \ii|{\bf p}_{\rm T}|^2 \mu {\rm t} \right)}{4 {\rm j} m \, m_{\rm T}^2 \tau \omega_+},\nonumber
\\
&\quad \left. \frac{(p_x - \ii p_y) \mu}{2 m \, \tau \omega_+}
- \frac{(p_x - \ii p_y) \Omega\left( -m \, m_{\rm T} \mathfrak{s} \tau {\rm j} {\rm s} + \ii (m_{\rm L} + m \, m_{\rm T} \mathfrak{s} \tau \mathfrak{h}) {\rm t} \right)}{4 m \, m_{\rm T} \mathfrak{s} \omega_+ {\rm j}}
\right\}\;
\end{align*}   
and 
\begin{align*}
    |-\omega_{-}\rangle=\mathcal{N}_{--}&\left\{
- \frac{(p_x - \ii p_y) \mu}{2{\rm j} m \, m_{\rm T}^2 \tau^2}
+ \frac{(p_x - \ii p_y) \mu \mathfrak{h}}{2{\rm j} m \, \tau \omega_-}
\right.\nonumber \\
&- \frac{(p_x - \ii p_y) \Omega \left[ m_{\rm T}^2 \tau^2 (-m_{\rm L} - m \, m_{\rm T} \mathfrak{s} \tau \mathfrak{h}) {\rm s} - m \mathfrak{s} (\mu + \ii m_T^3 \tau^3 {\rm j}^* {\rm t}) \right]}{4{\rm j} m \, m_{\rm T}^3 \mathfrak{s} \tau^2 \omega_-},
\nonumber\\
&\quad \frac{m \left(m^2 - m_{\rm L}^2 - m_{\rm T}^2\right) \mathfrak{s} + m_{\rm L} m_{\rm T}^3 \tau \mathfrak{h} }{2{\rm j} m \, m_{\rm T}^2 \mathfrak{s} \tau \omega_-}
+ \frac{-m_{\rm L} + m \, m_{\rm T} \mathfrak{s} \tau \mathfrak{h}}{2{\rm j} m \, m_{\rm T} \mathfrak{s} \tau}
\nonumber \\
&+ \frac{\Omega \left[ m \, m_{\rm L} - m^2 m_{\rm T} \mathfrak{s} \tau \mathfrak{h} + |{\bf p}_{\rm T}|^2 \mathfrak{s} \mu {\rm s} \right)}{4{\rm j} m \, m_{\rm T}^2 \mathfrak{s} \tau \omega_-},\nonumber
\\
&\quad \frac{1}{2} + \frac{m_{\rm L} m_{\rm T}}{2 m \mathfrak{s} \omega_-} - \frac{\Omega \left( m^2 m_{\rm T} \tau {\rm j} - \ii|{\bf p}_{\rm T}|^2 \mu {\rm t}\right)}{4{\rm j} m \, m_{\rm T}^2 \tau \omega_-},\nonumber
\\
&\quad \left. \frac{(p_x - \ii p_y) \mu}{2 m \, \tau \omega_-}
- \frac{(p_x - \ii p_y) \Omega \left( -m \, m_{\rm T} \mathfrak{s} \tau {\rm j} {\rm s} - \ii (m_{\rm L} - m \, m_{\rm T} \mathfrak{s} \tau \mathfrak{h}) {\rm t} \right)}{4{\rm j} m \, m_{\rm T} \mathfrak{s} \omega_-}
\right\}\;,
\end{align*}
where the normalization factors $\mathcal{N_{\pm\pm}}$ are left implicit and are fixed by $\langle \pm\omega_\pm|\pm \omega_\pm\rangle=1$.


\bibliographystyle{JHEP}
\bibliography{main}

@article{Becattini:2023ouz,
    author = "Becattini, Francesco and Daher, Asaad and Sheng, Xin-Li",
    title = "{Entropy current and entropy production in relativistic spin hydrodynamics}",
    eprint = "2309.05789",
    archivePrefix = "arXiv",
    primaryClass = "nucl-th",
    doi = "10.1016/j.physletb.2024.138533",
    journal = "Phys. Lett. B",
    volume = "850",
    pages = "138533",
    year = "2024"
}

@article{Singh:2026ytd,
    author = "Singh, Sejal and Dey, Sourav and Das, Arpan and Mishra, Hiranmaya and Jaiswal, Amaresh",
    title = "{Dissipative spin hydrodynamics in Bjorken flow and thermal dilepton production}",
    eprint = "2604.04533",
    archivePrefix = "arXiv",
    primaryClass = "nucl-th",
    month = "4",
    year = "2026"
}

@article{Braguta:2025ddq,
    author = "Braguta, V. V. and Chernodub, M. N. and Roenko, A. A.",
    title = "{Chiral and deconfinement thermal transitions at finite quark spin polarization in lattice QCD simulations}",
    eprint = "2503.18636",
    archivePrefix = "arXiv",
    primaryClass = "hep-lat",
    doi = "10.1103/xptn-qgfl",
    journal = "Phys. Rev. D",
    volume = "111",
    number = "11",
    pages = "114508",
    year = "2025"
}

@article{Singha:2025bda,
    author = "Singha, Pracheta and Ambrus, Victor E. and Busuioc, Sergiu and Bandyopadhyay, Aritra and Chernodub, Maxim N.",
    title = "{Chiral restoration temperature at finite spin density in QCD}",
    eprint = "2508.20237",
    archivePrefix = "arXiv",
    primaryClass = "nucl-th",
    month = "8",
    year = "2025"
}

@article{Palermo:2025imv,
    author = "Palermo, Andrea and Shokri, Masoud",
    title = "{Freeze-out at constant energy density and spin polarization in heavy-ion collisions}",
    eprint = "2507.18761",
    archivePrefix = "arXiv",
    primaryClass = "hep-ph",
    doi = "10.1103/2qsx-kgwr",
    journal = "Phys. Rev. D",
    volume = "112",
    number = "7",
    pages = "076011",
    year = "2025"
}

@manual{mpmath,
  key     = {mpmath},
  author  = {The mpmath development team},
  title   = {mpmath: a {P}ython library for arbitrary-precision floating-point arithmetic (version 1.4.0)},
  note    = {{\tt http://mpmath.org/}},
  year    = {2026},
}

@misc{repo,
  author = {Palermo, Andrea and Roselli, Daniele},
  title = {DiracFermions-in-MilneSpacetime},
  year = {2025},
  publisher = {GitHub},
  journal = {GitHub repository},
  howpublished = {\url{https://github.com/AndrePalermo/DiracFermions-in-MilneSpacetime}}
}

@book{Kapusta:2006pm,
    author = "Kapusta, J. I. and Gale, Charles",
    title = "{Finite-temperature field theory: Principles and applications}",
    doi = "10.1017/CBO9780511535130",
    isbn = "978-0-521-17322-3, 978-0-521-82082-0, 978-0-511-22280-1",
    publisher = "Cambridge University Press",
    series = "Cambridge Monographs on Mathematical Physics",
    year = "2011"
}

@article{Alzhrani:2022dpi,
    author = "Alzhrani, Sahr and Ryu, Sangwook and Shen, Chun",
    title = "{{\ensuremath{\Lambda}} spin polarization in event-by-event relativistic heavy-ion collisions}",
    eprint = "2203.15718",
    archivePrefix = "arXiv",
    primaryClass = "nucl-th",
    doi = "10.1103/PhysRevC.106.014905",
    journal = "Phys. Rev. C",
    volume = "106",
    number = "1",
    pages = "014905",
    year = "2022"
}

@article{Becattini:2025twu,
    author = "Becattini, F. and Hoyos, C.",
    title = "{Pseudo-gauge invariant non-equilibrium density operator}",
    eprint = "2507.09249",
    archivePrefix = "arXiv",
    primaryClass = "nucl-th",
    month = "7",
    year = "2025"
}

@article{Buzzegoli:2024mra,
    author = "Buzzegoli, Matteo and Palermo, Andrea",
    title = "{Emergent Canonical Spin Tensor in the Chiral-Symmetric Hot QCD}",
    eprint = "2407.14345",
    archivePrefix = "arXiv",
    primaryClass = "hep-ph",
    doi = "10.1103/PhysRevLett.133.262301",
    journal = "Phys. Rev. Lett.",
    volume = "133",
    number = "26",
    pages = "262301",
    year = "2024"
}

@article{Dey:2023hft,
    author = "Dey, Sourav and Florkowski, Wojciech and Jaiswal, Amaresh and Ryblewski, Radoslaw",
    title = "{Pseudogauge freedom and the SO(3) algebra of spin operators}",
    eprint = "2303.05271",
    archivePrefix = "arXiv",
    primaryClass = "hep-th",
    doi = "10.1016/j.physletb.2023.137994",
    journal = "Phys. Lett. B",
    volume = "843",
    pages = "137994",
    year = "2023"
}

@article{Singh:2026wvf,
    author = "Singh, Rajeev and Soloviev, Alexander",
    title = "{Spin hydrodynamics on a hyperbolic expanding background}",
    eprint = "2603.02296",
    archivePrefix = "arXiv",
    primaryClass = "nucl-th",
    month = "3",
    year = "2026"
}

@article{Wang:2021ngp,
    author = "Wang, Dong-Lin and Fang, Shuo and Pu, Shi",
    title = "{Analytic solutions of relativistic dissipative spin hydrodynamics with Bjorken expansion}",
    eprint = "2107.11726",
    archivePrefix = "arXiv",
    primaryClass = "nucl-th",
    doi = "10.1103/PhysRevD.104.114043",
    journal = "Phys. Rev. D",
    volume = "104",
    number = "11",
    pages = "114043",
    year = "2021"
}

@article{Biswas:2022bht,
    author = "Biswas, Rajesh and Daher, Asaad and Das, Arpan and Florkowski, Wojciech and Ryblewski, Radoslaw",
    title = "{Boost invariant spin hydrodynamics within the first order in derivative expansion}",
    eprint = "2211.02934",
    archivePrefix = "arXiv",
    primaryClass = "nucl-th",
    doi = "10.1103/PhysRevD.107.094022",
    journal = "Phys. Rev. D",
    volume = "107",
    number = "9",
    pages = "094022",
    year = "2023"
}

@article{Drogosz:2024lkx,
    author = "Drogosz, Zbigniew and Florkowski, Wojciech and {\L}ygan, Natalia and Ryblewski, Radoslaw",
    title = "{Boost-invariant spin hydrodynamics with spin feedback effects}",
    eprint = "2411.06154",
    archivePrefix = "arXiv",
    primaryClass = "hep-ph",
    doi = "10.1103/PhysRevC.111.024909",
    journal = "Phys. Rev. C",
    volume = "111",
    number = "2",
    pages = "024909",
    year = "2025"
}

@article{Ambrus:2025dca,
    author = "Ambrus, Victor E. and Geci{\'c}, Aleksandar",
    title = "{Thermodynamics of rotating fermions}",
    eprint = "2509.17640",
    archivePrefix = "arXiv",
    primaryClass = "hep-th",
    month = "9",
    year = "2025"
}

@article{Armas:2026bmw,
    author = "Armas, Jay and Jain, Akash",
    title = "{Thermodynamics of ideal spin fluids and pseudo-gauge ambiguity}",
    eprint = "2601.14421",
    archivePrefix = "arXiv",
    primaryClass = "hep-th",
    month = "1",
    year = "2026"
}

@article{Florkowski:2026ofs,
    author = "Florkowski, Wojciech and Kar, Sudip Kumar and Mykhaylova, Valeriya",
    title = "{Spin alignment, tensor polarizabilities, and local equilibrium for spin-1 particles}",
    eprint = "2602.00819",
    archivePrefix = "arXiv",
    primaryClass = "nucl-th",
    month = "1",
    year = "2026"
}

@article{Niida:2024ntm,
    author = "Niida, Takafumi and Voloshin, Sergei A.",
    title = "{Polarization phenomenon in heavy-ion collisions}",
    eprint = "2404.11042",
    archivePrefix = "arXiv",
    primaryClass = "nucl-ex",
    doi = "10.1142/S0218301324300108",
    journal = "Int. J. Mod. Phys. E",
    volume = "33",
    number = "09",
    pages = "2430010",
    year = "2024"
}

@article{Becattini:2024uha,
    author = "Becattini, Francesco and Buzzegoli, Matteo and Niida, Takafumi and Pu, Shi and Tang, Ai-Hong and Wang, Qun",
    title = "{Spin polarization in relativistic heavy-ion collisions}",
    eprint = "2402.04540",
    archivePrefix = "arXiv",
    primaryClass = "nucl-th",
    doi = "10.1142/9789811294679_0005",
    journal = "Int. J. Mod. Phys. E",
    volume = "33",
    number = "06",
    pages = "2430006",
    year = "2024"
}

@article{Shi:2020htn,
    author = "Shi, Shuzhe and Gale, Charles and Jeon, Sangyong",
    title = "{From chiral kinetic theory to relativistic viscous spin hydrodynamics}",
    eprint = "2008.08618",
    archivePrefix = "arXiv",
    primaryClass = "nucl-th",
    doi = "10.1103/PhysRevC.103.044906",
    journal = "Phys. Rev. C",
    volume = "103",
    number = "4",
    pages = "044906",
    year = "2021"
}

@book{Peskin:1995ev,
    author = "Peskin, Michael E. and Schroeder, Daniel V.",
    title = "{An Introduction to quantum field theory}",
    doi = "10.1201/9780429503559",
    isbn = "978-0-201-50397-5, 978-0-429-50355-9, 978-0-429-49417-8",
    publisher = "Addison-Wesley",
    address = "Reading, USA",
    year = "1995"
}

@article{Akkelin:2018hpu,
    author = "Akkelin, S. V.",
    title = "{Quasi equilibrium state of expanding quantum fields and two-pion Bose-Einstein correlations in $pp$ collisions at the LHC}",
    eprint = "1812.03905",
    archivePrefix = "arXiv",
    primaryClass = "hep-ph",
    doi = "10.1140/epja/i2019-12755-9",
    journal = "Eur. Phys. J. A",
    volume = "55",
    number = "5",
    pages = "78",
    year = "2019"
}

@article{Rindori:2021quq,
    author = "Rindori, D. and Tinti, L. and Becattini, F. and Rischke, D. H.",
    title = "{Relativistic quantum fluid with boost invariance}",
    eprint = "2102.09016",
    archivePrefix = "arXiv",
    primaryClass = "hep-th",
    doi = "10.1103/PhysRevD.105.056003",
    journal = "Phys. Rev. D",
    volume = "105",
    number = "5",
    pages = "056003",
    year = "2022"
}

@article{Akkelin:2020cfs,
    author = "Akkelin, S. V.",
    title = "{Cosmological particle creation in the little bang}",
    eprint = "2008.13606",
    archivePrefix = "arXiv",
    primaryClass = "hep-ph",
    doi = "10.1103/PhysRevD.103.116014",
    journal = "Phys. Rev. D",
    volume = "103",
    number = "11",
    pages = "116014",
    year = "2021"
}

@article{Arcuri:1994kd,
    author = "Arcuri, R. C. and Svaiter, N. F. and Svaiter, B. F.",
    title = "{Is the Milne coordinate system a good one?}",
    doi = "10.1142/S0217732394000046",
    journal = "Mod. Phys. Lett. A",
    volume = "9",
    pages = "19--27",
    year = "1994"
}

@article{Padmanabhan:1990fm,
    author = "Padmanabhan, T.",
    title = "{Physical interpretation of quantum field theory in noninertial coordinate systems}",
    doi = "10.1103/PhysRevLett.64.2471",
    journal = "Phys. Rev. Lett.",
    volume = "64",
    pages = "2471--2474",
    year = "1990"
}

@book{Fulling_1989, place={Cambridge}, series={London Mathematical Society Student Texts}, title={Aspects of Quantum Field Theory in Curved Spacetime}, publisher={Cambridge University Press}, author={Fulling, Stephen A.}, year={1989}, collection={London Mathematical Society Student Texts}}

@book{gradshteyn2007,
  author = {Gradshteyn, I. S. and Ryzhik, I. M.},
  edition = {Seventh},
  isbn = {978-0-12-373637-6; 0-12-373637-4},
  mrclass = {00A22 (33-00 65-00 65A05)},
  mrnumber = {2360010 (2008g:00005)},
  pages = {xlviii+1171},
  publisher = {Elsevier/Academic Press, Amsterdam},
  title = {Table of integrals, series, and products},
  year = 2007
}

@misc{NIST:DLMF,
         key = "{\relax DLMF}",
       title = "{\it NIST Digital Library of Mathematical Functions}",
howpublished = "\url{https://dlmf.nist.gov/}, Release 1.2.4 of 2025-03-15",
         url = "https://dlmf.nist.gov/",
        note = "F.~W.~J. Olver, A.~B. {Olde Daalhuis}, D.~W. Lozier, B.~I. Schneider,
                R.~F. Boisvert, C.~W. Clark, B.~R. Miller, B.~V. Saunders,
                H.~S. Cohl, and M.~A. McClain, eds."}

@article{Becattini:2020sww,
    author = "Becattini, Francesco",
    title = "{Polarization in Relativistic Fluids: A Quantum Field Theoretical Derivation}",
    eprint = "2004.04050",
    archivePrefix = "arXiv",
    primaryClass = "hep-th",
    doi = "10.1007/978-3-030-71427-7_2",
    journal = "Lect. Notes Phys.",
    volume = "987",
    pages = "15--52",
    year = "2021"
}

@book{Collas:2018jfx,
    author = "Collas, Peter and Klein, David",
    title = "{The Dirac Equation in Curved Spacetime}: {A Guide for Calculations}",
    eprint = "1809.02764",
    archivePrefix = "arXiv",
    primaryClass = "gr-qc",
    doi = "10.1007/978-3-030-14825-6",
    isbn = "978-3-030-14825-6",
    publisher = "Springer",
    series = "SpringerBriefs in Physics",
    year = "2019"
}

@article{Becattini:2019dxo,
    author = "Becattini, F. and Buzzegoli, M. and Grossi, E.",
    title = "{Reworking the Zubarev's approach to non-equilibrium quantum statistical mechanics}",
    eprint = "1902.01089",
    archivePrefix = "arXiv",
    primaryClass = "cond-mat.stat-mech",
    doi = "10.3390/particles2020014",
    journal = "Particles",
    volume = "2",
    number = "2",
    pages = "197--207",
    year = "2019"
}

@article{Bjorken:1982qr,
    author = "Bjorken, J. D.",
    title = "{Highly Relativistic Nucleus-Nucleus Collisions: The Central Rapidity Region}",
    reportNumber = "FERMILAB-PUB-82-044-THY, FERMILAB-PUB-82-044-T",
    doi = "10.1103/PhysRevD.27.140",
    journal = "Phys. Rev. D",
    volume = "27",
    pages = "140--151",
    year = "1983"
}

@article{Becattini:2024vtf,
    author = "Becattini, F. and Roselli, D.",
    title = "{Negative pressure as a quantum effect in free-streaming in the cosmological background}",
    eprint = "2403.08661",
    archivePrefix = "arXiv",
    primaryClass = "gr-qc",
    doi = "10.1103/PhysRevD.111.085020",
    journal = "Phys. Rev. D",
    volume = "111",
    number = "8",
    pages = "085020",
    year = "2025"
}

@article{Becattini:2022bia,
    author = "Becattini, F. and Roselli, D.",
    title = "{Quantum field corrections to the equation of state of freely streaming matter in the Friedman{\textendash}Lema{\^\i}tre{\textendash}Robertson{\textendash}Walker space-time}",
    eprint = "2212.05518",
    archivePrefix = "arXiv",
    primaryClass = "gr-qc",
    doi = "10.1088/1361-6382/ace495",
    journal = "Class. Quant. Grav.",
    volume = "40",
    number = "17",
    pages = "175007",
    year = "2023"
}

@article{Tinti:2023mtv,
    author = "Tinti, Leonardo",
    title = "{Quantum free-streaming: Out of equilibrium expansion for the free scalar fields}",
    eprint = "2304.00109",
    archivePrefix = "arXiv",
    primaryClass = "hep-ph",
    doi = "10.1103/PhysRevD.108.076022",
    journal = "Phys. Rev. D",
    volume = "108",
    number = "7",
    pages = "076022",
    year = "2023"
}

@article{Palermo:2023cup,
    author = "Palermo, Andrea and Becattini, Francesco",
    title = "{Exact spin polarization of massive and massless particles in relativistic fluids at global equilibrium}",
    eprint = "2304.02276",
    archivePrefix = "arXiv",
    primaryClass = "nucl-th",
    doi = "10.1140/epjp/s13360-023-04169-w",
    journal = "Eur. Phys. J. Plus",
    volume = "138",
    number = "6",
    pages = "547",
    year = "2023"
}

@article{Becattini:2021suc,
    author = "Becattini, F. and Buzzegoli, M. and Palermo, A.",
    title = "{Spin-thermal shear coupling in a relativistic fluid}",
    eprint = "2103.10917",
    archivePrefix = "arXiv",
    primaryClass = "nucl-th",
    doi = "10.1016/j.physletb.2021.136519",
    journal = "Phys. Lett. B",
    volume = "820",
    pages = "136519",
    year = "2021"
}

@article{Sheng:2025cjk,
    author = "Sheng, Xin-Li and Becattini, Francesco and Roselli, Daniele",
    title = "{An improved formula for Wigner function and spin polarization in a decoupling relativistic fluid at local thermodynamic equilibrium}",
    eprint = "2509.14301",
    archivePrefix = "arXiv",
    primaryClass = "nucl-th",
    month = "9",
    year = "2025"
}

@article{Liu:2021uhn,
    author = "Liu, Shuai Y. F. and Yin, Yi",
    title = "{Spin polarization induced by the hydrodynamic gradients}",
    eprint = "2103.09200",
    archivePrefix = "arXiv",
    primaryClass = "hep-ph",
    doi = "10.1007/JHEP07(2021)188",
    journal = "JHEP",
    volume = "07",
    pages = "188",
    year = "2021"
}

@book{BecattiniBOOK,
    editor = "Becattini, Francesco and Liao, Jinfeng and Lisa, Michael Annan",
    title = "{Strongly Interacting Matter under Rotation}",
    doi = "10.1007/978-3-030-71427-7",
    publisher = "Springer",
    series = "Lecture Notes in Physics",
    volume = "987",
    year = "2021"
}

@article{Becattini:2013fla,
    author = "Becattini, F. and Chandra, V. and Del Zanna, L. and Grossi, E.",
    title = "{Relativistic distribution function for particles with spin at local thermodynamical equilibrium}",
    eprint = "1303.3431",
    archivePrefix = "arXiv",
    primaryClass = "nucl-th",
    doi = "10.1016/j.aop.2013.07.004",
    journal = "Annals Phys.",
    volume = "338",
    pages = "32--49",
    year = "2013"
}

@article{Becattini:2025oyi,
    author = "Becattini, Francesco and Singh, Rajeev",
    title = "{On the local thermodynamic relations in relativistic spin hydrodynamics}",
    eprint = "2506.20681",
    archivePrefix = "arXiv",
    primaryClass = "nucl-th",
    doi = "10.1140/epjc/s10052-025-15071-3",
    journal = "Eur. Phys. J. C",
    volume = "85",
    number = "11",
    pages = "1338",
    year = "2025"
}

@article{Florkowski:2024bfw,
    author = "Florkowski, Wojciech and Hontarenko, Mykhailo",
    title = "{Generalized Thermodynamic Relations for Perfect Spin Hydrodynamics}",
    eprint = "2405.03263",
    archivePrefix = "arXiv",
    primaryClass = "hep-ph",
    doi = "10.1103/PhysRevLett.134.082302",
    journal = "Phys. Rev. Lett.",
    volume = "134",
    number = "8",
    pages = "082302",
    year = "2025"
}

@article{Drogosz:2024gzv,
    author = "Drogosz, Zbigniew and Florkowski, Wojciech and Hontarenko, Mykhailo",
    title = "{Hybrid approach to perfect and dissipative spin hydrodynamics}",
    eprint = "2408.03106",
    archivePrefix = "arXiv",
    primaryClass = "hep-ph",
    doi = "10.1103/PhysRevD.110.096018",
    journal = "Phys. Rev. D",
    volume = "110",
    number = "9",
    pages = "096018",
    year = "2024"
}

@article{STAR:2017ckg,
    author = "STAR Collaboration",
    title = "{Global $\Lambda$ hyperon polarization in nuclear collisions}",
    journal = "Nature",
    volume = "548",
    pages = "62",
    year = "2017"
}

@article{STAR:2018gyt,
    author = "STAR Collaboration",
    title = "{Global polarization of $\Lambda$ hyperons at RHIC}",
    journal = "Phys. Rev. C",
    volume = "98",
    pages = "014910",
    year = "2018"
}

@article{Becattini:2020ngo,
    author = "Becattini, Francesco and Lisa, Michael A.",
    title = "{Polarization and Vorticity in the Quark{\textendash}Gluon Plasma}",
    eprint = "2003.03640",
    archivePrefix = "arXiv",
    primaryClass = "nucl-ex",
    doi = "10.1146/annurev-nucl-021920-095245",
    journal = "Ann. Rev. Nucl. Part. Sci.",
    volume = "70",
    pages = "395--423",
    year = "2020"
}

@article{Hattori:2019lfp,
    author = "Hattori, Koichi and Hongo, Masaru and Huang, Xu-Guang and Matsuo, Mamoru and Taya, Hidetoshi",
    title = "{Fate of spin polarization in a relativistic fluid: An entropy-current analysis}",
    eprint = "1901.06615",
    archivePrefix = "arXiv",
    primaryClass = "hep-th",
    reportNumber = "RIKEN-iTHEMS-Report-19, YITP-19-15",
    doi = "10.1016/j.physletb.2019.05.040",
    journal = "Phys. Lett. B",
    volume = "795",
    pages = "100--106",
    year = "2019"
}

@article{Bhadury:2020puc,
    author = "Bhadury, Samapan and Florkowski, Wojciech and Jaiswal, Amaresh and Kumar, Avdhesh and Ryblewski, Radoslaw",
    title = "{Relativistic dissipative spin dynamics in the relaxation time approximation}",
    eprint = "2002.03937",
    archivePrefix = "arXiv",
    primaryClass = "hep-ph",
    doi = "10.1016/j.physletb.2021.136096",
    journal = "Phys. Lett. B",
    volume = "814",
    pages = "136096",
    year = "2021"
}

@article{Florkowski:2017ruc,
    author = "Florkowski, Wojciech and Friman, Bengt and Jaiswal, Amaresh and Speranza, Enrico",
    title = "{Relativistic fluid dynamics with spin}",
    eprint = "1705.00587",
    archivePrefix = "arXiv",
    primaryClass = "nucl-th",
    doi = "10.1103/PhysRevC.97.041901",
    journal = "Phys. Rev. C",
    volume = "97",
    number = "4",
    pages = "041901",
    year = "2018"
}

@article{Florkowski:2018fap,
    author = "Florkowski, Wojciech and Kumar, Avdhesh and Ryblewski, Radoslaw",
    title = "{Relativistic hydrodynamics for spin-polarized fluids}",
    eprint = "1811.04409",
    archivePrefix = "arXiv",
    primaryClass = "nucl-th",
    doi = "10.1016/j.ppnp.2019.07.001",
    journal = "Prog. Part. Nucl. Phys.",
    volume = "108",
    pages = "103709",
    year = "2019"
}

@article{Florkowski:2019qdp,
    author = "Florkowski, Wojciech and Kumar, Avdhesh and Ryblewski, Radoslaw and Singh, Rajeev",
    title = "{Spin polarization evolution in a boost invariant hydrodynamical background}",
    eprint = "1901.09655",
    archivePrefix = "arXiv",
    primaryClass = "hep-ph",
    doi = "10.1103/PhysRevC.99.044910",
    journal = "Phys. Rev. C",
    volume = "99",
    number = "4",
    pages = "044910",
    year = "2019"
}

@article{Florkowski:2019voj,
    author = "Florkowski, Wojciech and Kumar, Avdhesh and Ryblewski, Radoslaw and Mazeliauskas, Aleksas",
    title = "{Longitudinal spin polarization in a thermal model}",
    eprint = "1904.00002",
    archivePrefix = "arXiv",
    primaryClass = "nucl-th",
    doi = "10.1103/PhysRevC.100.054907",
    journal = "Phys. Rev. C",
    volume = "100",
    number = "5",
    pages = "054907",
    year = "2019"
}

@article{Gallegos:2021bzp,
    author = {Gallegos, A. D. and G{\"u}rsoy, U. and Yarom, A.},
    title = "{Hydrodynamics of spin currents}",
    eprint = "2101.04759",
    archivePrefix = "arXiv",
    primaryClass = "hep-th",
    doi = "10.21468/SciPostPhys.11.2.041",
    journal = "SciPost Phys.",
    volume = "11",
    pages = "041",
    year = "2021"
}

@article{Becattini:2018duy,
    author = "Becattini, F. and others",
    title = "{Spin tensor and its role in non-equilibrium thermodynamics}",
    journal = "Phys. Rev. Lett.",
    volume = "120",
    pages = "012302",
    year = "2018"
}

@article{Buzzegoli:2021wlg,
    author = "Buzzegoli, M.",
    title = "{Pseudogauge dependence of the spin polarization and of the axial vortical effect}",
    eprint = "2109.12084",
    archivePrefix = "arXiv",
    primaryClass = "nucl-th",
    doi = "10.1103/PhysRevC.105.044907",
    journal = "Phys. Rev. C",
    volume = "105",
    number = "4",
    pages = "044907",
    year = "2022"
}

@article{Li:2020eon,
    author = "Li, Shiyong and Stephanov, Mikhail A. and Yee, Ho-Ung",
    title = "{Nondissipative Second-Order Transport, Spin, and Pseudogauge Transformations in Hydrodynamics}",
    eprint = "2011.12318",
    archivePrefix = "arXiv",
    primaryClass = "hep-th",
    doi = "10.1103/PhysRevLett.127.082302",
    journal = "Phys. Rev. Lett.",
    volume = "127",
    number = "8",
    pages = "082302",
    year = "2021"
}

@article{Fukushima:2020ucl,
    author = "Fukushima, Kenji and Pu, Shi",
    title = "{Spin hydrodynamics and symmetric energy-momentum tensors {\textendash} A current induced by the spin vorticity {\textendash}}",
    eprint = "2010.01608",
    archivePrefix = "arXiv",
    primaryClass = "hep-th",
    doi = "10.1016/j.physletb.2021.136346",
    journal = "Phys. Lett. B",
    volume = "817",
    pages = "136346",
    year = "2021"
}

@article{Drogosz:2024rbd,
    author = "Drogosz, Zbigniew and Florkowski, Wojciech and Hontarenko, Mykhailo and Ryblewski, Radoslaw",
    title = "{Dynamical constraints on pseudo-gauge transformations}",
    eprint = "2411.06249",
    archivePrefix = "arXiv",
    primaryClass = "hep-ph",
    doi = "10.1016/j.physletb.2025.139244",
    journal = "Phys. Lett. B",
    volume = "861",
    pages = "139244",
    year = "2025"
}

@article{Crispino:2007eb,
    author = "Crispino, Luis C. B. and Higuchi, Atsushi and Matsas, George E. A.",
    title = "{The Unruh effect and its applications}",
    eprint = "0710.5373",
    archivePrefix = "arXiv",
    primaryClass = "gr-qc",
    doi = "10.1103/RevModPhys.80.787",
    journal = "Rev. Mod. Phys.",
    volume = "80",
    pages = "787--838",
    year = "2008"
}

@article{Karpenko:2016jyx,
    author = "Karpenko, I. and Becattini, F.",
    title = "{Study of $\Lambda $ polarization in relativistic nuclear collisions at $\sqrt{s_\mathrm {NN}}=7.7$ {\textendash}200 GeV}",
    eprint = "1610.04717",
    archivePrefix = "arXiv",
    primaryClass = "nucl-th",
    doi = "10.1140/epjc/s10052-017-4765-1",
    journal = "Eur. Phys. J. C",
    volume = "77",
    number = "4",
    pages = "213",
    year = "2017"
}

@article{Li:2017slc,
    author = "Li, Hui and Pang, Long-Gang and Wang, Qun and Xia, Xiao-Liang",
    title = "{Global $\Lambda$ polarization in heavy-ion collisions from a transport model}",
    eprint = "1704.01507",
    archivePrefix = "arXiv",
    primaryClass = "nucl-th",
    doi = "10.1103/PhysRevC.96.054908",
    journal = "Phys. Rev. C",
    volume = "96",
    number = "5",
    pages = "054908",
    year = "2017"
}

@article{Xie:2017upb,
    author = "Xie, Yilong and Wang, Dujuan and Csernai, L{\'a}szl{\'o} P.",
    title = "{Global {\ensuremath{\Lambda}} polarization in high energy collisions}",
    eprint = "1703.03770",
    archivePrefix = "arXiv",
    primaryClass = "nucl-th",
    doi = "10.1103/PhysRevC.95.031901",
    journal = "Phys. Rev. C",
    volume = "95",
    number = "3",
    pages = "031901",
    year = "2017"
}

@article{Bagchi:2023ysc,
    author = "Bagchi, Arjun and Kolekar, Kedar S. and Shukla, Ashish",
    title = "{Carrollian Origins of Bjorken Flow}",
    eprint = "2302.03053",
    archivePrefix = "arXiv",
    primaryClass = "hep-th",
    reportNumber = "CPHT079.122022",
    doi = "10.1103/PhysRevLett.130.241601",
    journal = "Phys. Rev. Lett.",
    volume = "130",
    number = "24",
    pages = "241601",
    year = "2023"
}

@article{Xia:2018tes,
    author = "Xia, Xiao-Liang and Li, Hui and Tang, Ze-Bo and Wang, Qun",
    title = "{Probing vorticity structure in heavy-ion collisions by local $\Lambda$ polarization}",
    eprint = "1803.00867",
    archivePrefix = "arXiv",
    primaryClass = "nucl-th",
    doi = "10.1103/PhysRevC.98.024905",
    journal = "Phys. Rev. C",
    volume = "98",
    pages = "024905",
    year = "2018"
}

@article{Wu:2019eyi,
    author = "Wu, Hong-Zhong and Pang, Long-Gang and Huang, Xu-Guang and Wang, Qun",
    title = "{Local spin polarization in high energy heavy ion collisions}",
    eprint = "1906.09385",
    archivePrefix = "arXiv",
    primaryClass = "nucl-th",
    doi = "10.1103/PhysRevResearch.1.033058",
    journal = "Phys. Rev. Research.",
    volume = "1",
    pages = "033058",
    year = "2019"
}

@article{Becattini:2019ntv,
    author = "Becattini, Francesco and Cao, Gaoqing and Speranza, Enrico",
    title = "{Polarization transfer in hyperon decays and its effect in relativistic nuclear collisions}",
    eprint = "1905.03123",
    archivePrefix = "arXiv",
    primaryClass = "nucl-th",
    doi = "10.1140/epjc/s10052-019-7213-6",
    journal = "Eur. Phys. J. C",
    volume = "79",
    number = "9",
    pages = "741",
    year = "2019"
}

@article{Hongo:2021,
  author = {Hongo, M. and Huang, X.-G. and Kaminski, M. and Stephanov, M. and Yee, H.-U.},
  title = {Relativistic spin hydrodynamics with torsion and linear response theory for spin relaxation},
  journal = {JHEP},
  volume = {11},
  year = {2021},
  pages = {150},
  eprint = {2107.14231},
  archivePrefix = {arXiv},
  primaryClass = {hep-th}
}

@article{Peng:2021ago,
    author = "Peng, Hao-Hao and Zhang, Jun-Jie and Sheng, Xin-Li and Wang, Qun",
    title = "{Ideal Spin Hydrodynamics from the Wigner Function Approach}",
    eprint = "2107.00448",
    archivePrefix = "arXiv",
    primaryClass = "hep-th",
    doi = "10.1088/0256-307X/38/11/116701",
    journal = "Chin. Phys. Lett.",
    volume = "38",
    number = "11",
    pages = "116701",
    year = "2021"
}

@article{Singh:2020rht,
    author = "Singh, Rajeev and Sophys, Gabriel and Ryblewski, Radoslaw",
    title = "{Spin polarization dynamics in the Gubser-expanding background}",
    eprint = "2011.14907",
    archivePrefix = "arXiv",
    primaryClass = "hep-ph",
    doi = "10.1103/PhysRevD.103.074024",
    journal = "Phys. Rev. D",
    volume = "103",
    number = "7",
    pages = "074024",
    year = "2021"
}

@article{Chiarini:2024cuv,
    author = "Chiarini, Annamaria and Sammet, Julia and Shokri, Masoud",
    title = "{Semiclassical spin hydrodynamics in flat and curved spacetime: Covariance, linear waves, and Bjorken background}",
    eprint = "2412.19854",
    archivePrefix = "arXiv",
    primaryClass = "gr-qc",
    doi = "10.1103/9tmd-rnp1",
    journal = "Phys. Rev. D",
    volume = "113",
    number = "3",
    pages = "036009",
    year = "2026"
}

@article{Huang:2024ffg,
    author = "Huang, Xu-Guang",
    title = "{An introduction to relativistic spin hydrodynamics}",
    eprint = "2411.11753",
    archivePrefix = "arXiv",
    primaryClass = "nucl-th",
    doi = "10.1007/s41365-025-01784-3",
    journal = "Nucl. Sci. Tech.",
    volume = "36",
    number = "11",
    pages = "208",
    year = "2025"
}

@article{Olver:1954wc,
    author = "Olver, F. W. J.",
    title = "{On Bessel functions of large order}",
    doi = "10.1098/rsta.1954.0021",
    journal = "Phil. Trans. Roy. Soc. Lond. A",
    volume = "247",
    pages = "328--368",
    year = "1954"
}

@article{Fu:2022myl,
    author = "Fu, Baochi and Pang, Longgang and Song, Huichao and Yin, Yi",
    title = "{Signatures of the spin Hall effect in hot and dense QCD matter}",
    eprint = "2201.12970",
    archivePrefix = "arXiv",
    primaryClass = "hep-ph",
    month = "1",
    year = "2022"
}

@article{Li:2025pef,
    author = "Li, Youyu and Liu, Shuai Y. F.",
    title = "{Zubarev response approach to polarization phenomena in local equilibrium}",
    eprint = "2501.17861",
    archivePrefix = "arXiv",
    primaryClass = "nucl-th",
    doi = "10.1103/8klr-kq9k",
    journal = "Phys. Rev. C",
    volume = "113",
    number = "3",
    pages = "034911",
    year = "2026"
}


\end{document}